%% file: Streufert-dp.tex
\pdfoutput=1
\providecommand{\forpreamble}{ 
\newlabel{B149}{{{{[i]}}}{35}{}{Item.67}{}}
\newlabel{B150}{{{{[ii]}}}{35}{}{Item.68}{}}
\newlabel{B224}{{2.3}{5}{\small \capp }{figure.2.3}{}}
\newlabel{B225}{{2.2}{5}{\small \capp }{figure.2.2}{}}
\newlabel{B234}{{1}{1}{Introduction}{section.1}{}}
\newlabel{B250}{{2.1}{4}{\small \capp }{figure.2.1}{}}
\newlabel{B265}{{3.1}{14}{\small \capp }{figure.3.1}{}}
\newlabel{B266}{{3.2}{14}{\small \capp }{figure.3.2}{}}
\newlabel{B292}{{6}{4}{}{Hfootnote.6}{}}
\newlabel{B331}{{D.2}{53}{\bf implies Theorem~\rf {C691}}{lemma.D.2}{}}
\newlabel{B359}{{12}{11}{}{Hfootnote.12}{}}
\newlabel{B405}{{D.1}{50}{\bf implies Theorem~\rf {C689}}{lemma.D.1}{}}
\newlabel{B406}{{{{a}}}{50}{\bf implies Theorem~\rf {C689}}{Item.149}{}}
\newlabel{B407}{{8}{51}{}{Item.177}{}}
\newlabel{B408}{{9}{51}{}{Item.178}{}}
\newlabel{B418}{{6}{50}{}{Item.170}{}}
\newlabel{B419}{{{{a}}}{51}{}{Item.172}{}}
\newlabel{B420}{{{{b}}}{51}{}{Item.173}{}}
\newlabel{B421}{{{{c}}}{51}{}{Item.174}{}}
\newlabel{B422}{{{{d}}}{51}{}{Item.175}{}}
\newlabel{B423}{{{{e}}}{51}{}{Item.176}{}}
\newlabel{B424}{{4}{50}{}{Item.168}{}}
\newlabel{B426}{{1}{50}{}{Item.165}{}}
\newlabel{B427}{{3}{50}{}{Item.167}{}}
\newlabel{B428}{{5}{50}{}{Item.169}{}}
\newlabel{B462}{{D.4}{55}{}{lemma.D.4}{}}
\newlabel{B463}{{D.3}{55}{}{lemma.D.3}{}}
\newlabel{B475}{{2}{56}{}{Item.219}{}}
\newlabel{B476}{{3}{56}{}{Item.225}{}}
\newlabel{B481}{{{{[a]}}}{56}{}{Item.223}{}}
\newlabel{B482}{{{{[i]}}}{56}{}{Item.221}{}}
\newlabel{B489}{{1}{55}{}{Item.210}{}}
\newlabel{B493}{{2}{55}{}{Item.211}{}}
\newlabel{B494}{{6}{55}{}{Item.215}{}}
\newlabel{B495}{{4}{55}{}{Item.213}{}}
\newlabel{B496}{{5}{55}{}{Item.214}{}}
\newlabel{B4Lp}{{24}{53}{}{Item.193}{}}
\newlabel{B4L}{{{{o}}}{50}{\bf implies Theorem~\rf {C689}}{Item.163}{}}
\newlabel{B4Mp}{{25}{53}{}{Item.194}{}}
\newlabel{B4M}{{{{p}}}{50}{\bf implies Theorem~\rf {C689}}{Item.164}{}}
\newlabel{B4a}{{{{e}}}{50}{\bf implies Theorem~\rf {C689}}{Item.153}{}}
\newlabel{B4e}{{{{i}}}{50}{\bf implies Theorem~\rf {C689}}{Item.157}{}}
\newlabel{B4f}{{{{g}}}{50}{\bf implies Theorem~\rf {C689}}{Item.155}{}}
\newlabel{B4h}{{{{c}}}{50}{\bf implies Theorem~\rf {C689}}{Item.151}{}}
\newlabel{B4i}{{{{b}}}{50}{\bf implies Theorem~\rf {C689}}{Item.150}{}}
\newlabel{B4pp}{{{{k}}}{50}{\bf implies Theorem~\rf {C689}}{Item.159}{}}
\newlabel{B4r}{{{{j}}}{50}{\bf implies Theorem~\rf {C689}}{Item.158}{}}
\newlabel{B4sp}{{{{m}}}{50}{\bf implies Theorem~\rf {C689}}{Item.161}{}}
\newlabel{B4wp}{{{{${ \ell }$}}}{50}{\bf implies Theorem~\rf {C689}}{Item.160}{}}
\newlabel{B4w}{{{{d}}}{50}{\bf implies Theorem~\rf {C689}}{Item.152}{}}
\newlabel{B4x}{{{{h}}}{50}{\bf implies Theorem~\rf {C689}}{Item.156}{}}
\newlabel{B4y}{{{{f}}}{50}{\bf implies Theorem~\rf {C689}}{Item.154}{}}
\newlabel{B4z}{{{{n}}}{50}{\bf implies Theorem~\rf {C689}}{Item.162}{}}
\newlabel{B503}{{3}{55}{}{Item.212}{}}
\newlabel{B552}{{{{[e]}}}{56}{}{Item.224}{}}
\newlabel{B560p}{{D.5}{56}{\bf for Theorem~\rf {B560}}{lemma.D.5}{}}
\newlabel{B560}{{5.4}{26}{\bf Main theorem}{lemma.5.4}{}}
\newlabel{B561}{{3}{6}{Pentaform Games}{section.3}{}}
\newlabel{B562}{{{\normalfont  3.6}}{13}{Pentaform Games}{subsection.3.6}{}}
\newlabel{B565}{{{\normalfont  5.2}}{24}{``Pentaforming'' a \ct {Gm} game}{subsection.5.2}{}}
\newlabel{B566}{{C}{40}{For Pentaform Games}{appendix.C}{}}
\newlabel{B567}{{D}{50}{For Equivalence with \ct {Gm} Games}{appendix.D}{}}
\newlabel{B570}{{{\normalfont  3.4}}{9}{Pentaforms}{subsection.3.4}{}}
\newlabel{B572}{{{\normalfont  1.1}}{1}{New concepts and main result}{subsection.1.1}{}}
\newlabel{B573}{{{\normalfont  1.2}}{2}{Motivation}{subsection.1.2}{}}
\newlabel{B574}{{{\normalfont  1.3}}{3}{Literature}{subsection.1.3}{}}
\newlabel{B575}{{{\normalfont  1.4}}{3}{Organization}{subsection.1.4}{}}
\newlabel{B598}{{2}{1}{}{Hfootnote.2}{}}
\newlabel{B977}{{2}{4}{Initial Intuition}{section.2}{}}
\newlabel{C273}{{10}{9}{}{Hfootnote.10}{}}
\newlabel{C366}{{C.8}{44}{}{lemma.C.8}{}}
\newlabel{C372p}{{C.7}{44}{\bf for Proposition~\rf {C372}}{lemma.C.7}{}}
\newlabel{C372}{{4.1}{15}{}{lemma.4.1}{}}
\newlabel{C457}{{a}{44}{}{Item.108}{}}
\newlabel{C458}{{b}{44}{}{Item.109}{}}
\newlabel{C492}{{3.1}{10}{\small A pentaform is implicitly accompanied by its derivatives (\protect \rotatebox [origin=c]{180}{$\Lsh $}). Definitions are in the sections in brackets {\tiny [˙]}}{table.3.1}{}}
\newlabel{C519}{{c}{44}{}{Item.110}{}}
\newlabel{C550}{{4.2}{20}{The definition (\rf {D177}) of perfect-recall assumes the solid lines and then requires the dashed lines}{figure.4.2}{}}
\newlabel{C563}{{4}{14}{Some Pentaform Tools and Applications}{section.4}{}}
\newlabel{C564}{{{\normalfont  4.2}}{16}{Unions of blocks}{subsection.4.2}{}}
\newlabel{C576p}{{C.10}{45}{\bf for Proposition~\rf {C576}}{lemma.C.10}{}}
\newlabel{C576}{{4.3}{16}{}{lemma.4.3}{}}
\newlabel{C577}{{{\normalfont  4.1}}{14}{Subsets of Pentaforms}{subsection.4.1}{}}
\newlabel{C578}{{{{b}}}{15}{}{Item.11}{}}
\newlabel{C579}{{4.2}{15}{}{lemma.4.2}{}}
\newlabel{C585}{{C.9}{45}{}{lemma.C.9}{}}
\newlabel{C587}{{2}{45}{\bf for Proposition~\rf {C576}}{Item.113}{}}
\newlabel{C590}{{1}{45}{\bf for Proposition~\rf {C576}}{Item.112}{}}
\newlabel{C591}{{3}{46}{}{Item.119}{}}
\newlabel{C593}{{1}{45}{}{Item.114}{}}
\newlabel{C594}{{{\normalfont  3.1}}{6}{The components of quintuples}{subsection.3.1}{}}
\newlabel{C595}{{{\normalfont  3.2}}{7}{Quintuple sets and their slices}{subsection.3.2}{}}
\newlabel{C596}{{{\normalfont  3.3}}{8}{Projections}{subsection.3.3}{}}
\newlabel{C601p}{{C.4}{41}{\bf for Proposition~\rf {C601}}{lemma.C.4}{}}
\newlabel{C601}{{3.3}{11}{}{lemma.3.3}{}}
\newlabel{C602}{{a}{11}{}{Item.4}{}}
\newlabel{C603}{{c}{11}{}{Item.6}{}}
\newlabel{C604}{{d}{11}{}{Item.7}{}}
\newlabel{C605}{{b}{11}{}{Item.5}{}}
\newlabel{C614}{{{\normalfont  4.3}}{19}{Perfect-recall in terms of pentaforms}{subsection.4.3}{}}
\newlabel{C628}{{{\normalfont  B.2}}{29}{Definition of out-trees}{subsection.B.2}{}}
\newlabel{C629}{{{\normalfont  B.5}}{36}{Edge-trees}{subsection.B.5}{}}
\newlabel{C632p}{{C.6}{43}{\bf for Proposition~\rf {C632}}{lemma.C.6}{}}
\newlabel{C632}{{3.4}{13}{}{lemma.3.4}{}}
\newlabel{C636}{{b}{31}{}{Item.41}{}}
\newlabel{C637}{{B.8}{31}{}{lemma.B.8}{}}
\newlabel{C638}{{a}{32}{}{Item.44}{}}
\newlabel{C639}{{b}{32}{}{Item.45}{}}
\newlabel{C641}{{{{$[$a$]$}}}{39}{}{Item.89}{}}
\newlabel{C642}{{{{$[$b$]$}}}{39}{}{Item.90}{}}
\newlabel{C643}{{{{$[$c$]$}}}{39}{}{Item.91}{}}
\newlabel{C644}{{9}{39}{}{Item.88}{}}
\newlabel{C645}{{2}{37}{}{Item.75}{}}
\newlabel{C646}{{10}{40}{}{Item.92}{}}
\newlabel{C649}{{31}{33}{}{Hfootnote.31}{}}
\newlabel{C654}{{4}{38}{}{Item.77}{}}
\newlabel{C655}{{B.15}{37}{}{lemma.B.15}{}}
\newlabel{C656}{{B}{28}{Rooted Trees, Out-trees, and Edge-trees}{appendix.B}{}}
\newlabel{C657}{{{\normalfont  3.5}}{13}{Out-trees}{subsection.3.5}{}}
\newlabel{C659}{{A}{27}{The Category {\bf Gm}}{appendix.A}{}}
\newlabel{C660}{{a}{31}{}{Item.40}{}}
\newlabel{C661}{{c}{31}{}{Item.42}{}}
\newlabel{C663}{{B.1}{32}{The conclusions of Lemmas~\rf {C637} and \rf {C796}(\rf {C638}). The points depict nodes}{figure.B.1}{}}
\newlabel{C664}{{d}{31}{}{Item.43}{}}
\newlabel{C665}{{{\normalfont  5.1}}{21}{Definition of \ct {Gm} games}{subsection.5.1}{}}
\newlabel{C666}{{15}{13}{}{Hfootnote.15}{}}
\newlabel{C667}{{9}{9}{}{Hfootnote.9}{}}
\newlabel{C668}{{3.5}{14}{\bf Pentaform game}{lemma.3.5}{}}
\newlabel{C669}{{3.1}{9}{\bf Pentaform}{lemma.3.1}{}}
\newlabel{C670}{{5.1}{22}{\small {\em Right-hand Side:} Out-trees and \ct {Gm} games are implicitly accompanied by their derivatives (\protect \rotatebox [origin=c]{180}{$\Lsh $}). Definitions are in the sections in brackets~{\tiny [˙]}. {\em Left-hand Side:} Pentaform equivalents in $(\hq {Q},\hq {u}) = \PB (X,E,\HH ,λ,τ,u)$, from Theorem~\rf {C699} (recall from (\rf {C825}) that $\hq {I}$, $\hq {J}$, $\hq {W}$, $\hq {A}$, and $\hq {Y}$ abbreviate $π_I(\hq {Q})$, $π_J(\hq {Q})$, $π_W(\hq {Q})$, $π_A(\hq {Q})$, and $π_Y(\hq {Q})$)}{table.5.1}{}}
\newlabel{C671}{{21}{21}{}{Hfootnote.21}{}}
\newlabel{C673}{{22}{22}{}{Hfootnote.22}{}}
\newlabel{C674}{{24}{23}{}{Hfootnote.24}{}}
\newlabel{C675}{{25}{24}{}{Hfootnote.25}{}}
\newlabel{C677}{{32}{25}{``Standardizing'' a pentaform game}{equation.5.32}{}}
\newlabel{C678}{{30}{24}{``Pentaforming'' a \ct {Gm} game}{equation.5.30}{}}
\newlabel{C679}{{30a}{24}{``Pentaforming'' a \ct {Gm} game}{equation.5.30a}{}}
\newlabel{C680}{{30b}{24}{``Pentaforming'' a \ct {Gm} game}{equation.5.30b}{}}
\newlabel{C689}{{5.2}{25}{}{lemma.5.2}{}}
\newlabel{C690}{{{\normalfont  5.3}}{25}{``Standardizing'' a pentaform game}{subsection.5.3}{}}
\newlabel{C691}{{5.3}{26}{}{lemma.5.3}{}}
\newlabel{C692}{{{\normalfont  5.4}}{26}{Bijection}{subsection.5.4}{}}
\newlabel{C693}{{32a}{25}{``Standardizing'' a pentaform game}{equation.5.32a}{}}
\newlabel{C694}{{32b}{25}{``Standardizing'' a pentaform game}{equation.5.32b}{}}
\newlabel{C695}{{32c}{25}{``Standardizing'' a pentaform game}{equation.5.32c}{}}
\newlabel{C696}{{32d}{25}{``Standardizing'' a pentaform game}{equation.5.32d}{}}
\newlabel{C697}{{32e}{25}{``Standardizing'' a pentaform game}{equation.5.32e}{}}
\newlabel{C698}{{32f}{25}{``Standardizing'' a pentaform game}{equation.5.32f}{}}
\newlabel{C699p}{{D.6}{56}{\bf for Theorem~\rf {C699}}{lemma.D.6}{}}
\newlabel{C699}{{5.5}{27}{}{lemma.5.5}{}}
\newlabel{C714}{{29}{24}{Definition of \ct {Gm} games}{equation.5.29}{}}
\newlabel{C722}{{2}{50}{}{Item.166}{}}
\newlabel{C724}{{7}{51}{}{Item.171}{}}
\newlabel{C725}{{10}{51}{}{Item.179}{}}
\newlabel{C726}{{13}{51}{}{Item.182}{}}
\newlabel{C727}{{12}{51}{}{Item.181}{}}
\newlabel{C728}{{11}{51}{}{Item.180}{}}
\newlabel{C729}{{14}{52}{}{Item.183}{}}
\newlabel{C730}{{16}{52}{}{Item.185}{}}
\newlabel{C731}{{15}{52}{}{Item.184}{}}
\newlabel{C732}{{17}{52}{}{Item.186}{}}
\newlabel{C733}{{18}{52}{}{Item.187}{}}
\newlabel{C734}{{19}{52}{}{Item.188}{}}
\newlabel{C735}{{23}{53}{}{Item.192}{}}
\newlabel{C736}{{22}{52}{}{Item.191}{}}
\newlabel{C740}{{20}{52}{}{Item.189}{}}
\newlabel{C741}{{21}{52}{}{Item.190}{}}
\newlabel{C744}{{3}{53}{}{Item.197}{}}
\newlabel{C745}{{5}{53}{}{Item.199}{}}
\newlabel{C749}{{1}{53}{}{Item.195}{}}
\newlabel{C750}{{6}{53}{}{Item.200}{}}
\newlabel{C751}{{7}{53}{}{Item.201}{}}
\newlabel{C752}{{8}{54}{}{Item.202}{}}
\newlabel{C753}{{9}{54}{}{Item.203}{}}
\newlabel{C755}{{10}{54}{}{Item.204}{}}
\newlabel{C757}{{13}{55}{}{Item.207}{}}
\newlabel{C758}{{12}{54}{}{Item.206}{}}
\newlabel{C760}{{14}{55}{}{Item.208}{}}
\newlabel{C761}{{15}{55}{}{Item.209}{}}
\newlabel{C763}{{4}{53}{}{Item.198}{}}
\newlabel{C770}{{{{[q]}}}{56}{}{Item.220}{}}
\newlabel{C771}{{{{$[$b$]$}}}{56}{}{Item.218}{}}
\newlabel{C774}{{4}{56}{}{Item.226}{}}
\newlabel{C777}{{5.6}{27}{}{lemma.5.6}{}}
\newlabel{C782}{{1}{6}{Initial Intuition}{equation.2.1}{}}
\newlabel{C783}{{3}{7}{Quintuple sets and their slices}{equation.3.3}{}}
\newlabel{C784}{{3a}{7}{Quintuple sets and their slices}{equation.3.3a}{}}
\newlabel{C785}{{3b}{7}{Quintuple sets and their slices}{equation.3.3b}{}}
\newlabel{C786}{{3c}{7}{Quintuple sets and their slices}{equation.3.3c}{}}
\newlabel{C788}{{4}{8}{Projections}{equation.3.4}{}}
\newlabel{C789}{{5}{8}{Projections}{equation.3.5}{}}
\newlabel{C790}{{6}{8}{Projections}{equation.3.6}{}}
\newlabel{C791}{{7}{8}{Projections}{equation.3.7}{}}
\newlabel{C794}{{9}{8}{Projections}{equation.3.9}{}}
\newlabel{C795}{{10}{8}{Projections}{equation.3.10}{}}
\newlabel{C796}{{B.9}{32}{}{lemma.B.9}{}}
\newlabel{C797}{{{\normalfont  B.3}}{31}{Projections of out-trees}{subsection.B.3}{}}
\newlabel{C798}{{B.11}{34}{}{lemma.B.11}{}}
\newlabel{C7L}{{{{k}}}{27}{}{Item.31}{}}
\newlabel{C7M}{{{{n}}}{27}{}{Item.34}{}}
\newlabel{C7a}{{{{${ \ell }$}}}{27}{}{Item.32}{}}
\newlabel{C7e}{{{{b}}}{27}{}{Item.22}{}}
\newlabel{C7f}{{{{m}}}{27}{}{Item.33}{}}
\newlabel{C7h}{{{{j}}}{27}{}{Item.30}{}}
\newlabel{C7i}{{{{o}}}{27}{}{Item.35}{}}
\newlabel{C7pp}{{{{f}}}{27}{}{Item.26}{}}
\newlabel{C7r}{{{{c}}}{27}{}{Item.23}{}}
\newlabel{C7sp}{{{{h}}}{27}{}{Item.28}{}}
\newlabel{C7u}{{{{p}}}{27}{}{Item.36}{}}
\newlabel{C7wp}{{{{g}}}{27}{}{Item.27}{}}
\newlabel{C7w}{{{{d}}}{27}{}{Item.24}{}}
\newlabel{C7x}{{{{a}}}{27}{}{Item.21}{}}
\newlabel{C7y}{{{{e}}}{27}{}{Item.25}{}}
\newlabel{C7z}{{{{i}}}{27}{}{Item.29}{}}
\newlabel{C801}{{{\normalfont  B.4}}{33}{Paths in out-trees}{subsection.B.4}{}}
\newlabel{C802}{{{{$[$1$]$}}}{37}{}{Item.69}{}}
\newlabel{C803}{{{{$[$2$]$}}}{37}{}{Item.70}{}}
\newlabel{C804}{{{{$[$3$]$}}}{37}{}{Item.71}{}}
\newlabel{C805}{{{{$[$4$]$}}}{37}{}{Item.72}{}}
\newlabel{C806}{{{{$[$5$]$}}}{37}{}{Item.73}{}}
\newlabel{C807}{{11}{40}{}{Item.93}{}}
\newlabel{C808}{{13}{40}{}{Item.96}{}}
\newlabel{C810}{{{{$[$1$]$}}}{33}{}{Item.47}{}}
\newlabel{C811}{{{{$[$2$]$}}}{33}{}{Item.48}{}}
\newlabel{C815}{{B.1}{29}{Diestel 2010, Theorem~1.5.1}{lemma.B.1}{}}
\newlabel{C816}{{B.2}{29}{}{lemma.B.2}{}}
\newlabel{C819}{{{{$[$1$]$}}}{29}{}{Item.37}{}}
\newlabel{C821}{{{{$[$4$]$}}}{33}{}{Item.50}{}}
\newlabel{C822}{{{{$[$5$]$}}}{33}{}{Item.51}{}}
\newlabel{C825}{{8}{8}{Projections}{equation.3.8}{}}
\newlabel{C826}{{5}{21}{Equivalence with \ct {Gm} Games}{section.5}{}}
\newlabel{C827}{{30}{31}{}{Hfootnote.30}{}}
\newlabel{C840}{{B.14}{36}{\bf Edge-tree}{lemma.B.14}{}}
\newlabel{C843}{{5.1}{24}{{\bf Gm game}, Streufert 2021Gm}{lemma.5.1}{}}
\newlabel{C844}{{{{a}}}{41}{}{Item.101}{}}
\newlabel{C847}{{{{c}}}{41}{}{Item.103}{}}
\newlabel{C848}{{{{b}}}{41}{}{Item.102}{}}
\newlabel{C980}{{11}{9}{Pentaforms}{equation.3.11}{}}
\newlabel{C981}{{12}{11}{Pentaforms}{equation.3.12}{}}
\newlabel{C982}{{13}{12}{Pentaforms}{equation.3.13}{}}
\newlabel{C983}{{14}{13}{Out-trees}{equation.3.14}{}}
\newlabel{C984}{{1}{43}{}{Item.104}{}}
\newlabel{C987}{{d}{44}{}{Item.111}{}}
\newlabel{D006}{{4}{2}{}{Hfootnote.4}{}}
\newlabel{D028}{{28}{28}{}{Hfootnote.28}{}}
\newlabel{D041}{{26}{27}{}{Hfootnote.26}{}}
\newlabel{D057}{{29}{29}{}{Hfootnote.29}{}}
\newlabel{D068}{{{{$[$3$]$}}}{33}{}{Item.49}{}}
\newlabel{D072}{{{{$[$b$]$}}}{35}{}{Item.65}{}}
\newlabel{D075}{{{{$[$c$]$}}}{35}{}{Item.66}{}}
\newlabel{D076}{{{{$[$1$]$}}}{35}{}{Item.63}{}}
\newlabel{D077}{{3}{38}{}{Item.76}{}}
\newlabel{D078}{{12}{40}{}{Item.95}{}}
\newlabel{D079}{{40}{37}{}{equation.B.40}{}}
\newlabel{D080}{{39}{37}{}{equation.B.39}{}}
\newlabel{D082}{{5}{38}{}{Item.82}{}}
\newlabel{D083}{{{{$[$a$]$}}}{40}{}{Item.94}{}}
\newlabel{D084}{{6}{38}{}{Item.83}{}}
\newlabel{D085}{{7}{38}{}{Item.84}{}}
\newlabel{D086}{{1}{37}{}{Item.74}{}}
\newlabel{D087}{{{{$[$a$]$}}}{38}{}{Item.78}{}}
\newlabel{D088}{{{{$[$b$]$}}}{38}{}{Item.79}{}}
\newlabel{D089}{{{{$[$c$]$}}}{38}{}{Item.80}{}}
\newlabel{D090}{{{{$[$d$]$}}}{38}{}{Item.81}{}}
\newlabel{D091}{{{{$[$a$]$}}}{38}{}{Item.85}{}}
\newlabel{D093}{{41}{39}{}{equation.B.41}{}}
\newlabel{D094}{{42}{39}{}{equation.B.42}{}}
\newlabel{D095}{{8}{38}{}{Item.87}{}}
\newlabel{D096}{{43}{39}{}{equation.B.43}{}}
\newlabel{D097}{{44}{39}{}{equation.B.44}{}}
\newlabel{D100}{{B.4}{30}{\bf Out-Tree}{lemma.B.4}{}}
\newlabel{D104}{{45}{43}{}{equation.C.45}{}}
\newlabel{D109}{{B.5}{30}{}{lemma.B.5}{}}
\newlabel{D114}{{{{$[$a$]$}}}{30}{}{Item.38}{}}
\newlabel{D115}{{{{$[$b$]$}}}{30}{}{Item.39}{}}
\newlabel{D116}{{B.6}{30}{}{lemma.B.6}{}}
\newlabel{D117}{{B.17}{40}{}{lemma.B.17}{}}
\newlabel{D118}{{{\normalfont  B.1}}{28}{Rooted trees}{subsection.B.1}{}}
\newlabel{D119}{{B.16}{37}{}{lemma.B.16}{}}
\newlabel{D120}{{B.3}{29}{\bf Rooted Tree}{lemma.B.3}{}}
\newlabel{D124}{{{{$[$a$]$}}}{35}{}{Item.64}{}}
\newlabel{D126}{{5.1}{26}{The operators $\PB $ and $\SB $}{figure.5.1}{}}
\newlabel{D127}{{{{$[$2$]$}}}{44}{\bf for Proposition~\rf {C372}}{Item.107}{}}
\newlabel{D128}{{{{$[$1$]$}}}{44}{\bf for Proposition~\rf {C372}}{Item.106}{}}
\newlabel{D130}{{33}{31}{Projections of out-trees}{equation.B.33}{}}
\newlabel{D133}{{{{a}}}{13}{}{Item.8}{}}
\newlabel{D134}{{{{b}}}{13}{}{Item.9}{}}
\newlabel{D138}{{2}{43}{}{Item.105}{}}
\newlabel{D139}{{B.7}{30}{\bf Tree and Root of an Out-Tree}{lemma.B.7}{}}
\newlabel{D141}{{1}{55}{}{Item.216}{}}
\newlabel{D144}{{{{$[$a$]$}}}{56}{}{Item.217}{}}
\newlabel{D145}{{{{[j]}}}{56}{}{Item.222}{}}
\newlabel{D146}{{B.10}{34}{}{lemma.B.10}{}}
\newlabel{D147}{{B.12}{35}{}{lemma.B.12}{}}
\newlabel{D148}{{{{a}}}{35}{}{Item.61}{}}
\newlabel{D149}{{{{b}}}{35}{}{Item.62}{}}
\newlabel{D156}{{{{a}}}{34}{}{Item.54}{}}
\newlabel{D157}{{{{b}}}{34}{}{Item.55}{}}
\newlabel{D158}{{{{$[$a$]$}}}{34}{}{Item.56}{}}
\newlabel{D159}{{{{$[$b$]$}}}{34}{}{Item.57}{}}
\newlabel{D161}{{{{$[$c$]$}}}{34}{}{Item.58}{}}
\newlabel{D162}{{{{$[$d$]$}}}{34}{}{Item.59}{}}
\newlabel{D163}{{{{$[$e$]$}}}{34}{}{Item.60}{}}
\newlabel{D165}{{B.13}{35}{}{lemma.B.13}{}}
\newlabel{D169}{{26}{23}{Definition of \ct {Gm} games}{equation.5.26}{}}
\newlabel{D175}{{16}{15}{}{Hfootnote.16}{}}
\newlabel{D176}{{{{a}}}{15}{}{Item.10}{}}
\newlabel{D177}{{23}{19}{Perfect-recall in terms of pentaforms}{equation.4.23}{}}
\newlabel{D178}{{4.6}{20}{}{lemma.4.6}{}}
\newlabel{D300}{{1}{32}{}{Item.46}{}}
\newlabel{D301}{{2}{33}{}{Item.52}{}}
\newlabel{D302}{{3}{33}{}{Item.53}{}}
\newlabel{D304}{{2}{7}{Quintuple sets and their slices}{equation.3.2}{}}
\newlabel{D305}{{16}{15}{Subsets of Pentaforms}{equation.4.16}{}}
\newlabel{D306}{{38}{37}{}{equation.B.38}{}}
\newlabel{D307}{{37}{36}{\bf Edge-tree}{equation.B.37}{}}
\newlabel{D310}{{C.5}{42}{}{lemma.C.5}{}}
\newlabel{D312}{{38}{43}{}{Hfootnote.38}{}}
\newlabel{D313}{{14}{40}{}{Item.97}{}}
\newlabel{D316}{{15}{15}{Subsets of Pentaforms}{equation.4.15}{}}
\newlabel{D318}{{14}{12}{}{Hfootnote.14}{}}
\newlabel{D319}{{31}{25}{``Pentaforming'' a \ct {Gm} game}{equation.5.31}{}}
\newlabel{D321}{{{{a}}}{21}{Definition of \ct {Gm} games}{Item.16}{}}
\newlabel{D322}{{{{b}}}{22}{Definition of \ct {Gm} games}{Item.17}{}}
\newlabel{D323}{{{{c}}}{22}{Definition of \ct {Gm} games}{Item.18}{}}
\newlabel{D324}{{{{d}}}{23}{Definition of \ct {Gm} games}{Item.19}{}}
\newlabel{D325}{{{{e}}}{23}{Definition of \ct {Gm} games}{Item.20}{}}
\newlabel{D326}{{2}{53}{}{Item.196}{}}
\newlabel{D327}{{36}{41}{}{Hfootnote.36}{}}
\newlabel{D328p}{{C.2}{41}{\bf for Proposition~\rf {D328}}{lemma.C.2}{}}
\newlabel{D328}{{3.2}{10}{}{lemma.3.2}{}}
\newlabel{D329}{{a}{10}{}{Item.1}{}}
\newlabel{D330}{{b}{10}{}{Item.2}{}}
\newlabel{D331}{{c}{10}{}{Item.3}{}}
\newlabel{D332}{{11}{11}{}{Hfootnote.11}{}}
\newlabel{D333}{{C.1}{40}{}{lemma.C.1}{}}
\newlabel{D338}{{11}{54}{}{Item.205}{}}
\newlabel{D343}{{34a}{35}{Paths in out-trees}{equation.B.34a}{}}
\newlabel{D344}{{34b}{35}{Paths in out-trees}{equation.B.34b}{}}
\newlabel{D345}{{34}{35}{Paths in out-trees}{equation.B.34}{}}
\newlabel{D346}{{2}{46}{}{Item.118}{}}
\newlabel{D347}{{25}{23}{Definition of \ct {Gm} games}{equation.5.25}{}}
\newlabel{D348}{{3.0}{7}{\small \capp }{figure.3.0}{}}
\newlabel{D349}{{35}{40}{}{Hfootnote.35}{}}
\newlabel{D350}{{27}{23}{Definition of \ct {Gm} games}{equation.5.27}{}}
\newlabel{D351}{{28}{23}{Definition of \ct {Gm} games}{equation.5.28}{}}
\newlabel{D352}{{24}{21}{Definition of \ct {Gm} games}{equation.5.24}{}}
\newlabel{D353}{{23}{23}{}{Hfootnote.23}{}}
\newlabel{D366}{{a}{40}{}{Item.98}{}}
\newlabel{D367}{{b}{40}{}{Item.99}{}}
\newlabel{D368}{{C.3}{41}{}{lemma.C.3}{}}
\newlabel{D388}{{C.1}{42}{Eight examples, each of which violates exactly one axiom.$^{\rf {D389}}$ This can be efficiently verified by inspecting each column}{table.C.1}{}}
\newlabel{D389}{{37}{42}{}{Hfootnote.37}{}}
\newlabel{D390}{{{{$[$b$]$}}}{38}{}{Item.86}{}}
\newlabel{D391}{{{{$[$3$]$}}}{45}{}{Item.117}{}}
\newlabel{D392}{{{{$[$2$]$}}}{45}{}{Item.116}{}}
\newlabel{D393}{{{{$[$1$]$}}}{45}{}{Item.115}{}}
\newlabel{D394}{{36}{36}{Paths in out-trees}{equation.B.36}{}}
\newlabel{D396}{{1}{48}{\bf for Proposition~\rf {D454}}{Item.134}{}}
\newlabel{D397}{{2}{49}{\bf for Proposition~\rf {D454}}{Item.135}{}}
\newlabel{D401}{{3}{49}{\bf for Proposition~\rf {D454}}{Item.136}{}}
\newlabel{D402}{{{{$[$a$]$}}}{49}{\bf for Proposition~\rf {D454}}{Item.137}{}}
\newlabel{D403}{{{{$[$b$]$}}}{49}{\bf for Proposition~\rf {D454}}{Item.138}{}}
\newlabel{D404}{{7}{50}{\bf for Proposition~\rf {D454}}{Item.147}{}}
\newlabel{D405}{{5}{49}{\bf for Proposition~\rf {D454}}{Item.142}{}}
\newlabel{D406}{{4}{49}{\bf for Proposition~\rf {D454}}{Item.141}{}}
\newlabel{D407}{{8}{50}{\bf for Proposition~\rf {D454}}{Item.148}{}}
\newlabel{D408}{{6}{49}{\bf for Proposition~\rf {D454}}{Item.144}{}}
\newlabel{D409}{{{{$[$1$]$}}}{50}{\bf for Proposition~\rf {D454}}{Item.145}{}}
\newlabel{D410}{{{{$[$2$]$}}}{50}{\bf for Proposition~\rf {D454}}{Item.146}{}}
\newlabel{D411}{{{{$[$c$]$}}}{49}{\bf for Proposition~\rf {D454}}{Item.139}{}}
\newlabel{D412}{{{{$[$d$]$}}}{49}{\bf for Proposition~\rf {D454}}{Item.140}{}}
\newlabel{D413}{{{{a}}}{15}{}{Item.12}{}}
\newlabel{D414}{{{{b}}}{15}{}{Item.13}{}}
\newlabel{D415}{{4.4}{17}{}{lemma.4.4}{}}
\newlabel{D416p}{{C.12}{46}{\bf for Proposition~\rf {D415}(\rf {D416})}{lemma.C.12}{}}
\newlabel{D416}{{{{a}}}{17}{}{Item.14}{}}
\newlabel{D417p}{{C.13}{47}{\bf for Proposition~\rf {D415}(\rf {D417})}{lemma.C.13}{}}
\newlabel{D417}{{{{b}}}{17}{}{Item.15}{}}
\newlabel{D418}{{18}{17}{Unions of blocks}{equation.4.18}{}}
\newlabel{D419}{{19}{17}{Unions of blocks}{equation.4.19}{}}
\newlabel{D420}{{22a}{18}{Unions of blocks}{equation.4.22a}{}}
\newlabel{D421}{{22b}{18}{Unions of blocks}{equation.4.22b}{}}
\newlabel{D422}{{4.1}{18}{\small \capp }{figure.4.1}{}}
\newlabel{D423}{{20}{17}{Unions of blocks}{equation.4.20}{}}
\newlabel{D424}{{21}{17}{Unions of blocks}{equation.4.21}{}}
\newlabel{D426}{{35}{36}{Paths in out-trees}{equation.B.35}{}}
\newlabel{D427}{{C.11}{45}{}{lemma.C.11}{}}
\newlabel{D428}{{4}{47}{\bf for Proposition~\rf {D415}(\rf {D416})}{Item.123}{}}
\newlabel{D429}{{5}{47}{\bf for Proposition~\rf {D415}(\rf {D416})}{Item.124}{}}
\newlabel{D430}{{1}{46}{\bf for Proposition~\rf {D415}(\rf {D416})}{Item.120}{}}
\newlabel{D431}{{6}{47}{\bf for Proposition~\rf {D415}(\rf {D416})}{Item.125}{}}
\newlabel{D432}{{3}{46}{\bf for Proposition~\rf {D415}(\rf {D416})}{Item.122}{}}
\newlabel{D433}{{2}{46}{\bf for Proposition~\rf {D415}(\rf {D416})}{Item.121}{}}
\newlabel{D434}{{46}{46}{\bf for Proposition~\rf {D415}(\rf {D416})}{equation.C.46}{}}
\newlabel{D435}{{47}{46}{\bf for Proposition~\rf {D415}(\rf {D416})}{equation.C.47}{}}
\newlabel{D436}{{2}{47}{\bf for Proposition~\rf {D415}(\rf {D417})}{Item.127}{}}
\newlabel{D437}{{1}{47}{\bf for Proposition~\rf {D415}(\rf {D417})}{Item.126}{}}
\newlabel{D438}{{48}{47}{\bf for Proposition~\rf {D415}(\rf {D417})}{equation.C.48}{}}
\newlabel{D439}{{4}{48}{\bf for Proposition~\rf {D415}(\rf {D417})}{Item.129}{}}
\newlabel{D440}{{{{$[$b$]$}}}{48}{\bf for Proposition~\rf {D415}(\rf {D417})}{Item.131}{}}
\newlabel{D441}{{{{$[$a$]$}}}{48}{\bf for Proposition~\rf {D415}(\rf {D417})}{Item.130}{}}
\newlabel{D442}{{{{$[$c$]$}}}{48}{\bf for Proposition~\rf {D415}(\rf {D417})}{Item.132}{}}
\newlabel{D443}{{5}{48}{\bf for Proposition~\rf {D415}(\rf {D417})}{Item.133}{}}
\newlabel{D444}{{3}{48}{\bf for Proposition~\rf {D415}(\rf {D417})}{Item.128}{}}
\newlabel{D445}{{c}{40}{}{Item.100}{}}
\newlabel{D447}{{{{$[$a$]$}}}{49}{\bf for Proposition~\rf {D454}}{Item.143}{}}
\newlabel{D454p}{{C.14}{48}{\bf for Proposition~\rf {D454}}{lemma.C.14}{}}
\newlabel{D454}{{4.5}{19}{}{lemma.4.5}{}}
\newlabel{D464}{{17}{16}{Unions of blocks}{equation.4.17}{}}
\newlabel{D471}{{49}{48}{\bf for Proposition~\rf {D454}}{equation.C.49}{}}
\newlabel{D478}{{13}{12}{}{Hfootnote.13}{}}
\newlabel{D479}{{17}{16}{}{Hfootnote.17}{}}
\newlabel{E1}{{[E1]}{37}{\bf Edge-tree}{AMS.32}{}}
\newlabel{E2}{{[E2]}{37}{\bf Edge-tree}{AMS.33}{}}
\newlabel{E3}{{[E3]}{37}{\bf Edge-tree}{AMS.34}{}}
\newlabel{Gm1}{{[Gm1]}{24}{{\bf Gm game}, Streufert 2021Gm}{AMS.23}{}}
\newlabel{Gm2}{{[Gm2]}{24}{{\bf Gm game}, Streufert 2021Gm}{AMS.24}{}}
\newlabel{Gm3}{{[Gm3]}{24}{{\bf Gm game}, Streufert 2021Gm}{AMS.25}{}}
\newlabel{Gm4}{{[Gm4]}{24}{{\bf Gm game}, Streufert 2021Gm}{AMS.26}{}}
\newlabel{Gm5}{{[Gm5]}{24}{{\bf Gm game}, Streufert 2021Gm}{AMS.27}{}}
\newlabel{Gm6}{{[Gm6]}{24}{{\bf Gm game}, Streufert 2021Gm}{AMS.28}{}}
  }

\providecommand{\ifmain}[1]{#1}
\ifmain{
  \documentclass[12pt]{amsart} 
  \setlength{\topmargin}{0in}\setlength{\textheight}{8.6in}
  \setlength{\oddsidemargin}{.25in}\setlength{\evensidemargin}{.25in}\setlength{\textwidth}{6in}
  \newcommand{\tpath}{.}

\input{common-dp}

  \providecommand{\forpreamble}{} \forpreamble 
  \usepackage{hyperref} \hypersetup{colorlinks=true,linkcolor=black}
  \begin{document}
  }
\ifmain{
  \numberwithin{lemma}{section}
  \numberwithin{figure}{section}
  \numberwithin{table}{section}
  \newcommand{\payoa}[1]{{%
    $\mi{\mathit{#1}}\mo$}}
  \newcommand{\payob}[2]{{ %
    $\mi{\pxRed{#1}}\\{\pxGreen{#2}}\mo$}}
  \newcommand{\payoc}[3]{{ %
    $\mi{\pxRed{#1}}\\{\pxGreen{#2}}\\{\pxBlue{#3}}\mo$}}
  \newcommand{\dve}{\dra\,\,}
  \newcommand{\pj}[1]{π_{{\mathit{#1}}}}
  \setlength{\unitlength}{1cm} 
   \newcommand{\fromPicFive}[1]{%
    \noindent\begin{pspicture} \end{pspicture} } 
  \newcommand{\showit}{\ifdraft{\vspace{-.7cm}{⋅}\vspace{.3cm}}}
  }

\title[]{Dynamic Programming for \\
Pure-Strategy Subgame Perfection \\ in an Arbitrary Game}
\date{March 10, 2023.  Revises February version by providing better non-convergent examples in Section~\rf{C256}. 
{\em Classifications:} MSC 90C39, 91A18; JEL C61, C73. 
{\em Contact information:} pstreuf@uwo.ca, 519-661-2111x85384, Economics Department, Western University, London, Ontario, N6A 5C2, Canada.}  

\maketitle

\iffinal{\vspace{-5mm}}
\begin{centering}
Peter A. Streufert \\[-.5mm]
Economics Department \\[-.5mm]
Western University \\
\end{centering} 
\vspace{2mm}

\begin{abstract} This paper uses value functions to characterize the pure-strategy \linebreak subgame-perfect equilibria of an arbitrary, possibly infinite-horizon game.  It specifies the game's extensive form as a pentaform (Streufert 2023p, arXiv:2107.10801v4), which is a set of quintuples formalizing the abstract relationships between nodes, actions, players, and situations (situations generalize information sets).  Because a pentaform is a set, this paper can explicitly partition the game form into piece forms, each of which starts at a (Selten) subroot and contains all subsequent nodes except those that follow a subsequent subroot.  Then the set of subroots becomes the \linebreak domain of a value function, and the piece-form partition becomes the framework for a value recursion which generalizes the Bellman equation from dynamic programming.  The main results connect the value recursion with the subgame-perfect equilibria of the original game, under the assumptions of upper- and lower-convergence.  Finally, a corollary characterizes subgame perfection as the absence of an improving one-piece deviation.
\end{abstract}

\section{Introduction}\label{C380}\showit
\markb{\sc \rf{C380}. Introduction}

\nssec{Selten subroots and value functions}{C487}

This is the first paper to use value functions to characterize the pure-strategy subgame-perfect equilibria of arbitrary games.  As might be expected, a value function assigns a value profile (that is, a vector of utility-like values indexed by players) to each Selten subroot (that is, to each root of a nontrivial subgame as defined in Selten 1975).  Now by analogy with the Bellman equation from dynamic programming, one might hope to calculate the value profile at each subroot from the value profiles at immediately subsequent subroots.  This endeavour is relatively simple when there is perfect information, for then decision nodes and subroots are identical.  \nocite{Selte75} 

However, in general, [1] the path leading from a subroot to an immediately subsequent subroot can include one or more intermediate nodes which are not subroots.  Further, [2] there may be a path leading away from a subroot, which never reaches a subsequent subroot, but instead reaches an endnode of the entire game.  Still further, [3] there may be a path leading away from a subroot, which never reaches a subsequent subroot, and which is infinite.  Note that all three types of paths can follow a single subroot.
\enlargethispage*{\baselineskip}

Such complexity might be called ``combinatoric'' in the sense that it involves graph theory and the consideration of special cases.  To address the problem, this paper first specifies a game's extensive form as a pentaform (Streufert 2023p), which is a set of quintuples formalizing the abstract relationships between nodes, actions, players, and situations (``situations'' generalize information sets; ``form'' routinely abbreviates ``pentaform''). \nocite{2301-5f-as-2023p} 
Then a result from Streufert 2023p is used to partition the whole (penta)form into a collection of ``piece'' (penta)forms.  Each piece form starts at a subroot and includes all subsequent nodes except those that follow a subsequent subroot.  As a consequence, each piece form has runs (that is, ``plays'' or ``maximal paths'') which start at a subroot and have no intermediate subroots.  Such a piece run can be [1] a finite run going from the subroot to an immediately subsequent subroot, [2] a finite run going from the subroot to a whole-form endnode, and [3] an infinite run going from the subroot.  As a whole, the piece-form partition is the basis upon which the value profiles at different subroots will be interconnected.  In other words, it is the basis on which the ``piecewise'' (intuitively ``recursive'') properties of value functions will be defined.

\nssec{Two Theorems}{D386}

The paper culminates in two theorems about value functions.  This paragraph states the two theorems very superficially.  Theorem~\rf{C290} assumes two conditions called ``upper-convergence'' and ``lower-convergence'', and concludes that a value function, with certain essentially piecewise properties, characterizes the whole-game utilities generated by a given grand pure strategy (a grand pure strategy is equivalent to a strategy profile listing a pure strategy for each player).  Theorem~\rf{C292} assumes only ``lower-convergence'', and concludes that the existence of a value function, with other essentially piecewise properties, is equivalent to a given grand pure strategy being a subgame-perfect equilibrium.  Examples suggest that the theorems' conclusions can easily fail when their assumptions fail.

\nssec{Relation to dynamic programming}{C490}

\newcommand{\notedp}{\footnotetext{\label{D179}[from page 2] (a) To be clear, dynamic programming characterizes optima and proves their existence.  In contrast, the present paper characterizes equilibria but does not prove their existence (it is well-known that a game can fail to have a pure-strategy equilibrium).  (b) Many economists understand dynamic programming via contraction mappings, as in Denardo 1967, Stokey and Lucas 1989, Boyd 1990, and Becker and Boyd 1997.  A more general approach using convergence appears in Blackwell 1965, Strauch 1966, Sobel 1975, Kreps 1977, Blair 1984, Ozaki and Streufert 1996, and Streufert 1998.  Streufert 1993 can provide a useful bridge from this convergence literature, via consistent intergenerational games, to the present paper.
\nocite{Denar67}\nocite{StokeL89}\nocite{Boyd90}\nocite{BeckeB97}
\nocite{Black65}\nocite{Strau66}\nocite{Sobel75}\nocite{Kreps77}\nocite{Blair84}
\nocite{HO1v96}\nocite{INT2-93}\nocite{HSv98} (c) This Section~\rf{C490} uses ``nonstationary'' in an extremely general sense.}}

This section discusses Theorems \rf{C290} and \rf{C292} in somewhat more detail by showing how they generalize fundamental dynamic-programming theorems.  A one-player perfect-information game is equivalent to a nonstationary deterministic dynamic optimization problem.  When this paper's two theorems are applied to one-player perfect-information games, they collapse to the characterization theorems of nonstationary deterministic dynamic programming.  In this very special case, a subroot in this paper reduces to a (nonstationary) state in dynamic programming.  Similarly, a grand pure strategy here reduces to a (nonstationary) policy function there, and a profile-valued value function here reduces to a (nonstationary) scalar-valued value function there.{\footnotemark}  

From this perspective, it is possible to develop a number of parallels between this paper and dynamic programming.  To begin, in many formulations of dynamic programming, the Bellman equation encompasses two separate ideas, namely \ttz{(i)}{D381} that the policy today is optimal given the value tomorrow, and \ttz{(ii)}{D382} that the value today can be derived from the policy today and the value tomorrow.  Here, the notion of a time period is generalized to Section~\rf{C487}'s notion of ``piece'', \rf{D381} is generalized to the idea of the grand strategy being ``piecewise-Nash'' for the value function, and \rf{D382} is generalized to the idea of the value function being ``persistent'' for the grand strategy.  Note that both piecewise-Nashness and persistence are properties of a grand-strategy/value-function pair. \notedp

Three more parallels remain. \ttz{(iii)}{D383} In dynamic programming, the value at a state may or may not equal the utility level generated by following the policy thereafter.  Similarly here, the value profile at a subroot may or may not equal the utility profile generated by following the grand strategy thereafter.  Exactly when it does, the value function is said to be ``authentic'' for the grand strategy.  \ttz{(iv)}{D384} As in dynamic programming, a player's value function is ``admissible'' iff it satisfies weak upper and lower bounds.  Finally, \ttz{(v)}{D385} the concept of a grand strategy being subgame perfect generalizes the concept of a policy function being optimal from any state.

The conclusions of the two theorems can now be expressed in more detail.  Theorem~\rf{C290} concludes that the combination of admissibility \rf{D384} and persistence \rf{D382} is equivalent to authenticity \rf{D383}.  Theorem~\rf{C292} concludes that the combination of authenticity \rf{D383} and piecewise-Nashness \rf{D381} is equivalent to subgame perfection \rf{D385}.  In both theorems, the forward direction of the equivalence is substantial and the reverse direction is easy.  Also, the two theorems can be combined to conclude that the combination of admissibility, persistence, and piecewise-Nashness is equivalent to subgame perfection.

To reach its conclusions, Theorem~\rf{C290} assumes both upper- and lower-convergence.  Meanwhile, Theorem~\rf{C292} assumes only lower-convergence.  Upper-convergence means that conceivable utility increments vanish as one proceeds along a run.  Symmetrically, lower-convergence means that conceivable utility decrements vanish as one proceeds along a run.  This pair of assumptions generalizes a similar pair of assumptions from dynamic programming (footnote~\rf{D179} part (b)) to the broader context of games.

Later, Section~\rf{C245} will use example games to develop further intuition for the two theorems, and to suggest that the theorems' conclusions can easily fail when their assumptions fail.  Thereafter, Sections \rf{C849}--\rf{C291} will formally develop the two theorems and their underlying concepts.  Additionally, Section~\rf{C493} will develop Corollary~\rf{D211}, which assumes lower-convergence and shows that one-piece unimprovability is equivalent to subgame perfection (as in the two theorems, the forward direction is substantial and the reverse direction is easy).

\pagebreak
\nssec{Relation to game theory}{D283}

As stated at the outset, this is the first paper to use value functions to characterize the pure-strategy subgame-perfect equilibria of arbitrary, possibly infinite-horizon games.  This section will explore the generality of the paper's ``arbitrary games''.

First, this paper imposes no informational assumptions, such as perfect information, perfect recall, or no-absentmindedness.  But to accurately assess the paper's generality, one should regard each subroot as a specialized informational assumption.  At one extreme, the root node is the only subroot, Section~\rf{C487}'s piece partition has only one member, and the paper's results are vacuous.  The opposite extreme is perfect information, where every decision node is a subroot and the piece partition is as fine as possible.  There the paper's results are especially powerful, but mostly already present in the literature (Filar and Vrieze 1997).  It is the many intermediate cases to which this paper is addressed.  In this middle ground, the piece partition is nontrivial in the sense of being neither extremely coarse nor extremely fine.  Here there appear to be no general results concerning value functions, and limited results concerning one-shot and one-piece unimprovability (Hendon, Jacobsen, and Sloth 1996, Kaminski 2019).  Section~\rf{C270} further discusses informational assumptions, and Section~\rf{C493} further discusses unimprovability. \nocite{FilarV97} \nocite{HendoJS96} \nocite{Kamin19} \nocite{AlosfR17-GEB} 

\newcommand{\notediscrete}{\footnote{This excludes games in continuous time, games with simultaneous moves by infinitely many players, and the non-discrete games in Al\'os-Ferrer and Ritzberger 2016, Chapter~5.}}  

\newcommand{\noteGm}{\footnote{A standard game is called a ``\ct{Gm} game'' in Streufert 2023p.  \ct{Gm} stands for the category of standard extensive-form games (Streufert 2021Gm).  A brief introduction to this category is in Streufert 2023p, Appendix~\rf{C659}.\nocite{gm-2105-as-2021Gm}}}

Second, this paper studies arbitrary pentaform games.  These are general enough to encompass all the finite- or infinite-horizon games in which each decision node has a finite number of predecessors.{\notediscrete}   To explore this, recall that a standard game is specified as a tree decorated with information sets, actions, players, and utility functions.  Streufert 2023p shows that there is an intuitive and constructive bijection between the collection of standard games{\noteGm} and the collection of pentaform games that have information-set situations.  Therefore any standard game can be explicitly transformed into a pentaform game.  Some examples include standard games in which nodes have no special structure (Selten 1975), standard games in which nodes are sequences of past actions (Osborne and Rubinstein 1994), standard games in which nodes are sets of past actions (Streufert 2019), and standard games in which nodes are sets of future outcomes (Al\'os-Ferrer and Ritzberger 2016, Chapter~6).  Note that there can be (i) decision nodes with uncountably many immediate successors and (ii) countably infinite runs (that is, an infinite horizon). \nocite{Selte75}\nocite{OsborR94}\nocite{AlosfR16}\nocite{five-1903}

\nssec{Organization of Paper}{C489}

Section~\rf{C245} uses examples to casually introduce the paper's results.  Sections~\rf{C265} and \rf{C875} review and adapt the formal definitions of pentaform game and subgame-perfect equilibrium.  Then Section~\rf{C275} introduces the concepts of piece form and piece game, and states the paper's results.  Appendices \rf{C378}, \rf{C377}, and \rf{C406} contain lemmas and proofs.

\section{Examples}\label{C245}\showit
\markb{\sc \rf{C245}. Examples}

This Section~\rf{C245} uses examples to build intuition, step by step.  The section is casual, and presumes some familiarity with tree diagrams, subgame perfection, and dynamic programming (if the first example is unfamiliar, a good starting place would be Osborne 2004 Chapter 5).  In contrast, Sections \rf{C265}--\rf{C275} will be formal and logically self-contained.

\nssec{A familiar example}{C918}

Figure~\rf{C916}(a) begins with a well-known game between a potential firm called the ``entrant'' and an existing firm called the ``incumbent''.  If the entrant chooses to enter (denoted $\fe$), the incumbent can choose to fight ($\ff$).  The utility profiles list the entrant's utility first and the incumbent's utility second.  Essentially, a fight is won by the incumbent and reduces total utility.  Nodes are labelled $\f5$, $\f6$, $\f7$, $\f8$, and $\f9$ in order to avoid confusion with utility numbers.

\begin{figure}[h] \newcommand{\hgth}{5.2}
\begin{picture}(0,\hgth) 
  \put(0,2.1){\makebox(0,0){\scalebox{.82}{
    \fromPicFive{fC916-Stack} }}} \end{picture}
\caption{\small (a) A game. (b) The same game with its only subgame-perfect equilibrium (shown by heavy edges) and the associated value function.} \label{C916} \end{figure}
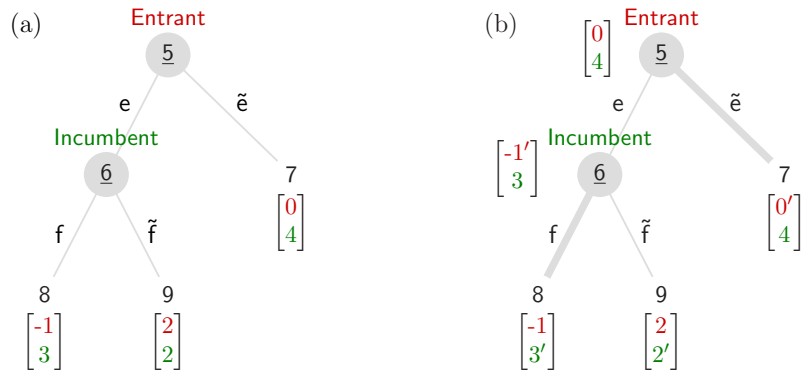

The subgame-perfect equilibrium of this game can be found by a well-known algorithm called ``backward induction'' or ``dynamic programming''.  This equilibrium and algorithm are shown in Figure~\rf{C916}(b).  The algorithm has two steps.  (1) Consider node $\f6$.  Here the incumbent would choose to fight because it gives the incumbent utility 3 from node $\f8$ rather than utility 2 from node $\f9$.  This choice is shown by the heavy edge from node $\f6$ to node $\f8$, and in accord with this choice, the utility profile from node $\f8$ is copied to node $\f6$.  (2) Consider node $\f5$.  Here the entrant would choose to not enter because it gives the entrant utility 0 from node $\f7$ rather than utility ${-}1$ from node $\f6$.  This choice is shown by the heavy edge from node $\f5$ to node $\f7$, and in accord with this choice, the utility profile from node $\f7$ is copied to node $\f5$.  (The utilities used in the above comparisons are marked with primes in the figure.) 

The two heavy edges together depict a ``grand strategy'', and the assignment of a utility profile to each of the two decision nodes is called a ``value function''.  The previous paragraph constructed this grand-strategy/value-function pair in two steps, the first being at node $\f6$ and the second being back at node $\f5$.  At each step, [a] the relevant player chose the better subsequent node on the basis of the utility profiles of the subsequent nodes, and [b] the utility profile of the better subsequent node was copied to the current node.  Call [a] the ``stepwise-optimality'' of the grand strategy given the value function, and call [b] the ``persistence'' of the value function given the grand strategy.  This [a] and [b] are two-player generalizations of \rf{D381} and \rf{D382} in Section~\rf{C490}'s initial discussion of dynamic programming.

It can be shown that the combination of stepwise-optimality and persistence is equivalent to subgame perfection in any finite game with perfect information (Osborne 2004 Proposition 172.1).  The purpose of this paper is to extend this equivalence to arbitrary games with possibly infinite horizon and possibly imperfect information.  To be somewhat more precise, the equivalence will evolve into Theorems \rf{C290} and \rf{C292}, and ``stepwise-optimality'' will become ``piecewise-Nashness''. \nocite{Osbor04}

\nssec{\sc The cry-wolf game}{C264}

Figure~\rf{C249} is a single-day version of a well-known fable.  Imagine that a wolf endangers a kid who lives in a town.  On the one hand, the wolf may attack ($\f{a}$).  Then the kid involuntarily cries {``Wolf!''\!} and the town either runs to the rescue ($\f{r}$) or not ($\f{\ggr}$).  The former is a big loss for the wolf, while the latter is a big win for the wolf.  This is reflected in the reward profiles beneath nodes $\f4$ and $\f5$, where the wolf's utility is listed first, the kid's second, and the town's last.  On the other hand, the wolf may not attack (denoted $\f{\ga}$).  Then the kid chooses whether to cry {``Wolf!''\!} ($\f{c}$) or not ($\f{\gc}$).  If the kid (untruthfully) cries out, the town either runs to the rescue ($\f{r}$) or not ($\f{\ggr}$), with the former being a small win for the kid, and the latter being a small win for the town.  If the kid (truthfully) remains quiet, the town enjoys a small win.  Importantly, the town cannot distinguish between an involuntary cry for help and a deliberate untruthful cry for help.  This is reflected by nodes $\f2$ and $\f3$ being in the same information set in the figure.

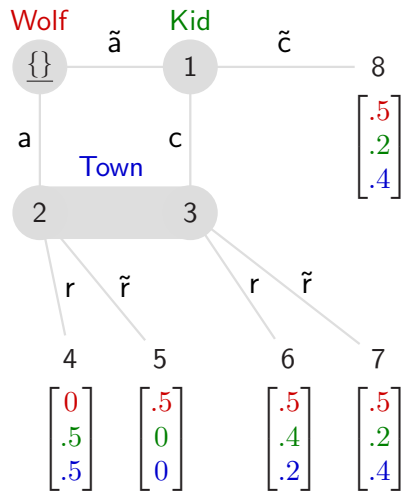
\begin{figure}[h] \newcommand{\hgth}{6.5}
\begin{picture}(0,\hgth) 
  \put(0.3,3.1){\makebox(0,0){\scalebox{1}{
    \fromPicFive{fC249-cw0} }}} \end{picture}
\caption{A single day with its rewards.} \label{C249} \end{figure}

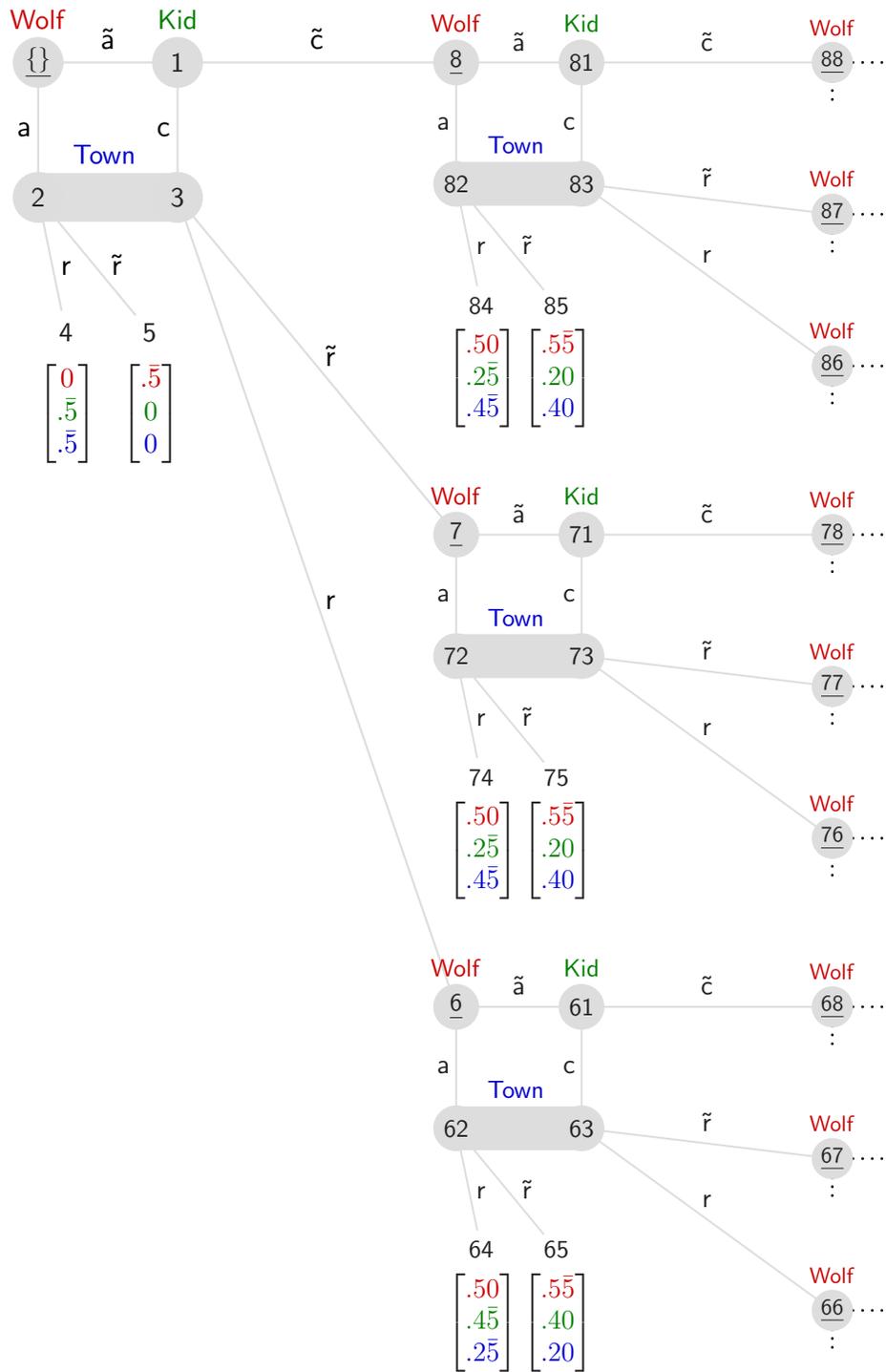
\begin{figure}[p] 
\newcommand{\hgth}{19.5}
\begin{picture}(0,\hgth) 
  \put(-.06,9.5){\makebox(0,0){\scalebox{1}{
    \noindent\begin{pspicture}(-.7,-20)(12.5,1) 
      \end{pspicture}
    }}} \end{picture}
\caption{The cry-wolf game.  Decision nodes are shaded, and subroots are underlined.  Utilities are shown for eight of the finite runs and none of the infinite runs.} \label{C243} \end{figure}
 
\newcommand{\notebarfive}{\footnote{\label{C494}To tell a story, if the wolf attacks, the single-day rewards from Figure~\rf{C249} are enjoyed permanently in Figure~\rf{C243}.  Accordingly, the positive single-day rewards from nodes $\f4$ and $\f5$ are changed from $.5$ in Figure~\rf{C249} to $∑^∞_{ℓ=0}(.1)^{ℓ-1}(.5) = .\bar{5}$ in Figure~\rf{C243}.}}

Figure~\rf{C243} is a multiple-day extension of Figure~\rf{C249}.  Notice that the multiple-day game ends if the wolf attacks.  Until that happens (if it ever does), Figure~\rf{C249}'s single-day rewards are accumulated with a discount factor of 0.1 (except that the positive single-day rewards in the event of wolf attack are changed from $.5$ to $.\bar{5}$){\notebarfive}.  For example, consider the run ending at node $\f{65}$.  Its utility profile is calculated as \begin{align}
\zz
&\mi.5\\.4\\.2\mo + .1\mi.\bar{5}\\0\\0\mo = \mi.5\bar{5}\\.40\\.20\mo, \notag
\zz
\end{align} where the first profile is $\f6$'s single-day reward from the first day (when the kid fooled the town) and the second profile is $\f5$'s single-day reward from the second and final day (when the kid's real cry for help was ignored).  This (total discounted) utility appears next to node $\f{65}$ in Figure~\rf{C243}.

Figure~\rf{C243} might be called an ``infinitely partially repeated horse game''.  In a typical repeated game, (a) the stage game is simultaneous-move or perfect-information and (b) the stage game is repeated a predetermined number of times.  Here (a) fails because the stage game is Figure~\rf{C249}, which is neither simultaneous-move nor perfect-information.  Rather, Figure~\rf{C249}'s stage game
visually resembles Selten 1975's ``horse'' game.  In addition, (b) fails because the stage game stops repeating if the wolf attacks.  Thus Figure~\rf{C243} is a ``partially'' repeated game. 

Regardless of any comparison to conventional repeated games, Figure~\rf{C243} is ``stationary'' in the sense that many nodes strongly resemble one another.  Although the theorems in this paper are especially useful when there is some sort of stationarity, the theorems themselves do not assume any sort of stationarity.  Rather, this paper will define a ``subroot'' to be the root of a nontrivial subgame (Selten 1975), and will use these subroots to partition the game form (that is, the game without utilities) into ``piece'' forms, as first discussed in Section~\rf{C487}.  These pieces can be said to generalize stages.  For example, the subroots in Figure~\rf{C243} are underlined.  These subroots divide the figure's 32 edges into the four 8-edge piece forms that begin at subroots $⎨⎬$, $\f6$, $\f7$, and $\f8$.  Each of these four piece forms is very similar to the single-day form of Figure~\rf{C249}.

\nssec{Intuition for the two theorems}{C255}

This section uses the cry-wolf game to develop intuition for Theorems \rf{C290} and \rf{C292}.  (These two theorems were discussed with less detail in Sections \rf{D386}, \rf{C490}, and \rf{C918}.)

Figure~\rf{C250} shows a grand strategy for Figure~\rf{C243}'s game.  Specifically, at each information set, one or more thick edges show the action chosen by the player in control at that information set.  Such a grand strategy determines a player strategy{\pagebreak} for each player by means of restriction.  For instance, Figure~\rf{C250} shows the kid's player strategy by the thick edges leaving the singleton information sets containing nodes ending in~$\f1$ (in particular, the figure shows the kid choosing $\fc$ at every such information set).  For brevity, ``grand strategy'' will henceforth be abbreviated as ``strategy''. 

How could one prove that Figure~\rf{C250}'s (grand) strategy is a Nash equilibrium?  If one used the definition of Nash equilibrium, one would need to be prove that each of the three players does not have an alternative player strategy which increases the player's utility.  Unfortunately, this approach seems intractable since each of the three players has infinitely many alternative player strategies.  A different approach is provided by this paper's generalized dynamic-programming technique.  It can be used to prove not only Nashness, but also the stronger concept of subgame perfection.  

To illustrate this technique, Figure~\rf{C250} places a value profile next to each subroot.  Such a map from subroots to value profiles is called a ``value function''.  The figure's (grand) strategy and value function satisfy properties \rf{D261}--\rf{D264} below.  The paper's two theorems concern the logical relationships between these four properties and the property of subgame perfection.  Note that Section~\rf{C490} introduced \rf{D261}--\rf{D264} in a different order as \rf{D381}--\rf{D384}, that the combination of \rf{D262} and \rf{D264} can be regarded as a generalization of the Bellman equation from dynamic programming, and that only these two were discussed in connection with Section~\rf{C918}'s simple example.

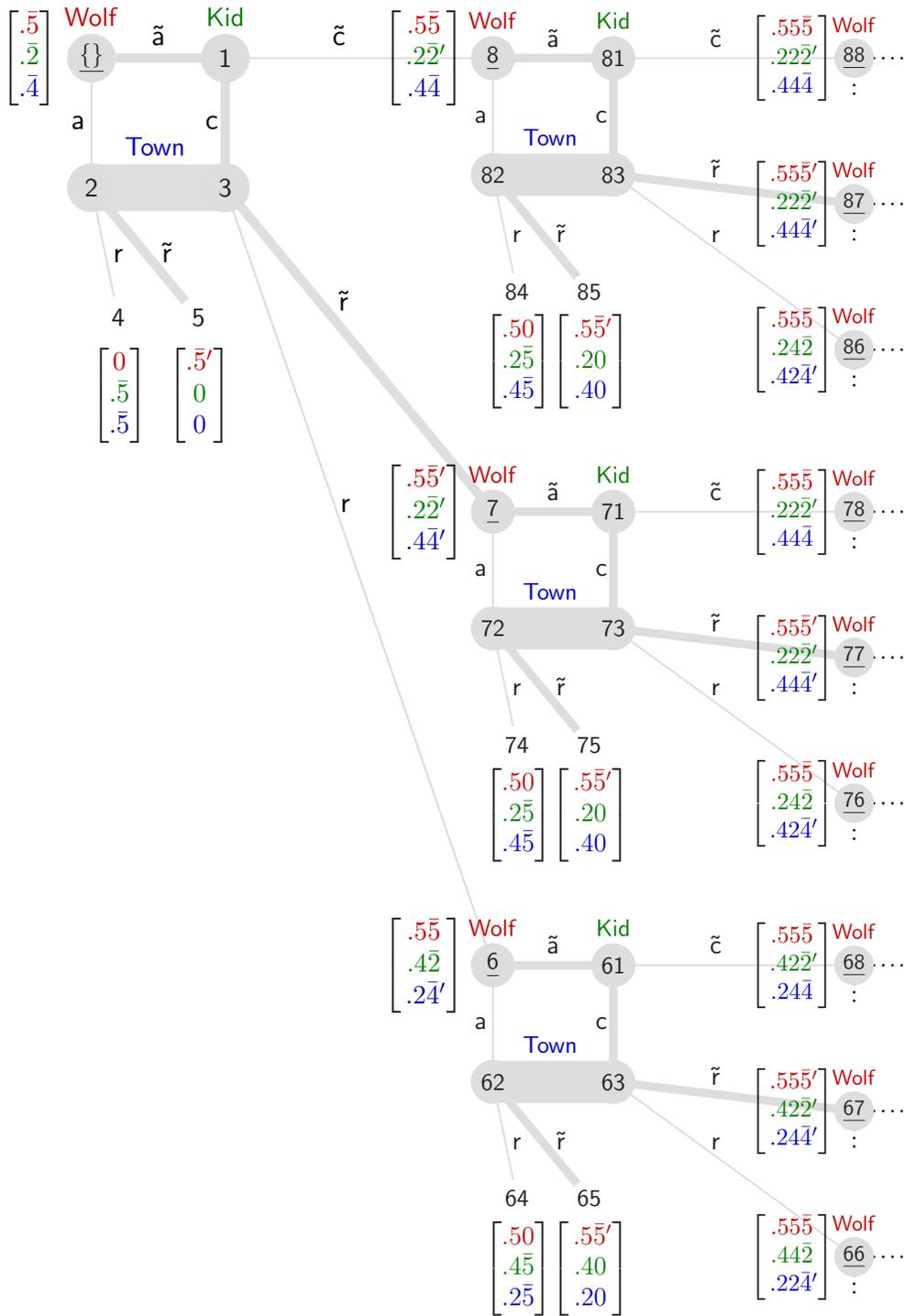
\begin{figure}[p] \newcommand{\hgth}{19.5}
\begin{picture}(0,\hgth) 
  \put(0,9.5){\makebox(0,0){\scalebox{1}{
    \fromPicFive{fC250-cwE} }}} \end{picture}
\caption{A strategy shown by the heavy edges, and a value function shown by the value profiles at the subroots (subroots are underlined).} \label{C250} 
\end{figure}

{\em Property \tts{1}{D261}}.  {\em The value function is ``admissible'', which means roughly that the players' values are neither overoptimistic nor overpessimistic.}  Somewhat more precisely, it means that for each subroot and each player, (a) there is a run through the subroot generating utility weakly higher than the value, and symmetrically, (b) there is a (typically different) run through the subroot generating utility weakly {\em lower} than the value.  To illustrate this, consider subroot $\f6$ in Figure~\rf{C250}.  There the kid's value is $.4\bar{2}$.  This is \begin{align}
\zz
&\text{below $.4\bar{5}$, which is the kid's utility from the run ending in $\f{64}$, and}
\notag \\[-.4ex]
&\text{above $.40$, which is the kid's utility from the run ending in $\f{65}$}.
\notag
\zz
\end{align}  Note that the property of admissibility does not concern the (grand) strategy.  In contrast, properties \rf{D262}--\rf{D264} will concern both the strategy and the value function.

{\em Property \tts{2}{D262}}.  {\em The value function is ``persistent'' for the strategy, in the sense that the value profile at each subroot is equal to (a) the value profile at the next subroot reached by obeying the strategy, if that subroot exists, or otherwise (b) the utility profile of the whole-game run that is completed by obeying the strategy.}  Because of the example strategy in Figure~\rf{C250}, case (b) never occurs in the figure.  As a result, persistence reduces to the property that, if two subroots are connected by thick edges, then the two share the same value profile.  For instance, the profiles at $⎨⎬$, $\f7$, and $\f{77}$ are equal, the profiles at $\f6$ and $\f{67}$ are equal, and the profiles at $\f8$ and $\f{87}$ are equal.

{\em Property \tts{3}{D263}}. {\em The value function is ``authentic'' for the strategy, in the sense that the value profile at each subroot is equal to the utility profile that results from obeying the strategy after the subroot.}  For example, in Figure~\rf{C250}, the run that results from obeying the strategy after subroot $\f6$ passes through the nodes $\f{⎨⎬}$, $\f6$, $\f{67}$, $\f{677}$, and so on.  Thus the utility profile that results from obeying the strategy after subroot $\f6$ is \begin{gather}
\zz
\mi.5\\.4\\.2\mo + .1\mi.5\\.2\\.4\mo + .01\mi.5\\.2\\.4\mo +˙...⋅ 
= \mi.5\bar{5}\\.4\bar{2}\\.2\bar{4}\mo, \notag
\zz
\end{gather} where the first profile is $\f6$'s single-day reward from the first day (when the kid fooled the town), the second profile is $\f7$'s single-day reward from the second day (when the town ignores the kid's untruthful cry in accord with the strategy), the third profile is $\f7$'s single-day reward from the third day (when again the town ignores the kid's untruthful cry in accord with the strategy), and so on.  This (discounted total) utility profile appears next to node $\f6$ in the figure.\footnote{Readers accustomed to stationary dynamic programming might not expect this utility to include the single-day reward from the first day.  It may be helpful to notice that the formulation here treats the examples of Figures \rf{C916}(b) and \rf{C250} in a unified way.  In both examples, the value function is authentic for the strategy.}

{\em Property \tts{4}{D264}}. {\em The strategy is ``piecewise-Nash'' for the value function, in the sense that it implies a Nash equilibrium in each piece game.}  By definition, each piece-game's utility function assigns to each piece-run (a) the value profile at the piece-run's endnode, if that endnode exists and is a subroot, or otherwise (b) the utility profile of the whole-game run that is completed by the piece-run.  In Figure~\rf{C250}, consider the piece that begins at subroot~$\f6$.  Here there are five piece-runs culminating in the five piece-endnodes $\f{64}$, $\f{65}$, $\f{66}$, $\f{67}$, and $\f{68}$ (these five form a backwards ``L'' to the south and east of $\f6$ in the figure).  Within this piece game, the strategy shown by the thick edges is a Nash equilibrium because \begin{align}
\zz
\text{the wolf weakly prefers}⋅& .55\bar{5}⋅\text{at}⋅\f{67}⋅\text{to}⋅.5\bar{5}⋅\text{at}⋅\f{65}, \notag \\[-.5ex]
\text{the kid weakly prefers}⋅& .42\bar{2}⋅\text{at}⋅\f{67}⋅\text{to}⋅.42\bar{2}⋅\text{at}⋅\f{68},⋅\text{and} \notag \\[-.5ex]
\text{the town strictly prefers}⋅& .24\bar{4}⋅\text{at}⋅\f{67}⋅\text{to}⋅.22\bar{4}⋅\text{at}⋅\f{66}.\notag
\zz
\end{align} (the six utility numbers used in these three utility comparisons are marked with primes in the figure).

This paper's two theorems concern logical relationships between properties \rf{D261}--\rf{D264} and the property of subgame perfection.  Two straightforward observations can be made without any restrictions.  First, \rf{D261} and \rf{D262} are implied by \rf{D263}.  In other words, admissibility and persistence are implied by authenticity.  Second, \rf{D263} and \rf{D264} are implied by subgame perfection.  In particular, consider a subgame-perfect equilibrium (this is a strategy) and derive its associated value function.  Then the value function is authentic for the strategies (by definition), and the strategies are piecewise-Nash for the value function (by a straightforward argument).  [These two observations can be strung together to show that \rf{D261}--\rf{D264} are satisfied by any subgame-perfect equilibrium and its associated value function.]  

Theorems \rf{C290} and \rf{C292} state these straightforward observations, and much more importantly, provide their converses.  Specifically, Theorem~\rf{C290} gives broad conditions under which admissibility \rf{D261} and persistence \rf{D262} imply authenticity \rf{D263}, and Theorem~\rf{C292} gives broad conditions under which authenticity \rf{D263} and piecewise-Nashness \rf{D264} imply subgame perfection.  [These two results can be strung together to show that there are broad conditions under which \rf{D261}, \rf{D262}, and \rf{D264} imply subgame perfection.]

\newcommand{\notefinite}{\footnote{Games with only finite runs are relatively easy.  There [1] upper- and lower-convergence hold vacuously (Lemma~\rf{C405}).  Also [2] persistence and authenticity are equivalent, and either implies admissibility (Lemma~\rf{C926}).  In this sense, Section~\rf{C918}'s discussion of its finite example is complete even though it does not mention upper-convergence, lower-convergence, authenticity, or admissibility.}}

The ``broad conditions'' in the previous paragraph are the assumptions of ``upper-convergence'' and ``lower-convergence''.{\notefinite}  More specifically, Theorem~\rf{C290} assumes both upper- and lower-convergence, and Theorem~\rf{C292} assumes only lower-convergence.

\nssec{Intuition for upper- and lower-convergence}{C256}

This Section~\rf{C256} uses examples to develop intuition for the assumptions of upper- and lower-convergence, and to suggest that the conclusions of Theorems \rf{C290} and \rf{C292} can easily fail when these assumptions fail.  [Thus far, these assumptions have been discussed in only a few sentences, toward the end of Section~\rf{C490}.]

Roughly, a player's utility function is ``upper-convergent'' at a run iff conceivable utility increments eventually vanish while moving along the run.  Since this trivially holds at finite runs (Lemma~\rf{C405}), only infinite runs will be considered here.  For example, each player's utility function is upper-convergent at each infinite run in Figure~\rf{C243}.  There the passage of each subroot irrevocably determines another digit in the decimal expansion of the player's utility.  Hence the conceivable utility increments dwindle to zero while moving along the run.  Symmetrically, a player's utility function is ``lower-convergent'' at a run iff conceivable utility {\em decrements} eventually vanish while moving along the run.  For example, each player's utility function is lower-convergent at any infinite run in Figure~\rf{C243}.  This holds because the preceding decimal-expansion argument applies to utility decrements just as it did to utility increments. 

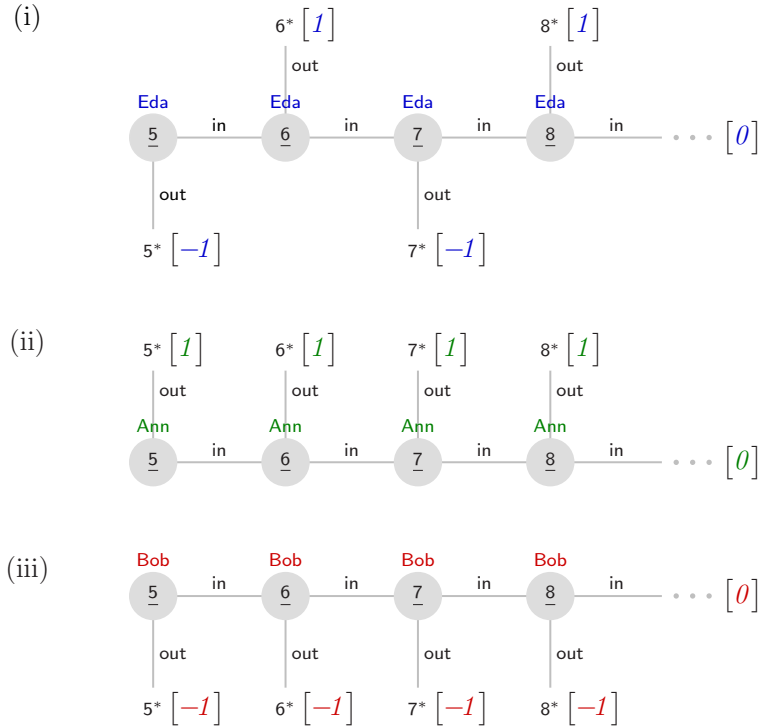
\begin{figure}[h] \newcommand{\hgth}{9.5}
\begin{picture}(0,\hgth) \myoutergrid{\hgth}
  \put(0,4.8){\makebox(0,0){\scalebox{1}{
    \fromPicFive{fD515-cent} }}} \end{picture}
\caption{Three one-player games.  (i) Eda's utility function violates both upper- and lower-convergence.  (ii) Ann's violates just upper-convergence.  (iii) Bob's violates just lower-convergence.} \label{D515} \end{figure}

In contrast, Figure~\rf{D515} shows (i) an example which violates both upper- and lower-convergence, (ii) an example which violates just upper-convergence, and (iii) an example which violates just lower-convergence.  All three examples have just one infinite run.  That run goes through the nodes $\f5$, $\f6$, $\f7$, and so on.  In all three examples, this infinite run gives the player utility $0$.  

In (i), the player Eda could conceivably reach utility 1 by choosing the action $\f{out}$ at any even-numbered node in the arbitrarily distant future while moving along the infinite run.  The perpetual plausibility of this utility increment from 0 to 1 violates upper-convergence at the run.  Symmetrically, Eda could conceivably reach utility $-1$ by choosing the action $\f{out}$ at any odd-numbered node along the run.  The perpetual plausibility of this utility decrement from 0 to $-1$ violates lower-convergence at the run.  Thus Eda's utility function violates both upper- and lower-convergence.

In (ii), the player Ann could conceivably reach utility 1 by choosing $\f{out}$ at any node along the infinite run.  The perpetual plausibility of this utility increment from 0 to 1 violates upper-convergence at the run.  Meanwhile, Ann's utility cannot fall below 0, which is the utility of the infinite run.  This implies the absence of utility decrements, which easily implies lower-convergence.  Finally in (iii), Bob could conceivably reach utility $-1$ by choosing $\f{out}$ at any node along the infinite run, and the perpetual existence of this utility decrement violates lower-convergence.  Meanwhile, Bob's utility cannot rise above 0, which easily implies upper-convergence.

\begin{figure}[h] \newcommand{\hgth}{10.2}
\begin{picture}(0,\hgth) \myoutergrid{\hgth}
  \put(0,5.2){\makebox(0,0){\scalebox{1}{
    \fromPicFive{fD517-centA} }}} \end{picture}
\caption{Three strategy/value-function pairs for which admissibility and persistence do not imply authenticity.  In each case, the strategy is shown by the heavy edges, and the value function is shown by the numbers at the decision nodes.  In (i), $ε$ is any number in $[-1,0)⋃(0,1]$.  In (ii), $α$ is any number in $(0,1]$.  In (iii), $β$ is any number in $[-1,0)$.} \label{D517} \end{figure}
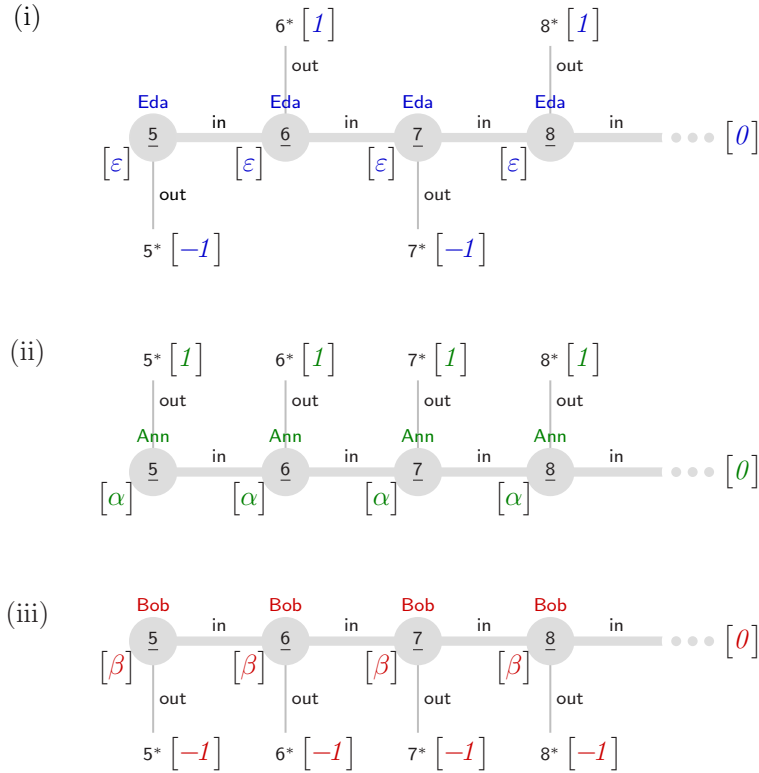

Theorem~\rf{C290}\footnote{\label{D387}In this informal context, the last two paragraphs of Section~\rf{C255} suffice for the statements of Theorems \rf{C290} and \rf{C292}.} assumes upper- and lower-convergence, and concludes that admissibility \rf{D261} and persistence \rf{D262} imply authenticity \rf{D263}.  Figure~\rf{D517} shows three strategy/{\linebreak}value-function pairs which satisfy admissibility and persistence but not authenticity.  Thus each violates the conclusion of Theorem~\rf{C290}.  Yet each is consistent with Theorem~\rf{C290} because, in each case, Figure~\rf{D515} showed that the corresponding game violates upper- or lower-convergence (or both).  

To explore this in more detail, first consider Figure~\rf{D517}(ii).  The strategy is to always play $\f{in}$, as shown by the heavy edges.  The value function is constant at some $α⋅∈⋅(0,1]$, as shown by the number $α$ at the decision nodes $\f5$, $\f6$, $\f7$, and so on.  The value function is admissible \rf{D261} because, at each decision node, (a) $α$ is greater than 0, which is the utility of the infinite run going through the decision node, and (b) $α$ is no greater than 1, which is the utility of the finite run resulting from choosing $\f{out}$ at the decision node.  Further, the value function is persistent \rf{D262} for the strategy because each heavy edge connects two decision nodes with the same value.  Yet, the value function is not authentic \rf{D263} for the strategy because, at each decision node, the value $α$ is greater than the utility 0 that results from following the strategy after that node.  In other words, the value function overestimates the utility of following the strategy.  

Intuitively, this overestimation suggests that there is a non-vanishing utility increment along the infinite run, which in turn suggests that Ann's utility function is not upper-convergent.  Indeed, Figure~\rf{D515}(ii) showed that Figure~\rf{D517}(ii)'s game violates upper-convergence.  In general, upper-convergence prevents an admissible persistent value function from overestimating the utility of following a strategy.  This is ``half'' of Theorem~\rf{C290} (loosely speaking).  The other ``half'' is similar:\ {\em lower}\,-convergence prevents an admissible persistent value function from {\em under}\,estimating the utility of following a strategy.  

To build intuition for this other ``half'', consider Figure~\rf{D517}(iii).  The strategy is to always play $\f{in}$, as shown by the heavy edges.  The value function is constant at some $β⋅∈⋅[-1,0)$, as shown by the number $β$ at the decision nodes.  As in the previous example, the value function is both admissible \rf{D261} and persistent \rf{D262} for the strategy.  Yet the value function is not authentic \rf{D263} for the strategy because, at each decision node, the value $β$ is less than the utility 0 that results from following the strategy after the node.  In other words, the value function underestimates the utility of following the strategy.  This underestimation can occur in spite of admissibility and persistence because the player Bob's utility function is not lower-convergent, as shown in Figure~\rf{D515}(iii).

To conclude this discussion of Theorem~\rf{C290}, Figure~\rf{D515}(i) shows that Eda's utility function is neither upper- nor lower-convergent, and Figure~\rf{D517}(i) shows that both overestimation ($ε⋅{>}⋅0$) and underestimation ($ε⋅{<}⋅0$) are possible, in spite of admissibility and persistence.  This suggests that, without upper- or lower-convergence, the utility of following a strategy is essentially unrelated to the admissible value functions that are persistent for the strategy.  In short, the assumptions of upper- and lower-convergence play a critical role in Theorem~\rf{C290}.

\begin{figure}[h] \newcommand{\hgth}{2.6}
\begin{picture}(0,\hgth) \myoutergrid{\hgth}
  \put(0.6,1.5){\makebox(0,0){\scalebox{1}{
    \fromPicFive{fD518-centE} }}} \end{picture}
\caption{A strategy/value-function pair for which authenticity and piecewise-Nashness do not imply subgame perfection.} \label{D518} \end{figure}
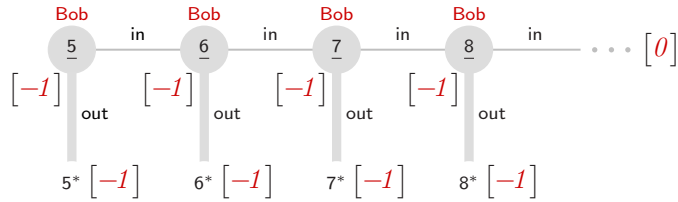

Theorem~\rf{C292} (see footnote~\rf{D387}) assumes lower-convergence, and concludes that authenticity \rf{D263} and piecewise-Nashness \rf{D264} imply subgame perfection.  Figure~\rf{D518} shows a strategy/value-function pair which satisfies authenticity and piecewise-Nashness but not subgame perfection.  Thus it violates the conclusion of Theorem~\rf{C292}.  Yet this is consistent with Theorem~\rf{C292} because Figure~\rf{D515}(iii) showed that the surrounding game violates lower-convergence.

In more detail, the strategy in Figure~\rf{D518} is to always choose $\f{out}$, as shown by the heavy edges.  The value function in Figure~\rf{D518} is constant at $-1$, as shown by the $-1$ at each decision node.  This value function is authentic \rf{D263} for the strategy because, at each decision node, the value $-1$ equals the utility $-1$ from the endnode that is immediately reached by following the strategy's choice of $\f{out}$ (as always, admissibility \rf{D261} and persistence \rf{D262} follow from authenticity \rf{D263}).  Further, the strategy is piecewise-Nash \rf{D264} for the value function because, at each decision node, the utility $-1$ resulting from the strategy's $\f{out}$ equals the value $-1$ resulting from the alternative $\f{in}$.  Yet, the strategy is not optimal for the player Bob, because he can obtain utility $0$ by always choosing $\f{in}$.  Hence subgame perfection fails.

Essentially, the strategy of always choosing $\f{out}$ conceals the very-long-run benefit of always choosing $\f{in}$.  Intuitively, this is accomplished by perpetually ``wrecking'' the game's infinite run, which suggests that there are non-vanishing utility decrements along this run, which suggests that lower-convergence fails at this run.  Indeed, Figure~\rf{D515}(iii) showed that lower-convergence fails at this run.  Thus it seems that ``wrecking'' implies lower-convergence fails.  This intuition is confirmed by the deepest part of Theorem~\rf{C292}'s proof, which shows that lower-convergence prevents the perpetual ``wrecking'' of a better infinite run (for details see footnote~\rf{D519} on page~\pageref{D519}).

\newcommand{\mysize}{\SMALL}
\newcommand{\tablewhole}{
\begin{table}[t]
{\small 
\begin{tabular}{clr} 

\multicolumn{2}{l}{Pentaform $Q$} & \mysize [\rf{C865}]\\ \hline
\\[-1.9ex]
$Q$ & set of quintuples $⁅i˛j˛w˛a˛y⁆$ & \mysize [\rf{C849}]\\
$I˙{=}˙\pj{I}(Q)$ & $\dve$ set of players $i$ &\mysize [\rf{C849},\rf{C854}]\\
$J˙{=}˙\pj{J}(Q)$ & $\dve$ set of situations $j$ &\mysize [\rf{C849},\rf{C854}]\\
$W˙{=}˙\pj{W}(Q)$ & $\dve$ set of decision nodes $w$&\mysize [\rf{C849},\rf{C854}] \\
$A˙{=}˙\pj{A}(Q)$ & $\dve$ set of actions $a$ &\mysize [\rf{C849},\rf{C854}] \\
$Y˙{=}˙\pj{Y}(Q)$ & $\dve$ set of successor nodes $y$ &\mysize [\rf{C849},\rf{C854}] \\%
[2mm]
$Q_j˙⊆˙Q$ & $\dve$ situation $j$'s slice of $Q$ &\mysize [\rf{C864}] \\
$W_j˙{=}˙\pj{W}(Q_j)$ & $\dve$ situation $j$'s information set &\mysize [\rf{C854}] \\
$A_j˙{=}˙\pj{A}(Q_j)$ & $\dve$ situation $j$'s (feasible) action set &\mysize [\rf{C854}] \\%
[2mm]
$p˙{=}˙\pj{YW}(Q)$ & $\dve$ immediate-predecessor function &\mysize [\rf{C865}] \\%
$X˙{=}˙W⋃Y$ & $\dve$ set of nodes $x$ &\mysize [\rf{C865}]\\
$⎨r⎬˙{=}˙W⧷Y$ & $\dve$ root node $r$ &\mysize [\rf{C865}]\\%
[2mm]
$≺$ & $\dve$ strict precedence relation &\mysize [\rf{C866}]\\  
$≼$ & $\dve$ weak precedence relation &\mysize [\rf{C866}]\\  
$R$ & $\dve$ weak-predecessor correspondence &\mysize [\rf{C866}]\\
$Y⧷W$ & $\dve$ set of endnodes $y$ &\mysize [\rf{C866}]\\
$\ZZf$ & $\dve$ collection of finite runs $Z$ &\mysize [\rf{C866}]\\
$\ZZi$ & $\dve$ collection of infinite runs $Z$ &\mysize [\rf{C866}]\\
$\ZZ$ & $\dve$ collection of runs $Z$ &\mysize [\rf{C866}]\\%
[2mm]
$S$ & $\dve$ set of (grand) strategies $s$ &\mysize [\rf{C269}] \\
$J_i$ & $\dve$ set of player $i$'s situations &\mysize [\rf{C269}]\\
$s_i˙{=}˙s|_{J_i}$ & $\dve$ player $i$'s restriction of $s˙∈˙S$ &\mysize [\rf{C269}]\\
$n$ & $\dve$ next-node function &\mysize [\rf{C269}]\\
$O$ & $\dve$ outcome function &\mysize[\rf{C269}]\\
[2mm]
$T$ & $\dve$ set of (Selten) subroots $t$ 
  &\mysize[\rf{C881}]\\%
  [2mm]
\multicolumn{2}{l}{Pentaform game $(Q,u)$} & \mysize [\rf{C867}]\\ \hline \\[-2ex]
$u{:}\ZZ→\eR^K$ & (grand) utility function &\mysize [\rf{C867}]\\
$K$ & $\dve$ set of stakeholders $k$ &\mysize [\rf{C867}]\\
$K⧷I$ & $\dve$ set of bystanders $k$ &\mysize [\rf{C867}]\\%
[3mm]
\end{tabular} }
\caption{\small Pentaforms and pentaform games are implicitly accompanied by their derivatives (\protect\rotatebox[origin=c]{180}{$\Lsh$}).  Definitions are in the sections in brackets {\mysize [˙]}.} \label{C870}
\end{table}
}

\section{Definitions for Pentaform Games} \label{C265}\showit
\markb{\sc \rf{C265}.\ Definitions for Pentaform Games}

This\begin{picture}(0,0)
  \put(1.5,1.4){\color{white} \rule{35ex}{2.5ex}}   
  \put(3.47,1.58){\SMALL\sc \rf{C265}. Definitions for Pentaform Games} \end{picture}
Section~\rf{C265} reviews and slightly extends the definition of pentaform game from Streufert 2023p.  The slight extension is the introduction of stakeholders in Section~\rf{C867}.  (Figure~\rf{C243}'s example will continue to be used for illustrative purposes.  Simpler examples can be found in Streufert 2023p.)

\nssec{Quintuple sets}{C849}

A arbitrary quintuple will be denoted $⁅i˛j˛w˛a˛y⁆$.  The first component $i$ is called the {\em player}, the second component $j$ is called the {\em situation}, the third component $w$ is called the {\em decision node}, the fourth component $a$ is called the {\em action}, and the fifth component $y$ is called the {\em successor node}.  These five terms have no formal meaning.  They merely name the five positions in a quintuple.  For example, in the quintuple $⁅\f{B1},\f{B2},\f{B3},\f{B4},\f{B5}⁆$, the player is $\f{B1}$, the situation is $\f{B2}$, the decision node is $\f{B3}$, the action is $\f{B4}$, and the successor node is $\f{B5}$.  Further, as would be expected, let the {\em nodes} of a quintuple be its decision node and its successor node, so that the nodes of $⁅\f{B1},\f{B2},\f{B3},\f{B4},\f{B5}⁆$ are $\f{B3}$ and $\f{B5}$.

A quintuple can specify an edge in the tree diagram of an extensive form.  For example, consider the tree diagram of Figure~\rf{C243}.  Within that diagram, consider the edge $⁅\f{63},\f{67}⁆$ from the decision node $\f{63}$ to the successor node $\f{67}$.  First, the action $\f{\ggr}$ labels this edge, and this data can be encoded in the triple $⁅\f{63},\f{\ggr},\f{67}⁆$.  Second, the information set $⎨\f{62}˛\f{63}⎬$ contains the decision node $\f{63}$, and this (self-evident) data can be encoded in the quadruple $⁅⎨\f{62}˛\f{63}⎬,\f{63},\f{\ggr},\f{67}⁆$ (information sets are a special kind of situation).  Finally, the player $\f{Town}$ makes the decision at the information set $⎨\f{62}˛\f{63}⎬$, and this data can be encoded in the quintuple $⁅\f{Town},⎨\f{62}˛\f{63}⎬,\f{63},\f{\ggr},\f{67}⁆$.  

In this fashion, a set of quintuples can specify an entire tree diagram.  For example, consider the tree diagram of Figure~\rf{C243}.  Let \begin{gather}
\zz
\rT = ⋃^∞_{ℓ=0}⎨\f6,\f7,\f8⎬^ℓ \label{D270} 
\zz
\end{gather} Thus $\rT$ is the set consisting of the strings $⎨⎬$, $\f6$, $\f7$, $\f8$, $\f{66}$, $\f{67}$, $\f{68}$, and so on.  Each such string is understood to be a sequence of characters (digits in this case), and $⎨⎬$ stands for the empty string.  In Figure~\rf{C243}, these strings are the nodes that do not end in $\f1$, $\f2$, $\f3$, $\f4$, or $\f5$.  Then for each $t⋅∈⋅\rT$, let \begin{align}
\zz
\rQ^t =⋅⎨⋅
⁅\f{Wolf},⎨t⎬&,t,\f{\ga},t⊕\f1⁆, \label{D271} \\
⁅\f{Wolf},⎨t⎬&,t,\fa,t⊕\f2⁆, \nt
⁅\f{Kid},⎨t⊕\f1⎬&,t⊕\f1,\fc,t⊕\f3⁆, \nt
⁅\f{Kid},⎨t⊕\f1⎬&,t⊕\f1,\f{\gc},t⊕\f8⁆, \nt
⁅\f{Town},⎨t⊕\f2˛t⊕\f3⎬&,t⊕\f2,\f{r},t⊕\f4⁆, \nt
⁅\f{Town},⎨t⊕\f2˛t⊕\f3⎬&,t⊕\f2,\f{\ggr},t⊕\f5⁆, \nt
⁅\f{Town},⎨t⊕\f2˛t⊕\f3⎬&,t⊕\f3,\f{r},t⊕\f6⁆, \nt
⁅\f{Town},⎨t⊕\f2˛t⊕\f3⎬&,t⊕\f3,\f{\ggr},t⊕\f7⁆⋅⎬,\notag
\zz
\end{align} where $⊕$ is the concatenation operator for strings.   Finally, let \begin{gather}
\zz
\rQ = ⋃_{t∈\rT}˙\rQ^t. \label{C855} 
\zz
\end{gather} The eight edges in Figure~\rf{C249} depict the eight quintuples in $\rQ^{⎨⎬}$.  The thirty-two edges in Figure~\rf{C243} depict the thirty-two quintuples in $\rQ^{⎨⎬}⋃\rQ^{\f6}⋃\rQ^{\f7}⋃\rQ^{\f8}$.  Finally, the quintuple of the previous paragraph appears in (\rf{D271}) as the last quintuple in $\rQ^{\f6}$ (set $t = \f6$).

\nssec{Slices}{C864}

A quintuple set will usually be denoted by the letter $Q$.  Relatedly, different quintuple sets will be distinguished from one another by means of markings around the letter $Q$.  Here are four instances of this notational principle:\ the example of equation (\rf{C855}) was denoted $\rQ$, the next paragraph will define slices $Q_j$, Section~\rf{C881} will define Selten subforms $\bQ{t}$, and Section~\rf{C276} will define piece-forms $Q^t$.  
  
This paragraph defines the slices $Q_j⋅⊆⋅Q$ of a quintuple set $Q$.  Specifically, consider an arbitrary quintuple set $Q$, and let $J$ denote its set of situations $j$.  In other words, let $J$ be the projection of $Q$ on its second coordinate.  Then for each situation $j⋅∈⋅J$, define \begin{gather}
\zz
Q_j = ⎨˙⁅i_*˛˙j˛˙w_*˛a_*˛y_*⁆∈Q˙⎬. \label{D212}
\zz
\end{gather} Thus $Q_j$ is the set of quintuples in $Q$ that list situation $j$.  Call $Q_j$ the {\em slice of $Q$ for situation $j$}.  By inspection $⁅Q_j⁆_{j∈J}$ is an injectively indexed partition of $Q$.  For instance, in example $\rQ$, definitions (\rf{D271})--(\rf{C855}) imply that the situation set $\rJ$ is equal to the collection of information sets $⋃_{t∈\rT}˙⎨⎨t⎬,⎨t⊕\f1⎬,⎨t⊕\f2˛t⊕\f3⎬⎬$.  Thus one  example situation $j⋅∈⋅\rJ$ is the information set $⎨\f{62}˛\f{63}⎬$ (set $t = \f6$).  Then definition (\rf{D271}) (via the last four rows at $t = \f6$) implies that the slice for $⎨\f{62}˛\f{63}⎬$ is  \begin{align}
\zz
\rQ_{⎨\f{62}˛\f{63}⎬} = ⎨⋅
⁅\f{Town},⎨\f{62}˛\f{63}⎬&,\f{62},\f{r},\f{64}⁆, \label{C871} \\
⁅\f{Town},⎨\f{62}˛\f{63}⎬&,\f{62},\f{\ggr},\f{65}⁆, \nt
⁅\f{Town},⎨\f{62}˛\f{63}⎬&,\f{63},\f{r},\f{66}⁆, \nt
⁅\f{Town},⎨\f{62}˛\f{63}⎬&,\f{63},\f{\ggr},\f{67}⁆⋅⎬.\notag
\zz
\end{align} These four quintuples are illustrated by the four edges at the bottom of Figure~\rf{C243}.

\nssec{Projections}{C854}

\newcommand{\notespeaking}{\footnote{When speaking aloud, it may be helpful to read  $\pj{W}(Q)$ as ``the $W$ of $Q$'' (abbreviation (\rf{C858}) shortens this to ``$W$''). Similarly, it may be helpful to read $\pj{JI}(Q)$ as ``the $\mathit{JI}$ of $Q$''.}}

Any quintuple set can be projected onto any sequence of the five coordinates.  Such a projection is denoted by the symbol $π$ followed by some sequence of the letters $I$, $J$, $W$, $A$, and $Y$.  For example,\notespeaking \subi \begin{gather}
\zz
\pj{W}(Q) = ⎨˙w˙|˙(∃i_*˛j_*˛a_*˛y_*)⋅⁅i_*˛j_*˛˙w˛˙a_*˛y_*⁆∈Q˙⎬⋅\text{and} \nt
\pj{JI}(Q) = ⎨˙⁅j˛i⁆˙|˙(∃w_*˛a_*˛y_*)˙⁅˙i˛˙j˛˙w_*˛a_*˛y_*⁆∈Q˙⎬. \notag
\zz
\end{gather} \subo Note that the example $\pj{JI}(Q)$ re-orders the coordinates.  Also note that projections of slices are well-defined because slices are quintuple sets (slicing will always come before projecting).  An example is \begin{align}
\zz
\pj{W}(Q_j) =&⋅⎨˙w˙|˙(∃i_*˛j_*˛a_*˛y_*)˙⁅i_*˛j_*˛˙w˛˙a_*˛y_*⁆∈Q_j˙⎬ \nt 
=&⋅⎨˙w˙|˙(∃i_*˛a_*˛y_*)˙⁅i_*˛˙j˛˙w˛˙a_*˛y_*⁆∈Q˙⎬, \notag
\notag
\zz
\end{align} where the second equality holds since every quintuple in $Q_j$ has situation $j$ by the slice definition (\rf{D212}). 

The notation for a single-coordinate projection will often be abbreviated by replacing the letter $Q$ with the single subscript.  In particular, define the five abbreviations $I$, $J$, $W$, $A$, and $Y$ by \label{C876} \begin{gather}
\zz
I = \pj{I}(Q),⋅J = \pj{J}(Q),⋅W = \pj{W}(Q),⋅A = \pj{A}(Q),⋅\text{and}⋅Y = \pj{Y}(Q). \label{C858}
\zz
\end{gather} These abbreviations inherit any markings on the letter $Q$.  For instance, in example $\rQ$ from (\rf{C855}), we have that the player set $\rI = \pj{I}(\rQ)$ is $⎨\f{Wolf},\f{Kid},\f{Town}⎬$.  

Two important applications of the same notational principle are \begin{gather}
\zz
W_j = \pj{W}(Q_j)⋅\text{and}⋅A_j = \pj{A}(Q_j). \label{C877} 
\zz
\end{gather} The former is called the {\em information set in situation $j$}, and the latter is called the {\em feasible action set in situation $j$}.  For instance in the example, equation (\rf{C871}) implies \begin{gather}
\zz
\rW_{⎨\f{62}˛\f{63}⎬} = ⎨\f{62},\f{63}⎬⋅\text{and}⋅
\rA_{⎨\f{62}˛\f{63}⎬} = ⎨\f{r},\f{\ggr}⎬. \notag
\zz
\end{gather} The former states that the information set in situation $⎨\f{62},\f{63}⎬$ is $⎨\f{62},\f{63}⎬$ (it is common but not necessary that a situation be identical to its information set, as discussed near Streufert 2023p equations (\rf{C794}) and (\rf{C795})).  Meanwhile, the latter states that the feasible set in situation $⎨\f{62},\f{63}⎬$ is $⎨\f{r},\f{\ggr}⎬$ (this is illustrated by the two actions assigned to the four edges at the bottom of Figure~\rf{C243}).

\nssec{Pentaforms}{C865}

For a quintuple set $Q$, let\begin{gather}
\zz
p = \pj{YW}(Q). \label{D374} 
\zz
\end{gather} Axiom \rf{pwy} below assumes that $p$ is a function (footnote \rf{C868} explains that this paper regards a function as a set of pairs).  Given this axiom, the statements $w = p(y)$, and $⁅y˛w⁆⋅∈⋅\pj{YW}(Q)$, and $⁅w˛y⁆⋅∈⋅\pj{WY}(Q)$ are equivalent.  Call $p$ the {\em immediate-predecessor function}.  

\newcommand{\noteaxioms}{\footnote{\label{C869}The label \rf{pij} can be read ``$i$ is a function of $j$''.  The labels \rf{pjw}, \rf{pwy}, and \rf{pay} can be read similarly.  Meanwhile, the label \rf{pway} can be read as ``$w$ and $a$ determine $y$''.}}

\newcommand{\notefcn}{\footnote{\label{C868}In this paper, an arbitrary function $f$ is a set of pairs such that $(∀x∈π_1f)(∃!y∈π_2f)⋅⁅x˛y⁆⋅∈⋅f$, where $π_1f$ and $π_2f$ are the projections of $f$ on its first and second coordinates.  Call $π_1f$ the domain of $f$, and call $π_2f$ the range of $f$ (in this paper functions do not have codomains).  
Relatedly, a {\em surjection from $X$ to $Y$} is a function with domain $X$ and range $Y$, and a {\em bijection from $X$ to $Y$} is an injective surjection from $X$ to $Y$.  For example, the set $g = ⎨⁅x˛3x^2⁆|x∈Ṛ⎬$ is a surjection from $Ṛ$ to $Ṛ_+$.  Finally, ``$f{:}X→Z$'' is occasionally used to mean ``$f$ is a function such that $π_1f = X$ and $π_2f⋅⊆⋅Z$''.}}

\newcommand{\fcnof}{\hspace{0ex}$\shortleftarrow$\hspace{-.9ex}%
  {\color{white}\rule{1ex}{1ex}}\hspace{-1ex}}
\newcommand{\Pway}{\mbox{[\hspace{3.5ex}$\shortrightarrow$\hspace{-1.8ex}%
  {\color{white}\rule{0.8ex}{1ex}}\hspace{-4.3ex}Pwa\hspace{1ex}y]}}

\begin{ndef}[{\bf Pentaform}, Streufert 2023p Definition~\rf{C669}]\label{D375} A {\em (penta)form} is a set $Q$ of quintuples $⁅i˛j˛w˛a˛y⁆$ such that\qquad\noteaxioms$^,$\notefcn%
\begin{picture}(0,0)
  \put(-.68,0){\color{white}\ifdraft{\red}\rule{4ex}{2ex}} \end{picture} 
\begin{gather} 
\zz
\ttt{[Pi{\fcnof}j]}{pij}%
  \begin{picture}(0,0)
    \put(-4.64,0){$^{\text{\normalfont\rf{C869}}}$} \end{picture}
  \pj{JI}(Q)⋅\text{is a function}, ^{\text{\normalfont\rf{C868}}} \\
\ttt{[Pj{\fcnof}w]}{pjw} \pj{WJ}(Q)⋅\text{is a function}, \\
\ttt{[Pwa]}{pwa} (∀j∈J)⋅\pj{WA}(Q_j)⋅\text{is a Cartesian product}, \\
\ttt{\Pway}{pway} \pj{WAY}(Q)⋅\text{is a function from its first two coordinates}, \\
\ttt{[Pw\hspace{-.1ex}{\fcnof}y]}{pwy} \pj{YW}(Q)⋅\text{is a function}, \\
\ttt{[Pa{\fcnof}y]}{pay} \pj{YA}(Q)⋅\text{is a function}, \\
\ttt{[Py]}{py} (∀y∈Y)(∃ℓ≥1)⋅p^ℓ(y)⋅∉⋅Y,⋅\text{and} \\
\ttt{[Pr]}{pr} W⧷Y⋅\text{is a singleton},
\zz
\end{gather} (where $Q$ determines $J$, $W$, $Y$, $p$, and each $Q_j$, as summarized in Table~\rf{C870}).\end{ndef}

Streufert 2023p Section~\rf{B570} interprets each of the eight pentaform axioms with the help of finite-horizon examples.  Meanwhile, this paper's Lemma~\rf{D297} shows that the $\rQ$ from (\rf{C855}) and Figure~\rf{C243} is an example of an infinite-horizon pentaform.  Further, pentaforms in general, and this paper's theorems in particular, can accommodate nodes with a continuum of immediate successors (an example is not provided).  The remainder of this Section~\rf{C865} will briefly discuss the eight axioms.

The first three axioms concern situations.  Axiom \rf{pij} requires that exactly one player $i$ is assigned to each situation $j$.  This is the player that controls the move at the situation.  Axiom \rf{pjw} requires that exactly one situation $j$ is assigned to each decision node $w$.  By Streufert 2023p Proposition~\rf{D328}, this is equivalent to $⁅W_j⁆_{j∈J}$ being an injectively indexed partition of $W$.  Each $W_j$ is called situation $j$'s information set (definition (\rf{C877})).  By Streufert 2023p Proposition~\rf{C601}(\rf{C602}$⟺$\rf{C605}), axiom \rf{pwa} is equivalent to requiring that, for each situation $j$, and for each decision node $w⋅∈⋅W_j$, the set of actions paired with $w$ is the set $A_j$.  This $A_j$ is called situation~$j$'s feasible set (definition (\rf{C877})), and likewise, it is called the feasible set of each decision node $w⋅∈⋅W_j$.

\tablewhole

Next consider the combination of \rf{pway}, \rf{pwy}, and \rf{pay}.  This combination is equivalent to requiring that the assignment of a decision-node/action pair $⁅w˛a⁆$ to a successor node $y$ is a bijection.  Thus a decision node and one of its feasible actions determine the successor node, and conversely, any successor node determines its immediate-predecessor node and its immediately previous action.

Finally, Streufert 2023p Proposition~\rf{C632} shows that the combination of \rf{pwy}, \rf{py}, and \rf{pr} is equivalent to $(W⋃Y,\pj{WY}(Q))$ being a nontrivial out-tree (that is, the divergent orientation of a rooted tree with at least one edge, as defined in Streufert 2023p Definition \rf{D100}).  The proposition also shows that the root of the out-tree is the sole element of $W⧷Y$.  In accord with these results, define a pentaform's $X$ and $r$ by \begin{gather}
\zz
X = W⋃Y⋅\text{and} \label{D182} \\
⎨r⎬ = W⧷Y, \label{D186}
\zz
\end{gather} call $X$ the set of {\em nodes}, call $\pj{WY}(Q)$ the set of {\em edges}, and call $r$ the {\em root}.

\nssec{Paths in a pentaform's out-tree}{C866} 

\newcommand{\notenodeonly}{\footnote{Within the context of a fixed out-tree, this definition of path as a set of nodes is equivalent to the standard graph-theoretic definition of path as a pair listing [1] a set of nodes and [2] a set of edges.  (This equivalence is implied, for both finite and infinite paths, by Streufert 2023p Lemma~\rf{D165}.)}}

Consider a pentaform $Q$ and its out-tree $(X,\pj{WY}(Q))$.  Two types of path will be defined.  First, let a {\em path in $(X,\pj{WY}(Q))$ from $x_0$ to $x_ℓ$} be a set of the form $⎨x_0,x_1,...˙x_ℓ⎬$ such that distinct $i$ and $j$ satisfy $x_i˙≠⋅x_j$ and{\notenodeonly} \subi \begin{gather}
\zz
(∀m∈⎨1˛2˛...ℓ⎬)⋅⁅x_{m-1}˛x_m⁆⋅∈⋅\pj{WY}(Q). \label{D493}
\zz
\end{gather}  For instance, in example $\rQ$ of definition (\rf{C855}) and Figure~\rf{C243}, the path from $x_0 = \f3$ to $x_2 = \f{62}$ is $⎨\f3,\f6,\f{62}⎬$, and the path from $x_0 = \f3$ to $x_ℓ = \f3$ is $⎨\f3⎬$.  Second, let an {\em infinite path in $(X,\pj{WY}(Q))$ from $x_0$} be a set of the form $⎨x_0,x_1,...⎬$ such that distinct $i$ and $j$ satisfy $x_i˙≠⋅x_j$ and \begin{gather}
\zz
(∀m∈⎨1˛2˛...⎬)⋅⁅x_{m-1}˛x_m⁆⋅∈⋅\pj{WY}(Q). \label{D494}
\zz
\end{gather} \subo For instance, in the same example, one of many infinite paths from $x_0 = ⎨⎬$ is $⎨⎨⎬, \f1, \f3, \f7, \f{71}, \f{73}, \f{77}, ...˙⎬$.  This particular path happens to be the longest path marked by heavy edges in Figure~\rf{C250}. (Streufert 2023p Lemmas \rf{D146} and \rf{C798} provide some basic facts:\ [i] any node is reached by a unique path from the root, [ii] there is no more than one path from any node to another, and [iii] a path from a first node to a second distinct node precludes a path from the second to the first.)

\newcommand{\notecorresp}{\footnote{In this paper, a correspondence is simply a set of pairs.  The expression ``$F{:}X⇉Y$'' means the statement ``$F⋅⊆⋅X×Y$''.}}

Let $≼$ and $≺$ be the binary relations defined by \begin{gather}
\zz
x_*⋅≼⋅x⋅\text{iff there is a path from $x_*$ to $x$, and} \label{D359} \\
x_*⋅≺⋅x⋅\text{iff}⋅(x_*⋅≠⋅x⋅\text{and}⋅x_*⋅≼⋅x). \label{D360}
\zz
\end{gather} Call $≼$ and $≺$ the {\em weak} and {\em strict precedence orders}, respectively.  (Streufert 2023p Lemma~\rf{D147} shows that $≼$ is a partial order on $X$, and that $≺$ is the asymmetric part of $≼$.)  Finally, define the correspondence{\notecorresp} $R{:}X⇉X$ by \begin{gather}
\zz
R(x) = ⎨x_*|x_*≼x⎬. \label{D206} 
\zz
\end{gather} Call $R$ the {\em weak-predecessor correspondence}.  Lemma~\rf{D486} shows  that $R(x)$ is identical to the path from $r$ to $x$ (both are sets).

This paragraph defines a run (or play) to be a special kind of path.  Runs come in two flavours:\ finite and infinite.  First, call $Y⧷W$ (which equals $X⧷W$) the set of {\em endnodes}, let \subi\label{D282}\begin{gather}
\zz
\ZZf = ⎨˙R(y)˙|˙y∈Y⧷W˙⎬, \label{D279}
\zz
\end{gather} and call $\ZZf$ the collection of {\em finite runs}.  Thus a finite run is a path which goes from the root $r$ to some endnode.  For instance, in example $\rQ$ of (\rf{C855}) and Figure~\rf{C243}, we have that $\dot{r} = ⎨⎬$, that $\f{64}⋅∈⋅\rY⧷\rW$, and that $⎨⎨⎬, \f1,\f3,\f6,\f{62},\f{64}⎬⋅∈⋅\dot{\ZZ}_{\f{ft}}$.  Second, let \begin{gather}
\zz
\ZZi = ⎨˙\text{infinite paths in}⋅(X,\pj{WY}(Q))⋅\text{from}⋅r˙⎬, \label{D280}
\zz
\end{gather} and call $\ZZi$ the collection of {\em infinite runs}.  For instance, in the same example, we have $⎨⎨⎬,\f1,\f3,\f7,\f{71},\f{73},\f{77},...˙⎬⋅∈⋅\dot{\ZZ}_{\f{inft}}$.  Finally, let\begin{gather}
\zz
\ZZ = \ZZf⋃\ZZi, \label{D281}
\zz
\end{gather} \subo and call $\ZZ$ the set of {\em runs}.

A pentaform $Q$ must have at least one run (Lemma~\rf{C966}(\rf{C988})), and it can have all finite runs, all infinite runs, or a combination of the two.  In the first case, $\ZZ = \ZZf$ and $\ZZi = ∅$.  This happens if, but not only if, the set $Q$ is finite.  For explicit examples, please see Streufert 2023p.  In the second case, $\ZZ = \ZZi$ and $\ZZf = ∅$.  This occurs iff there are no endnodes.  In the third case, both $\ZZf$ and $\ZZi$ are nonempty, as in the cry-wolf example $\rQ$ from (\rf{C855}) and Figure~\rf{C243}.

Finally, the weak-predecessor correspondence $R$ can be used to express the runs in $\ZZ$ in a more convenient way.  To do this, first extend the correspondence $R$ to accept a set argument in the usual way.  In particular, for a set $N⋅⊆⋅X$, let $R(N) = ⨆⎨R(x)|x∈N⎬$.  Then Lemma~\rf{C907} shows that $R(N)⋅∈⋅\ZZ$ iff either (a) $\max˙N$ exists and is in $Y⧷W$ or (b) $N$ is an infinite subset of a path.  In this fashion, the extended correspondence $R$ can build an entire run $R(N)$ from certain sets $N⋅⊆⋅X$.  For instance, in the cry-wolf example $\rQ$ from (\rf{C855}) and Figure~\rf{C243},\begin{gather}
\zz
\dot{R}(\f{64}) = \dot{R}(⎨\f{64}⎬) = ⎨⎨⎬,\f1,\f3,\f6,\f{62},\f{64}⎬⋅∈⋅\dot{\ZZ}_{\f{ft}}⋅\text{and} \nt
\dot{R}(⎨\f7,\f{77},...⎬) = ⎨⎨⎬,\f1,\f3,\f7,\f{71},\f{73},\f{77},...⎬⋅∈⋅\dot{\ZZ}_{\f{inft}}.\notag
\zz
\end{gather}

\nssec{Pentaform games}{C867}

Consider a pentaform $Q$ with its player set $I$ and run collection $\ZZ$.  Then let $K$ be a superset of $I$.  Call $K$ the set of {\em stakeholders}, and call \begin{gather}
\zz
K⧷I \label{D484}
\zz
\end{gather} the set of {\em bystanders}.  Bystanders will play an essential role in this paper's dynamic-programming theorems.  To provide some initial motivation, consider [1] a game, with its players, and [2] one of the game's Selten subgames, with its players.  There may be game players who are not subgame players.  Such game players will appear as bystanders in the subgame.  Although these bystanders will have no decisions to make in the subgame, the utility that the bystanders get from the subgame will play an essential role in the dynamic-programming theorems about the game itself.  A simple instance of this is a Stackelberg game, where the leader is a bystander in the follower's subgame, and at the same time, the leader's utility in the follower's subgame is used to calculate the subgame-perfect equilibria of the entire game.  This is illustrated by Figure~\rf{C916}'s example, in which the entrant is the leader, and the incumbent is the follower.

\newcommand{\noteinfutil}{\footnotetext{\label{C880}[from page \pageref{C880}] Infinite utility numbers are included because, in economics, many popular utility functions generate $-∞$ utility when some consumption level is zero.  Such utility functions often appear in consumer dynamic optimization problems, and such problems can be regarded as one-player games.}}

Let a {\em (grand) utility function} be a function $u{:}\ZZ→\smash{\eR^K}$, where $\eR = Ṛ⋃⎨{-}∞,∞⎬$ is the extended real line.{\footnotemark}  Such a function $u$ maps each run $Z⋅∈⋅\ZZ$ to a utility profile $u(Z) = ⁅u_k(Z)⁆_{k∈K}$, which lists a utility level $u_k(Z)⋅∈⋅\eR$ for each stakeholder $k⋅∈⋅K$.

\begin{ndef}[{{\bf Pentaform game}}]\label{D207} A {\em (pentaform) game} is a pair $(Q,u)$ listing a (penta)form $Q$ and a function $u{:}\ZZ→\smash{\eR^K}$ such that $I⋅⊆⋅K$ (where $Q$ determines $\ZZ$ and $I$, as summarized in Table~\rf{C870}). \end{ndef}\noteinfutil

For example, Figure~\rf{C243} illustrates the cry-wolf game $(\rQ,\ru)$.  The form $\rQ$ is defined formally in (\rf{C855}), and the utility function $\ru$ is defined verbally in the second paragraph 
of Section~\rf{C264}.  In this game, the stakeholder set $\rK$ equals the player set $\rI$, so there are no bystanders.

\nssec{Nash equilibria}{C269}

Consider a pentaform game $(Q,u)$.  Let the set of {\em (grand) strategies} be \begin{gather}
\zz
S = ⎨˙s⋅\text{is function from}˙J\,|\,(∀j∈J)\,s(j)∈A_j\,⎬ \label{D376}
\zz
\end{gather}Thus a strategy names a feasible action $s(j)$ at each situation $j⋅∈⋅J$.  Equivalently, $S = ⎨˙s⋅\text{is function from}⋅J˙|˙s⊆\pj{JA}(Q)˙⎬$.  Mixed strategies are not considered.   

\newcommand{\notefcninv}{\footnote{\label{D256}Suppose $G$ is a set of pairs $⁅x˛y⁆$, let $X = π_1G$, and let $Y = π_2G$, and define $⁅X^y⁆_{y∈Y}$ by $(∀y∈Y)$ $X^y = ⎨x|⁅x˛y⁆∈G⎬$.  Call $X^y$ the {\em inverse image} of $y$.  Then the following are equivalent.  \ttr{a}{D364} $G$ is a function.  \ttr{b}{D363} Distinct $y_1$ and $y_2$ satisfy $X^{y_1}⋂X^{y_2} = ∅$.  \ttr{c}{D362} $⁅X^y⁆_{y∈Y}$ is an injectively indexed partition of $X$.}}

For each player $i⋅∈⋅I$, let \begin{gather}
\zz
J_i = ⎨\,j∈J\,|\,⁅i˛j⁆∈\pj{IJ}(Q)\,⎬, \label{D184}
\zz
\end{gather} and call $J_i$ the set of {\em player $i$'s situations}.  Axiom \rf{pij} and a general fact about functions{\notefcninv} imply that $J$ is injectively partitioned by $⁅J_i⁆_{i∈I}$.  Thus the domain $J$ of a strategy $s⋅∈⋅S$ is partitioned by the players' situation sets $⁅J_i⁆_{i∈I}$.  Relatedly, for each strategy $s⋅∈⋅S$ and each player $i⋅∈⋅I$, let \begin{gather}
\zz
s_i = s|_{J_i}⋅\text{and}⋅s_{-i} = s|_{J⧷J_i}. \label{D253}
\zz
\end{gather} Thus $s_i$ abbreviates the restriction of $s$ to player $i$'s situations, and $s_{-i}$ abbreviates the restriction of $s$ to the situations of player $i$'s opponents.  Let the restriction $s_i$ be called {\em player $i$'s strategy}.%
\footnote{\label{D258}At the expense of more notation, one could define $S_i = ⎨˙f{:}J_i→A˙|˙(∀j∈J_i)˙f(j)∈A_j˙⎬$, and call $S_i$ the set of player-$i$ strategies.  Then $S = ⎨⨆_{i∈I}s_i|⁅s_i⁆_{i∈I}∈∏_{i∈I}S_i⎬$.}  
This definition, when applied to the other players, implies that $s_{-i} = ⨆_{i_*∈I⧷⎨i⎬}s_{i_*}$ is the union of the strategies of the other players (footnote \rf{C868} explains functions are sets).  In the event $I = ⎨i⎬$ where there is only one player, the set $s_{-i}$ is empty.  In any event, $s_i⋃s_{-i} = s$.  This union $s_i⋃s_{-i}$ will often be written $(s_i,s_{-i})$ for the sake of readability and familiarity. 

For instance, consider example $\rQ$ from (\rf{C855}).  As noted in Section~\rf{C864}, the situation set $\rJ$ is the information-set collection $⨆_{t∈\rT}⎨⎨t⎬,⎨t⊕1⎬,⎨t⊕2,t⊕3⎬⎬$.  This is partitioned by the player situation sets\begin{gather}
\zz
\rJ_{\f{Wolf}} = ⎨⎨t⎬|t∈\rT⎬, \rJ_{\f{Kid}} = ⎨⎨t⊕\f1⎬|t∈\rT⎬,⋅\text{and}⋅\rJ_{\f{Town}} = ⎨⎨t⊕\f2˛t⊕\f3⎬|t∈\rT⎬.\notag
\zz
\end{gather} Further, Figure~\rf{C250}'s heavy edges show the (grand) strategy \begin{gather}
\zz
s = s_{\f{Wolf}}⋃s_{\f{Kid}}⋃s_{\f{Town}}⋅\text{where} \label{C941} \\
s_{\f{Wolf}} = ⎨⁅⎨t⎬˛\f{\ga}⁆|t∈\rT⎬,⋅
s_{\f{Kid}} = ⎨⁅⎨t⊕\f1⎬˛\fc⁆|t∈\rT⎬,⋅\text{and}⋅
s_{\f{Town}} = ⎨⁅⎨t⊕\f2˛t⊕\f3⎬˛\f{\ggr}⁆|t∈\rT⎬ \notag
\zz
\end{gather} (footnote \rf{C868} explains functions are sets of pairs).  Thus the functions $s_{\f{Wolf}}$, $s_{\f{Kid}}$, and $s_{\f{Town}}$ are both the components used to construct the strategy $s$ and also the restrictions derived from $s$.  In terms of Section~\rf{C264}'s story, the wolf never attacks ($\f{\ga}$), the kid always cries ($\fc$), and the town never rescues ($\f{\ggr}$).

This paragraph will show that each grand strategy $s⋅∈⋅S$ determines a run $Z⋅∈⋅\ZZ$.  First, let the {\em next-node function} be \begin{gather}
\zz
n = ⎨⋅⁅⁅w˛a⁆˛y⁆⋅|⋅⁅w˛a˛y⁆∈\pj{WAY}(Q)˙⎬. \label{D377}
\zz
\end{gather} Thus the function $n$ maps each decision-node/feasible-action pair $⁅w˛a⁆⋅∈⋅\pj{WA}(Q)$ to a successor node $n(w˛a)⋅∈⋅Y$.  The well-definition of $n$ is equivalent to axiom \rf{pway}.  Second, note that [a] a decision node $w⋅∈⋅W$ determines a situation $j_w$ by \rf{pjw}, which [b] via a (grand) strategy $s⋅∈⋅S$ determines an action $s(j_w)$, which [c] via the next-node function determines a successor node $n(w,s(j_w))$.  As a result, a strategy $s⋅∈⋅S$ determines the run consisting of $r$, and $x_1{=}n(r,s(j_{r}))$, and $x_2{=}n(x_1,s(j_{x_1}))$, and so on, either indefinitely or until an endnode $x_ℓ{=}n(x_{ℓ-1},s(j_{x_{ℓ-1}}))$ is reached.  By this process, each grand strategy $s⋅∈⋅S$ determines a run $Z⋅∈⋅\ZZ$.  In other words, this process defines a function $O{:}S→\ZZ$.  Call $O$ the {\em outcome function}.  For instance, consider the example strategy $s⋅∈⋅\dot{S}$ from (\rf{C941}).  Figure~\rf{C250} illustrates that its outcome is $\dot{O}(s) = \rR(⎨\f7˛\f{77}˛...⎬)⋅∈⋅\rZZ$.

\newcommand{\noteNash}{\footnote{\label{D260}Definition (\rf{D376}) implies that the domain of the alternative $σ⋅∈⋅S$ is the situation set $J$, and definition (\rf{D253}) implies that the domain of $σ_i = s|_{J_i}$ is player $i$'s situation set $J_i$.  Thus (\rf{D187}) is unaffected by the values of the alternative $σ$ over the other players' situations in $J⧷J_i$.  Accordingly, at the expense of more notation, one could define player $i$'s strategy set $S_i$ (footnote \rf{D258}), and then quantify (\rf{D187}) by $(∀i∈I,ρ∈S_i)$ with $ρ$ replacing $σ_i$ in the inequality.}}

A {\em Nash equilibrium} is a strategy $s⋅∈⋅S$ such that{\noteNash} \begin{gather}
\zz
(∀i∈I,σ∈S)⋅u_i(O(˙s˙))⋅≥⋅u_i(O(˙σ_i,˙s_{-i}˙)). \label{D187}
\zz 
\end{gather} Thus a Nash equilibrium is a (grand) strategy such that, for each player $i$, the restriction $s_i$ of player $i$ is optimal for player $i$ given the restriction $s_{-i}$ of player $i$'s opponents.  In other words, each player's strategy is a best response to the strategies of their opponents.  Note that the bystanders in $K⧷I$ play no role in a Nash equilibrium.  

Although definition (\rf{D187}) is conceptually compelling, it threatens to be intractable.  For instance, consider definition (\rf{D187}) in the example game $(\rQ,\ru)$, for the strategy $s⋅∈⋅\dot{S}$ from (\rf{C941}), and for the player $i = \f{Kid}⋅∈⋅\rI$.  In this circumstance, the player's alternative strategy $σ_i = σ_{\f{Kid}}$ is some function from $\rJ_{\f{Kid}} = ⎨⎨t⊕1⎬|t∈\rT⎬$ to $⎨\fc,\f{\gc}⎬$.  Since $\rT$ is countably infinite, the number of such alternative strategies is uncountably infinite.

\newcommand{\tablesub}{
\begin{table}[t]
{\small 
\begin{tabular}{clr} 
\multicolumn{2}{l}{Subform $\tQ$ of a form $Q$ at subroot $t˙∈˙T$} 
  \\\hline \\[-3.6mm] 
$\tQ˙⊆˙Q$ & set of quintuples $⁅i˛j˛w˛a˛y⁆$ 
  &\mysize[\rf{C881}]\\
$\tI˙{=}˙\pj{I}(\tQ)$ & $\dve$ set of players $i$ 
  &\mysize[\rf{C854},\rf{C881}]\\
$\tJ˙{=}˙\pj{J}(\tQ)$ & $\dve$ set of situations $j$ 
  &\mysize[\rf{C854},\rf{C881}]\\
$\tW˙{=}˙\pj{W}(\tQ)$ & $\dve$ set of decision nodes $w$ 
  &\mysize[\rf{C854},\rf{C881}]\\
$\tA˙{=}˙\pj{A}(\tQ)$ & $\dve$ set of actions $a$ 
  &\mysize[\rf{C854},\rf{C881}]\\
$\tY˙{=}˙\pj{Y}(\tQ)$ & $\dve$ set of successor nodes $y$ 
  &\mysize[\rf{C854},\rf{C881}]\\
$⎨t⎬˙{=}˙\tW⧷\tY$ & $\dve$ root node $t$
  &\mysize[\rf{C881}]\\
$\tY⧷\tW$ & $\dve$ set of endnodes $y$
  &\mysize[\rf{C866},\rf{C881}]\\
$\tZZ$ & $\dve$ collection of runs $N$ (not $Z$)
  &\mysize[\rf{C866},\rf{C881}]\\
$\tSS$ & $\dve$ set of (grand) strategies
  &\mysize[\rf{C269},\rf{C881}]\\
$\tJ_i$ & $\dve$ player $i$'s set of situations $j$ 
  &\mysize[\rf{C269},\rf{C881}]\\
$\tOO$ & $\dve$ outcome function 
  &\mysize[\rf{C269},\rf{C881}]\\%
  [2mm]
$\tss˙{=}˙s|_{\tJ}˙∈˙\tSS$ & $\dve$ subform restriction of $s˙∈˙S$ 
  &\mysize[\rf{C884}]\\
$\tss_i˙{=}˙s|_{\tJ_i}$ & $\dve$ player $i$'s subform restriction of $s˙∈˙S$ 
  &\mysize[\rf{C884}]\\%
  [2.5mm]
\multicolumn{2}{l}{Subgame $(\tQ,\tu)$ of a game $(Q,u)$ at subroot $t˙∈˙T$} 
    \\\hline \\[-2ex]
$\tu{:}\tZZ→Ṛ^K$ & (grand) utility function derived from $u$ &\mysize[\rf{C884}]\\
$K$ & $\dve$ set of stakeholders $k$ &\mysize[\rf{C867},\rf{C884}]\\
$K⧷\tI$ & $\dve$ set of bystanders $k$ &\mysize[\rf{C867},\rf{C884}]\\%
[3mm]
\end{tabular} }
\caption{\small (Selten) subforms and subgames are implicitly accompanied by their derivatives (\protect\rotatebox[origin=c]{180}{$\Lsh$}).  Definitions are in the sections in brackets {\mysize [˙]}.} \label{C885}
\end{table}
}

\section{Subroots, Subforms, and Subgames}\label{C875}\showit
\markb{\sc \rf{C875}. Subroots, Subforms, and Subgames}

Section~\rf{C881} defines the subroots and subforms of a pentaform.  Then Section~\rf{C884} adapts the standard concept of subgame perfection (Selten 1975) to pentaform games.  Finally, Section~\rf{C270} discusses the relationships between the set of subroots and standard informational assumptions.

\nssec{Subroots and subforms}{C881}

Consider a form $Q$.  For any $w⋅∈⋅W$, define\begin{gather}
\zz
\bQ{w} = ⎨\,⁅i_*,j_*,w_*,a_*,y_*⁆\,|\,w≼w_*\,⎬. \label{D190}
\zz
\end{gather} To put this in other words, say that a quintuple is {\em weakly after} $w$ iff its decision node weakly succeeds $w$.  Then $\bQ{w}$ is the set of quintuples weakly after $w$.  A {\em (Selten) subroot} is a member of
\begin{gather}
\zz
T = ⎨⋅t∈W⋅|⋅\tJ⋅\text{and}⋅\pj{J}(Q⧷\tQ)⋅\text{are disjoint}⋅⎬, \label{D188}
\zz
\end{gather} where $\tJ$ abbreviates $π_J(\tQ)$ by the sentence following (\rf{C858}).  In other words, $t⋅∈⋅W$ is a subroot iff each situation listed in a quintuple weakly after $t$ is not listed in a quintuple anywhere else.  Note that the pentaform's root $r$ is a subroot.  In other words,\begin{gather}
\zz
r⋅∈⋅T. \label{D266}
\zz
\end{gather} To see this, note $\bQ{r} = Q$, which implies $Q⧷\bQ{r} = ∅$, which implies $\bJ{r} = J$ and $\pj{J}(Q⧷\bQ{r}) = ∅$ are disjoint. 

For instance, recall that equations (\rf{D270})--(\rf{C855}) defined the example $\rQ$ by first defining $\rT$ and then defining $\rQ = ⨆_{t∈\rT}\rQ^t$.  When applied to example $\rQ$, the previous paragraph's definitions imply that $\rQ$'s set of subroots is equal to the $\rT$ used to define $\rQ$.  Thus the set $\rT$ is both the initial step in the definition of $\rQ$ and the subroot set derived from $\rQ$.  The elements of $\rT$ are underlined in Figure~\rf{C243}.  (In addition, the subroots are underlined in the entrant-and-incumbent example of Figure~\rf{C916}, and in the non-convergent examples of Figure~\rf{D515}.)

{\tablesub}

\begin{prop}[Streufert 2023p, Proposition~\rf{C576}]\label{C892} Suppose $Q$ is a (penta)form with its $T$ (\rf{D188}).  Then $(∀t∈T)$ $\tQ$ (\rf{D190}) is a (penta)form with root $t$.  \end{prop}

In accord with the proposition, call $\tQ$ the {\em  (Selten) subform at $t$}.  For instance, in example $\rQ$ of (\rf{C855}) and Figure~\rf{C243}, the subform at each subroot $t⋅∈⋅\rT$ is $\bbQ{t}{\rQ} = ⨆_{t_*≽t}\rQ^{t_*}$.  In the figure, the subroots are the underlined nodes, and each subroot's subform is the set of edges that follow the subroot.

Consider an arbitrary form $Q$ and subroot $t⋅∈⋅T$.  As with any form, use the general definitions in Sections \rf{C854}, \rf{C866}, and \rf{C269} to derive from the subform $\tQ$ its player set $\tI$, its situation set $\tJ$, its decision-node set $\tW$, its action set $\tA$, its successor-node set $\tY$, its endnode set $\tY⧷\tW$, its run collection $\tZZ = \tZZ_{\f{ft}}⋃\ZZ_{\f{inft}}$, its strategy set $\tSS$, its player situation sets $⁅\tJ_i⁆_{i∈\tI}$, and its outcome function $\tOO{:}\tSS→\tZZ$.  These derivatives and the sections defining them are listed in Table~\rf{C885}.  

Note $\tI⋅⊆⋅I$, and relatedly, an arbitrary member of $\tI$ is denoted $i$.  A similar remark can be made for $\tJ⋅⊆⋅J$, for $\tW⋅⊆⋅W$, for $\tA⋅⊆⋅A$, for $\tY⋅⊆⋅Y$, for each $\tJ_i⋅⊆⋅J_i$, and less trivially, for $\tY⧷\tW⋅⊆⋅Y⧷W$ (Lemma~\rf{D200}(\rf{D201})).  However, $\tZZ$ is typically not a subset of $\ZZ$ because a subform run in $\tZZ$ begins at node $t$ while a whole-form run in $\ZZ$ begins at node $r$.  Relatedly, an arbitrary subform run in $\tZZ$ will be denoted by something other than $Z$ (an example is the $N$ in the following paragraph).  Likewise, $\tSS$ is typically not a subset of $S$ because a subform strategy in $\tSS$ has domain $\tJ$ while a whole-form strategy has domain $J$.  Fortunately, the paper seldom needs to denote an arbitrary subform strategy in $\tSS$ (instead, (\rf{D191}) will define a suitable restriction of a whole-form strategy in $S$).

\nssec{Subgames and subgame perfection}{C884}

Now consider a game $(Q,u)$ with its stakeholder set $K$ (Definition~\rf{D207}) and its subroot set $T$ (\rf{D188}).  Then consider a particular subroot $t⋅∈⋅T$.  The previous paragraphs have defined the subform $\tQ$ with its player set $\tI$ and its run collection $\tZZ$.  Now assign the whole game's stakeholder set $K$ to the subform so that [a] there is no need for the symbol $^t\mspace{-2mu}K$ and [b] $K⧷\tI$ is the subform's bystander set by general definition (\rf{D484}).  As suggested in Section~\rf{C867}, the subform bystanders will play an essential role in the dynamic-programming results of Section~\rf{C275}.  

Further, define the (grand) utility function $\tu = \tZZ→\smash{\eR^K}$ by\begin{gather}
\zz
(∀N∈\tZZ)⋅\tu(N) = u(R(N)), \label{D189}
\zz
\end{gather} where $R$ is the weak-predecessor correspondence (\rf{D206}).  This construction is well-defined because each subform run $N⋅∈⋅\tZZ$ satisfies $R(N)⋅∈⋅\ZZ$ (Lemma~\rf{D200}(\rf{D202})) and because the domain of $u$ is $\ZZ$ (Definition~\rf{D207}).  In this fashion, each subform run $N⋅∈⋅\tZZ$ is assigned the utility of the whole-form run $R(N)⋅∈⋅\ZZ$ that it finishes.

\begin{ndef}[{{\bf Subgame}}]\label{D208} Suppose $(Q,u)$ is a game (Definition~\rf{D207})  with its subroot set $T$ (\rf{D188}).  Then, at each subroot $t⋅∈⋅T$, the {\em (Selten) subgame at $t$} is the pair $(\tQ,\tu)$ listing the subform $\tQ$ (\rf{D190}) and the utility function $\tu$ (\rf{D189}).   \end{ndef}

For each whole-form strategy $s⋅∈⋅S$ and each subroot $t⋅∈⋅T$, define the restriction \begin{gather}
\zz
\tss = s|_{\tJ}˙. \label{D191}
\zz
\end{gather} It is easy to show%
\footnote{\label{D483}$S$'s general definition (\rf{D376}) and $s⋅∈⋅S$ imply that $s$ is a function from $J$ such that $(∀j∈J)$ $s(j)⋅∈⋅A_j$.  Thus $\tJ⋅⊆⋅J$ and definition (\rf{D191}) imply that the restriction $\tss$ is a function from $\tJ$ such that $(∀j∈\tJ)$ $\tss(j)⋅∈⋅A_j$.  Meanwhile, the general definition (\rf{D376}) of strategy set, applied to the subform $\tQ$, implies $\tSS = ⎨˙f$ is a function from $\tJ˙|˙(∀j∈\tJ)˙f(j)∈A_j˙⎬$.  The previous two sentences imply that the restriction $\tss$ belongs to the strategy set $\tSS$.} 
that $\tss⋅∈⋅\tSS$.  Hence $\tss$ is the subform-$\tQ$ strategy within the whole-form strategy $s⋅∈⋅S$.  Further, for each $i⋅∈⋅I$, define \begin{gather}
\zz
\tss_i = s|_{\tJ_i}⋅\text{and}⋅\tss_{-i} = s|_{\tJ⧷\tJ_i}. \label{D259}
\zz
\end{gather}These restrictions are the player's subform strategy, and the player's opponents' subform strategies, within the whole-form strategy $s⋅∈⋅S$.%
\footnote{Definition (\rf{D259}) is consistent with the general definition (\rf{D253}) for player strategies.  Specifically, since $\tss⋅∈⋅\tSS$, general definition (\rf{D253}) would set $(\tss)_i$ equal to $\tss|_{\tJ_i}$, which by (\rf{D191}) is $(s|_{\tJ})|_{\tJ_i}$ which by $\tJ_i⋅⊆⋅\tJ$ reduces to $s|_{\tJ_i}$.  The argument for $(\tss)_{-i}$ is similar.}

\newcommand{\notespe}{\footnote{\label{C889}At the expense of more notation, definition (\rf{D234}) could be quantified by $(∀t∈T,i∈\tI,φ∈\tSS)$ with $φ|_{\tJ_i}$ replacing $\tsig_i$ in the inequality.  This is equivalent because the inequality holds vacuously as an equality for each player $i$ in $I⧷\tI$, and because the inequality is unaffected by the actions of the alternative strategy~$σ$ over situations $j$ in $J⧷\tJ$.  Further, at the expense of still more notation, one could define player $i$'s strategy set $\tSS_i$ (as in footnote~\rf{D258}), and then quantify (\rf{D234}) by $(∀t∈T,i∈\tI,λ∈\tSS_i)$ with $λ$ replacing $\tsig_i$ in the inequality (this additional construction is like footnote~\rf{D260}).}}

A {\em subgame-perfect equilibrium} in a game $(Q,u)$ is a strategy $s⋅∈⋅S$ such that, at each subroot $t⋅∈⋅T$, the restriction $\tss$ is a Nash equilibrium (\rf{D187}) in the subgame $(\tQ,\tu)$.  In other words, a subgame-perfect equilibrium is a strategy $s⋅∈⋅S$ such that{\notespe} \begin{gather}
\zz
(∀t∈T,i∈I,σ∈S)⋅\tu_i(\tOO(⋅\tss⋅))⋅≥⋅\tu_i(\tOO(⋅\tsig_i,˙\tss_{-i}˙)). \label{D234}
\zz
\end{gather} Incidentally, a subgame-perfect equilibrium is necessarily a Nash equilibrium because $r⋅∈⋅T$ by (\rf{D266}).

Finally, the end of Section~\rf{C269} observed that the definition of Nash equilibrium (\rf{D187}) threatens to be intractable in the example $(\rQ,\ru)$.  The same remark also applies in each of the infinitely many subgames of $(\rQ,\ru)$.  Hence the definition of subgame perfection (\rf{D234}) threatens to be intractable in examples like this one.

\nssec{Subroots and informational assumptions}{C270}

Section~\rf{C275} will present this paper's main results.  None of the standard informational assumptions will be imposed.  In particular, imperfect information, imperfect recall, and even absentmindedness will be allowed.

Nonetheless, each subroot should be regarded as a very specialized informational assumption.  To start exploring this, imagine games with few subroots.  In the extreme, the only subroot is the root node $r$ (that is $T = ⎨r⎬$).  In such games, Nash equilibrium and subgame perfection are equivalent concepts, and the results of this paper are vacuous.  In brief, no assumptions no results.  Note that there are interesting games with only one subroot.  Some examples are poker (Zhang and Sandholm 2021), repeated games with imperfect monitoring (Abreu, Pearce, and Stachetti 1990), and an unusual example in footnote~\rf{C418} below.\nocite{ZhangS21}\nocite{AbreuPS90}

Next imagine games with many subroots.  In the extreme, every decision node is a subroot (that is $T = W$).  This extreme contingency is known as perfect information.  Here the concept of subgame perfection is most restrictive, and the results of this paper are most powerful.  In brief, strong assumptions strong results.  This extremity is intimately connected with dynamic programming and is accordingly rather well-developed (Filar and Vrieze 1997).  (In this paper, the entrant-and-incumbent example of Section~\rf{C918}, and the non-convergent examples of Section~\rf{C256}, have perfect information.  Such examples are given limited attention here.)\nocite{FilarV97}

This paper's main contribution is to explore the middle ground between these two extremes.  Some examples which fit within this paper's theory are Rubinstein's bargaining model (Shaked and Sutton 1984), repeated games with observed actions (Rubinstein and Wolinsky 1995), and the cry-wolf example $\rQ$ from (\rf{C855}) and Figure~\rf{C243}.  Other related examples fall outside this paper's theory because of chance moves, mixed strategies, or simultaneous moves with uncountably many players.  An example is block-recursive search equilibria (Menzio and Shi 2010).\nocite{ShakeS84}\nocite{RubinW95}\nocite{MenziS10}

\newcommand{\tablepiece}{
\begin{table}[t]
{\small 
\begin{tabular}{clr} 
\multicolumn{2}{l}{Piece form $Q^t$ of form $Q$ at subroot $t$} \\\hline \\[-3.6mm] 
$Q^t˙⊆˙Q$ & set of quintuples $⁅i˛j˛w˛a˛y⁆$ 
  &\mysize[\rf{C276}]\\
$I^t˙{=}˙\pj{I}(Q^t)$ & $\dve$ set of players $i$ 
  &\mysize[\rf{C854},\rf{C276}]\\
$J^t˙{=}˙\pj{J}(Q^t)$ & $\dve$ set of situations $j$ 
  &\mysize[\rf{C854},\rf{C276}]\\
$W^t˙{=}˙\pj{W}(Q^t)$ & $\dve$ set of decision nodes $w$ 
  &\mysize[\rf{C854},\rf{C276}]\\
$A^t˙{=}˙\pj{A}(Q^t)$ & $\dve$ set of actions $a$ 
  &\mysize[\rf{C854},\rf{C276}]\\
$Y^t˙{=}˙\pj{Y}(Q^t)$ & $\dve$ set of successor nodes $y$ 
  &\mysize[\rf{C854},\rf{C276}]\\
$⎨t⎬˙{=}˙W^t⧷Y^t$ & $\dve$ root node $t$
  &\mysize[\rf{C276}]\\
$Y^t⧷W^t$ & $\dve$ set of endnodes $y$
  &\mysize[\rf{C866},\rf{C895}]\\
$\ZZ^t$ & $\dve$ collection of runs $N$ (not $Z$)
  &\mysize[\rf{C866},\rf{C895}]\\
$S^t$ & $\dve$ set of (grand) strategies
  &\mysize[\rf{C269},\rf{C896}]\\
$J^t_i$ & $\dve$ player $i$'s set of situations $j$ 
  &\mysize[\rf{C269},\rf{C896}]\\
$O^t$ & $\dve$ outcome function 
  &\mysize[\rf{C269},\rf{C896}]\\%
[2mm]
$s^t˙{=}˙s|_{J^t}˙∈˙S^t$ & $\dve$ piece restriction of $s˙∈˙S$
  &\mysize[\rf{C896}]\\
$s^t_i˙{=}˙s|_{J^t_i}$ & $\dve$ player $i$'s piece restriction of $s˙∈˙S$ 
  &\mysize[\rf{C896}]\\%
[2mm]
\multicolumn{3}{l}{Piece game $(Q^t,u^t_v)$ of game $(Q,u)$ at subroot $t$ and value function $v$} \\\hline \\[-3.6mm]
$u^t_v{:}\ZZ^t→Ṛ^K$ & utility function derived from $u$ and $v$ &\mysize[\rf{C291}]\\
$K$ & $\dve$ set of stakeholders $k$ &\mysize[\rf{C291}]\\
$K⧷I^t$ & $\dve$ set of bystanders $k$ &\mysize[\rf{C291}]\\%
[3mm]
\end{tabular} }
\caption{\small Piece forms and games are implicitly accompanied by their derivatives (\protect\rotatebox[origin=c]{180}{$\Lsh$}).  Definitions are located in the sections in brackets {\mysize [˙]}.} \label{C886}
\end{table}
}

\section{Dynamic-Programming Results}\label{C275}\showit
\markb{\sc \rf{C275}.\ Dynamic-Programming Results}

The concepts and results in this Section~\rf{C275} are new.

\nssec{Piece forms}{C276}

Consider a form $Q$.  For each subroot $t⋅∈⋅T$, let\footnote{The subscripts $\hi$ and $\lo$ can be read as ``high'' and ``low'', respectively.} \begin{gather}
\zz
Q^t = \tQ˙⧷˙⋃_{t≺t\hi∈T}\,{\bQ{t\hi}}. \label{D192}
\zz
\end{gather} Thus each $Q^t$ is the set of quintuples that are weakly after $t$ but not weakly after a subroot successor of $t$ (this characterization is not quite obvious and is proved in Lemma~\rf{C463}(\rf{C464})).  The following proposition uses a general result for subsets of (penta)forms (Streufert 2023p, Corollary~\rf{C579}) to show that each $Q^t$ is a (penta)form.  In accord with this result, call each $Q^t$ the {\em piece form} at $t$.\footnote{The prominence of Selten 1975 has led the present author and game theorists in general to use the terms ``subtree'', ``subform'', and ``subgame'' more restrictively than graph theorists and category theorists would (for a full discussion see Streufert 2021Gm, Theorem~3.2).  Relatedly, graph theorists and category theorists would likely view this paper's ``piece form'' as another special kind of their less-restrictive notion of subform.} 

\begin{prop}\label{C282} Suppose $Q$ is a (penta)form and $t⋅∈⋅T$.  Then $Q^t$ is a \linebreak (penta)form with root $t$.  (Proof \rf{C282p}.) \end{prop}

For instance, equations (\rf{D270})--(\rf{C855}) defined the example $\rQ$ by first defining $\rT$ and then defining $\rQ = ⨆_{t∈\rT}\rQ^t$.  Section~\rf{C881} noted that $\rT$ is both the initial step in the definition of $\rQ$ and the subroot set derived from $\rQ$.  Now notice that the piece form at $t$ is equal to $\rQ^t$.  Thus the indexed collection $⁅\rQ^t⁆_{t∈\rT}$ is both the second step in the definition of $\rQ$ and the indexed collection of piece forms derived from $\rQ$.  

Proposition~\rf{C277}(\rf{C278}) shows that an arbitrary form $Q$ is partitioned by its piece forms $⁅Q^t⁆_{t∈T}$.  Further, use the general definitions in Section \rf{C854} to derive from each $Q^t$ its player set $I^t$, its situation set $J^t$, its decision-node set $W^t$, its action set $A^t$, and its successor-node set $Y^t$.  The proposition's part (\rf{C279}) shows that the situation set $J$ of a form is partitioned by the situation sets $⁅J^t⁆_{t∈T}$ of its piece forms.  Then part (\rf{C280}) does the same for decision nodes, and part (\rf{C281}) does the same for successor nodes.  Note that the proposition has no results about players and actions.  In fact, piece forms do not typically partition players and actions.  Rather, the same players can move in several piece forms, and similarly, the same actions can label edges in several piece forms.  This happens frequently in example $\rQ$ from equation (\rf{C855}).

\begin{prop}\label{C277} Suppose $Q$ is a form.  Then the following hold. \begin{tlist}
\yl{C278} $⁅Q^t⁆_{t∈T}$ is an injectively indexed partition of $Q$. 
\yl{C279} $⁅J^t⁆_{t∈T}$ is an injectively indexed partition of $J$.
\yl{C280} $⁅W^t⁆_{t∈T}$ is an injectively indexed partition of $W$.
\yl{C281} $⁅Y^t⁆_{t∈T}$ is an injectively indexed partition of $Y$.
(Proof~\rf{C277p}.) \end{tlist} \end{prop}

\tablepiece

\nssec{Piece endnodes and piece runs}{C895}

Consider a form $Q$ and its piece forms $⁅Q^t⁆_{t∈T}$.  For each $Q^t$, the set $Y^t⧷W^t$ consists of the piece's endnodes (this follows from the general definition of endnode near (\rf{D279})).  Then, for the whole form $Q$, call $⨆_{t∈T}(Y^t⧷W^t)$ the set of {\em piece endnodes}, and call $Y⧷W$ the set of {\em final endnodes}.  The following proposition relates piece endnodes to final endnodes.

\begin{prop}\label{C390} Suppose $Q$ is a form.  Then $⎨⎨r⎬⎬⋃⎨Y^t⧷W^t≠∅|t∈T⎬$ partitions $T˙⋃˙(Y⧷W)$. (Proof~\rf{C390p}.) \end{prop}

First, the proposition implies that each piece endnode $y$ in $⨆_{t∈T}(Y^t⧷W^t)$ is either a subroot in $T$ or a final endnode in $Y⧷W$, but not both.  Both is impossible because $T⋅⊆⋅W$ by $T$'s definition (\rf{D188}).

Conversely, the proposition implies that each final endnode is also a piece endnode in exactly one piece.  To be clear, consider a final endnode $y⋅∈⋅Y⧷W$.  The proposition implies that $y$ is [a] equal to $r$ or [b] equal to a piece endnode in exactly one $Y^t⧷W^t$.  Contingency [a] cannot hold because $y⋅∉⋅W$ (by assumption) and $r⋅∈⋅W$ (by $r$'s definition (\rf{D186})).  Hence the final endnode $y$ is a piece endnode in exactly one $Y^t⧷W^t$.

Now consider runs instead of endnodes.  The new topic is more expansive in the sense that endnodes concern only finite runs.  In preparation for the next proposition, consider a piece form $Q^t$ in a form $Q$.  Then use the general definition of runs (\rf{D281}) to derive the piece's run collection $\ZZ^t$.  Typically a piece run is not a whole-form run, and thus $\ZZ^t$ is typically not a subset of $\ZZ$.  In order to reserve $Z$ for a whole-form run in $\ZZ$, the symbol $N$ is used for a piece run in $\ZZ^t$. 
 
\begin{prop}\label{C391} Suppose $Q$ is a form and $t⋅∈⋅T$.  Then, for all $N⋅∈⋅\ZZ^t$, exactly one of the following holds. \begin{tlist}
\yl{C897} $R(N)⋅∉⋅\ZZ$, $N$ is finite, and $\max˙N$\! exists and is in $T$.
\yl{C898} $R(N)⋅∈⋅\ZZf$, $N$ is finite, and $\max˙N$\! exists and is in $Y⧷W$.
\yl{C899} $R(N)⋅∈⋅\ZZi$, $N$ is infinite, and $\max˙N$\! does not exist.
\end{tlist}  (Proof~\rf{C391p}.) \end{prop}

Proposition~\rf{C391} shows that a piece run $N$ either (\rf{C897}) terminates at a subsequent subroot, (\rf{C898}) terminates at a final endnode and completes a finite full run, or (\rf{C899}) completes an infinite full run.  For instance, consider the piece form $\rQ^{\f6}$ in example $\rQ$ of (\rf{C855}) and Figure~\rf{C243}.  That piece form has five piece runs.  The piece run $N = ⎨\f6,\f{61},\f{68}⎬⋅∈⋅\ZZ^{\f6}$ terminates at the subsequent subroot $\max˙N = \f{68}⋅∈⋅T$, in accord with the proposition's contingency (\rf{C897}).  Similarly, the piece runs $⎨\f6,\f{61},\f{63},\f{67}⎬$, and $⎨\f6,\f{61},\f{63},\f{66}⎬$ terminate at the subsequent subroots $\f{67}$ and $\f{66}$, respectively.  Meanwhile, the piece run $N = ⎨\f6,\f{62},\f{64}⎬⋅∈⋅\ZZ^{\f6}$ terminates at the final endnode $\f{64}⋅∈⋅Y⧷W$ and completes the finite full run $R(N) = ⎨⎨⎬,\f1,\f3,\f6,\f{62},\f{64}⎬⋅∈⋅\ZZf$, in accord with contingency (\rf{C898}).  Similarly, $⎨\f6,\f{62},\f{65}⎬$ completes the finite full run $⎨⎨⎬,\f1,\f3,\f6,\f{62},\f{65}⎬$.  Lastly, the entire example $\rQ$ has no infinite piece runs.  Thus the example has no piece runs in contingency~(\rf{C899}).%
\footnote{\label{C418}For an extreme example with infinite piece runs, imagine a one-player one-information-set form in which the player decides between $\f0$ and $\f1$ an infinite number of times.  To be specific, let $B = ⨆^∞_{ℓ=0}⎨\f0,\f1⎬^ℓ$ (so that $B$ is the set of all finite sequences of $\f0$'s and $\f1$'s), and let $Q = ⨆_{b∈B}⎨˙⁅\f{Joe},B,b,\f0,b⊕\f0⁆,$ $⁅\f{Joe},B,b,\f1,b⊕\f1⁆˙⎬$ (so that $B$ itself is the only information set). 
 In this extreme example, the only piece form $Q^r$ is identical to the whole form $Q$, all piece runs are infinite, and all whole-form runs are infinite.  Thus there are no piece endnodes, there are no final endnodes, and Proposition~\rf{C390} holds trivially in the sense that [i] $T = ⎨r⎬$ and [ii] both $⎨Y^t⧷W^t≠∅|t∈T⎬$ and $Y⧷W$ are empty.}

\nssec{Piece strategies}{C896}

Consider a piece form $Q^t$ of a form $Q$.  Then use the general definitions of Section~\rf{C269} to derive the piece's strategy set $S^t$ [by (\rf{D376})], the piece's player situation sets $⁅J^t_i⁆_{i∈I^t}$ [by (\rf{D184})], and the piece's outcome function $O^t{:}S^t→\ZZ^t$. 

Note that the domain of a piece strategy (in $S^t$) is $J^t⋅⊆⋅J$, and that the domain of a whole-form strategy (in $S$) is $J$.  Thus $S^t$ is typically not a subset of $S$.  Relatedly, for each whole-form strategy $s⋅∈⋅S$, define the restriction \begin{gather}
\zz
s^t = s|_{J^t}˙. \label{D193}
\zz
\end{gather} It is easy to show%
\footnote{\label{D510}(The argument here is similar to that of footnote~\rf{D483}, and also plays a role in a later proof.) $S$'s general definition (\rf{D376}) and $s⋅∈⋅S$ imply that $s$ is a function from $J$ such that $(∀j∈J)$ $s(j)⋅∈⋅A_j$.  Thus $J^t⋅⊆⋅J$ and definition (\rf{D193}) imply that the restriction $s^t$ is a function from $J^t$ such that $(∀j∈J^t)$ $s^t(j)⋅∈⋅A_j$.  Meanwhile, the general definition (\rf{D376}) of strategy set, applied to the piece form $Q^t$, implies $S^t = ⎨˙f$ is a function from $J^t˙|˙(∀j∈J^t)˙f(j)∈A_j˙⎬$.  The previous two sentences imply that the restriction $s^t$ belongs to the piece strategy set $S^t$.} 
that the restriction $s^t$ satisfies $s^t⋅∈⋅S^t$.  Hence $s^t$ is the piece strategy from within the whole-form strategy $s$.

Further, consider a player $i⋅∈⋅I$.  Note that $J^t_i⋅⊆⋅J^t⋅⊆⋅J$, and define the restrictions \begin{gather}
\zz
s^t_i = s|_{J^t_i}⋅\text{and}⋅s^t_{-i} = s|_{J^t⧷J^t_i}.  \label{D267}
\zz
\end{gather} These restrictions are the player's piece strategy, and the player's opponents' piece strategies, from within the whole-form strategy $s⋅∈⋅S$.%
\footnote{\label{D369}Definition (\rf{D267}) is consistent with the general definition (\rf{D253}) of player strategy sets.  Specifically, since $s^t⋅∈⋅S^t$, general definition (\rf{D253}) would set $(s^t)_i$ equal to $s^t|_{J^t_i}$, which by (\rf{D193}) is $(s|_{J^t})|_{J^t_i}$, which by $J^t_i⋅⊆⋅J^t$ reduces to $s|_{J^t_i}$.  The argument for $(s^t)_{-i}$ is similar.}

\nssec{Value functions}{C287}

Consider a game $(Q,u)$.  Let a {\em (grand) value function} be a function of the form $v{:}T→\smash{\eR^K}$.  Accordingly, a value function maps each subroot $t$ to an extended-real-valued profile of the form $v(t) = ⁅v_k(t)⁆_{k∈K}$.  Note that a value function may or may not be meaningful.  It is merely a function from $T$ which assumes values in $\smash{\eR^K}$.

\newcommand{\notepicayuneadmissibility}{\footnote{This paraphrase of admissibility is slightly stronger than the actual concept because the actual concept uses infimum and supremum.}}

``Admissibility'' was informally introduced in Section~\rf{C255} as property \rf{D261}.  Roughly,{\notepicayuneadmissibility} a value function is admissible iff, for each subroot $t$ and each stakeholder $k$, [a] the value $v_k(t)$ is weakly higher than the utility from some ``lower'' run $Z\lo$ going through $t$, and at the same time, [b] the value $v_k(t)$ is weakly lower than the utility from some  ``higher'' run $Z\hi$ going through $t$ (the runs $Z\lo$ and $Z\hi$ are typically different).  Formally, a value function $v$ is {\em admissible} iff \begin{gather}
\zz
(∀t∈T,k∈K)⋅\inf⎨u_k(Z\lo)|t∈Z\lo∈\ZZ⎬⋅≤⋅v_k(t)⋅≤⋅\sup⎨u_k(Z\hi)|t∈Z\hi∈\ZZ⎬. 
\label{C921}
\zz
\end{gather} In this sense, $v$ is admissible iff at each subroot, each stakeholder is neither overly pessimistic nor overly optimistic.

``Persistence'' was informally introduced in Section~\rf{C255} as property~\rf{D262}.  Informally, a value function is persistent for a strategy iff, at each subroot, the value is equal to [a] the value at the next subroot determined by the strategy, if that subroot exists, or otherwise [b] the utility of the full run determined by the strategy.  Formally, a value function $v$ is {\em persistent for a strategy} $s$ iff\begin{gather}
\zz
(∀t∈T)⋅v(t) = \mip
v(\max˙O^t(s^t))⋅\text{if}⋅\max˙O^t(s^t)⋅\text{exists and is in $T$} \\[.6ex]
\rule{1.1ex}{0ex}u(R(O^t(s^t)))⋅\text{otherwise}\rule{21.5ex}{0ex} \mop.
\label{C922}
\zz
\end{gather} 
In the first of the two cases, $v(\max˙O^t(s^t))⋅∈⋅\smash{\eR^K}$ is well-defined because $v{:}T→\smash{\eR^K}$.  This $v(\max˙O^t(s^t))$ is the value profile at the subsequent subroot determined by $t$ and $s^t$.  Meanwhile in the ``otherwise'' case, Proposition~\rf{C391} implies $R(O^t(s^t))⋅∈⋅\ZZ$, and thus, $u(R(O^t(s^t)))⋅∈⋅\smash{\eR^K}$ is well-defined since $u{:}\ZZ→\smash{\eR^K}$.  This $u(R(O^t(s^t)))$ is the utility profile of the full run determined by $t$ and $s^t$. 

``Authenticity'' was informally introduced in Section~\rf{C255} as property \rf{D263}.  Informally, a value function is authentic for a strategy iff, at each subroot, the value is equal to the utility of the run that results from following the strategy.  Formally, a value function $v$ is {\em authentic for a strategy} $s$ iff\begin{gather}
\zz
(∀t∈T)⋅v(t) = u(R(\tOO(\tss))). \label{C923}
\zz
\end{gather}

\nssec{Upper- and lower-convergence, and Theorem~\rf{C290}}{C289}

As suggested in Section~\rf{C255}, both admissibility and persistence are easily implied by authenticity (Lemma~\rf{C431}).  Conversely, this section's Theorem~\rf{C290} will show broad conditions under which admissibility and persistence together imply authenticity. These ``broad conditions'' are upper- and lower-convergence.\footnote{In fact, upper- and lower-convergence are slightly stronger than necessary for the results of this paper.  These assumptions will be used only at runs $Z⋅∈⋅\ZZ$ for which $Z⋂T$ is infinite.}

\newcommand{\notelim}{\footnote{\label{D232}The general definition of convergence over a directed set is more complicated than its application here.  In particular, if the utility $u_k(Z)$ is finite, the equality in equation (\rf{C924}) holds iff, for all $ε > 0$, there is a node $x⋅∈⋅Z$, such that for all $x_+⋅∈⋅Z$ satisfying $x⋅≼⋅x_+$, the number $\sup˙⎨u_k(Z\hi)|x_+∈Z\hi∈\ZZ⎬$ belongs to the interval $(u_k(Z){-}ε,u_k(Z){+}ε)$.  If $u_k(Z) = {-}∞$, the equality holds iff, for all $b⋅∈⋅Ṛ$, there is $x⋅∈⋅Z$, such that for all $x_+⋅∈⋅Z$ satisfying $x⋅≼⋅x_+$, the number $\sup˙⎨u_k(Z\hi)|x_+∈Z\hi∈\ZZ⎬$ is less than $b$.  The case $u_k(Z) = ∞$ is similar.}}

Consider a game $(Q,u)$.  Section~\rf{C256} informally introduced ``upper-convergence'' as the property that conceivable utility increments eventually vanish along each run.  This was illustrated by Figure~\rf{C243}'s cry-wolf example $(\rQ,\ru)$ satisfying upper-convergence, and by Figure~\rf{D515}(ii)'s ``Ann'' example violating upper-convergence.

Formally, $u{:}\ZZ→\smash{\eR^K}$ is {\em upper-convergent} iff\begin{gather}
\zz
(∀Z∈\ZZ,k∈K)⋅\text{lim}_{x∈Z}⋅\text{sup}⎨u_k(Z\hi)|x∈Z\hi∈\ZZ⎬ = u_k(Z), 
\label{C924}
\zz
\end{gather} where the limit is taken with respect to the set $Z$ as directed by $≼$ (Kelley 1955, Chapter 2; Munkres 2000, page~187, Exercise 3).{\notelim} \nocite{Kelle55} \nocite{Munkr00} The remainder of this paragraph discusses (\rf{C924}) in a way which is consistent with Section~\rf{C256}'s informal introduction.  To begin, first consider a node $x⋅∈⋅X$, and note that $⎨Z\hi|x∈Z\hi∈\ZZ⎬$ is the set of runs that are still conceivable after reaching node~$x$.  Second consider a stakeholder $k⋅∈⋅K$, and note that  $\text{sup}˙⎨u_k(Z\hi)|x∈Z\hi∈\ZZ⎬$ is essentially the highest stakeholder-$k$ utility that is still conceivable after reaching node~$x$.  Third consider a run $Z⋅∈⋅\ZZ$ such that $x⋅∈⋅Z$.  In other words, consider a specific run $Z$ that actually reaches node~$x$.  Since $⎨Z\hi|x∈Z\hi∈\ZZ⎬$ contains $Z$, it must be that $\text{sup}˙⎨u_k(Z\hi)|x∈Z\hi∈\ZZ⎬$ weakly exceeds $u_k(Z)$.  The difference between the two can be called the {\em conceivable utility increment to $u_k(Z)$ at $x$}.  Upper-convergence at $Z$ for $k$ means that the conceivable utility increment to $u_k(Z)$ at $x$ eventually vanishes as the node $x$ moves away from the root along the specific run $Z$.  This is consistent with Lemma~\rf{D284}, which shows that the limit in (\rf{C924}) always exists, and that upper-convergence fails exactly when \begin{gather}
\zz
(∃Z∈\ZZ,k∈K)⋅\text{\normalfont lim}_{x∈Z}⋅\text{\normalfont sup}⎨u_k(Z\hi)|x∈Z\hi∈\ZZ⎬ > u_k(Z). \notag
\zz
\end{gather} 

Similarly, Section~\rf{C256} casually introduced ``lower-convergence'' as the property that conceivable utility decrements eventually vanish along each run.  This was illustrated by Figure~\rf{C243}'s cry-wolf example $(\rQ,\ru)$ satisfying lower-convergence, and by Figure~\rf{D515}(iii)'s ``Bob'' example violating lower-convergence.  

Formally, $u{:}\ZZ→\smash{\eR^K}$ is {\em lower-convergent} iff\begin{gather}
\zz
(∀Z∈\ZZ,k∈K)⋅\text{lim}_{x∈Z}⋅\text{inf}⎨u_k(Z\lo)|x∈Z\lo∈\ZZ⎬ = u_k(Z), 
\label{C925}
\zz
\end{gather} where, as in (\rf{C924}), the limit is taken with respect to the set $Z$ as directed by $≼$.  The remainder of this paragraph discusses (\rf{C925}) in a way which is consistent with Section~\rf{C256}'s informal introduction.  To begin, first consider a node $x⋅∈⋅X$, and note that $⎨Z\lo|x∈Z\lo∈\ZZ⎬$ is the set of runs that are still conceivable after reaching node~$x$.  Second consider a stakeholder $k⋅∈⋅K$, and note that  $\text{inf}˙⎨u_k(Z\lo)|x∈Z\lo∈\ZZ⎬$ is essentially the lowest stakeholder-$k$ utility that is still conceivable after reaching node~$x$.  Third consider a run $Z⋅∈⋅\ZZ$ such that $x⋅∈⋅Z$.  In other words, consider a specific run $Z$ which reaches node~$x$.  Since $⎨Z\lo|x∈Z\lo∈\ZZ⎬$ contains $Z$, it must be that $\text{inf}˙⎨u_k(Z\lo)|x∈Z\lo∈\ZZ⎬$ is weakly below $u_k(Z)$.  The difference between the two can be called the {\em conceivable utility decrement to $u_k(Z)$ at $x$}.  Lower-convergence at $Z$ for $k$ means that the conceivable utility decrement to $u_k(Z)$ at $x$ eventually vanishes as the node $x$ moves away from the root along the specific run $Z$.  This is consistent with Lemma~\rf{D292}, which shows that the limit in (\rf{C925}) always exists, and that upper-convergence fails exactly when \begin{gather}
\zz
(∃Z∈\ZZ,k∈K)⋅\text{\normalfont lim}_{x∈Z}⋅\text{\normalfont inf}⎨u_k(Z\lo)|x∈Z\lo∈\ZZ⎬ < u_k(Z). \notag
\zz
\end{gather}  

\begin{nthm}\label{C290} Suppose $(Q,u)$ is a game and $u$ is both upper- and lower-convergent.  Consider a strategy $s$ and a value function $v$.  Then [1] $v$ is admissible and persistent for $s$ if and only if [2] $v$ is authentic for $s$. (Proof~\rf{C290p}.) \end{nthm}

Consider the special case $\ZZ = \ZZf$.  In other words, consider finite-horizon games.  Here Theorem~\rf{C290} immediately applies because both upper- and lower-convergence hold trivially at finite runs (Lemma~\rf{C405}).  Further, this special case can be more easily addressed in another way:\ if $\ZZ = \ZZf$, then persistence and authenticity are equivalent, and either implies admissibility (Lemma~\rf{C926}).

Theorem~\rf{C290} is also consistent with Section~\rf{C245}'s infinite-horizon examples.  To be specific, consider the theorem's forward direction only (the reverse is easy).  Figure~\rf{C250} depicts a strategy and value function for Figure~\rf{C243}'s cry-wolf game $(\rQ,\ru)$.  There upper- and lower-convergence hold, and the combination of admissibility and persistence implies authenticity (in particular, admissibility, persistence, and authenticity all hold).  Meanwhile, Figure~\rf{D517}(ii) depicts a strategy/value-function pair for Figure~\rf{D515}(ii)'s ``Ann'' game.  There upper-convergence fails, and the combination of admissibility and persistence does not imply authenticity.  Symmetrically, Figure~\rf{D517}(iii) depicts a strategy/value-function pair for Figure~\rf{D515}(iii)'s ``Bob'' game.  There lower-convergence fails, and once again the combination of admissibility and persistence does not imply authenticity.

\nssec{Piece games, piecewise-Nashness, and Theorem~\rf{C292}}{C291}
 
Consider a game $(Q,u)$ with its stakeholder set $K$ (Definition~\rf{D207}) and its subroot set $T$ (\rf{D188}), as summarized in Table~\rf{C870}.  Then consider a particular subroot $t⋅∈⋅T$.  Sections \rf{C276} and \rf{C895} defined the piece form $Q^t$ with its player set $I^t$ and run collection $\ZZ^t$, as summarized in Table~\rf{C886}.  Now assign the whole game's stakeholder set $K$ to the piece so that [a] there is no need for the symbol $K^t$, and [b] $K⧷I^t$ is the piece's bystander set by general definition (\rf{D484}).  As suggested in Section~\rf{C867}, the piece's bystanders play an essential role in the dynamic-programming technique of Theorem~\rf{C292}.

Now consider a value function $v{:}T→\smash{\eR^K}$, and define the piece's (grand) utility function $u^t_v = u^t_v{:}\ZZ^t→\smash{\eR^K}$ by \begin{gather}
\zz
(∀N∈\ZZ^t)⋅u^t_v(N) = \mip
v(\max˙N)⋅\text{if}⋅\max˙N⋅\text{exists and is in $T$} \\[.6ex]
u(R(N))⋅\text{otherwise}\rule{16ex}{0ex} \mop.
\label{C972}
\zz
\end{gather} 
In the first case, $v(\max˙N)⋅∈⋅\eR^K$ is well-defined since $v{:}T→\smash{\eR^K}$.  This $v(\max˙N)$ is the value profile at the subsequent subroot reached by the piece run $N$.  Meanwhile in the ``otherwise'' case, Proposition~\rf{C391} implies $R(N)⋅∈⋅\ZZ$, and thus, $u(R(N))$ is well-defined since $u{:}\ZZ→\smash{\eR^K}$.  This $u(R(N))$ is the utility profile of the full run finished by the piece run $N$.

\pagebreak

\begin{ndef}[{{\bf Piece game}}]\label{D209} Suppose that $(Q,u)$ is a game with its stakeholder set $K$ (Definition~\rf{D207}) and subroot set $T$ (\rf{D188}).  Further suppose $v{:}T→\smash{\eR^K}$.  Then, at each subroot $t⋅∈⋅T$, the {\em piece game at $t$ and $v$} is the pair $(Q^t,u^t_v)$ listing the piece form $Q^t$ (\rf{D192}) and the utility function $u^t_v$ (\rf{C972}). \end{ndef}

\newcommand{\notePNash}{\footnote{\label{D268}At the expense of more notation, definition (\rf{D223}) could be quantified by $(∀t∈T,i∈I^t,{\psi}∈S^t)$ with ${\psi}|_{J^t_i}$ replacing $σ^t_i$ in the inequality.  This is equivalent because the inequality holds vacuously as an equality for each player in $I⧷I^t$, and because the inequality is unaffected by the actions of the alternative strategy $σ$ over the situations in $J⧷J^t$.  Further, at the expense of still more notation, one could define player $i$'s piece strategy set $S^t_i$ (as in footnote~\rf{D258}) and then quantify (\rf{D223}) by $(∀t∈T,i∈I^t,δ∈S^t_i)$ with $δ$ replacing $σ^t_i$ in the inequality (this additional construction is like footnote~\rf{D260}). }}

``Piecewise-Nashness'' has been informally introduced as ``stepwise-optimality'' in Section~\rf{C918}, and as ``property \rf{D264}'' in  Section~\rf{C255}.  Formally, a strategy-value pair $(s,v)$ is said to be {\em piecewise-Nash} iff, at each subroot $t$, the restriction $s^t$ (from (\rf{D193})) is a Nash equilibrium in the piece game $(Q^t,u^t_v)$.  In other words, $(s,v)$ is piecewise-Nash iff{\notePNash} \begin{gather}
\zz
(∀t∈T,i∈I,σ∈S)⋅u^t_{v,i}(O^t(⋅s^t⋅))⋅≥⋅u^t_{v,i}(O^t(⋅σ^t_i,⋅s^t_{-i}⋅)).
\label{D223}
\zz
\end{gather} Note that piecewise-Nashness is a property of $(s,v)$ rather than $s$ alone because $v$ is used to construct the utility function $u^t_v$ of each piece game.  (Similarly, authenticity and persistence are properties of $(s,v)$.  In contrast, admissibility is a property of $v$ alone.) 

\begin{nthm}\label{C292} Suppose $(Q,u)$ is a game and $u$ is lower-convergent.  Consider a strategy $s$.  Then [1] there is a value function $v$ such that $(s,v)$ satisfies authenticity and piecewise-Nashness if and only if [2] $s$ is a subgame-perfect equilibrium. (Proof~\rf{C292p}.) \end{nthm}
 
\begin{ncrly}[{{\bf combines Theorems \rf{C290} and \rf{C292}}}]\label{C293} Suppose $(Q,u)$ is a game and $u$ is both upper- and lower-convergent.  Consider a strategy $s$.  Then [1] there is an admissible value function $v$ such that $(s,v)$ satisfies persistence and piecewise-Nashness if and only if [2] $s$ is a subgame-perfect equilibrium.\footnote{\label{C933}If $\ZZ = \ZZf$, then the combination of persistence and piecewise-Nashness is equivalent to subgame perfection.   This follows from Corollary~\rf{C293} because finiteness implies upper- and lower-convergence (Lemma~\rf{C405}) and because finiteness and persistence imply admissibility (Lemma~\rf{C926}).} (Follows immediately from Theorems \rf{C290} and \rf{C292}.) \end{ncrly}

For example, Figure~\rf{C250} depicts a strategy/value-function pair for Figure~\rf{C243}'s cry-wolf game $(\rQ,\ru)$.  As discussed in Sections \rf{C255} and \rf{C256}, lower-convergence, upper-convergence, admissibility, persistence, and piecewise-Nashness are all satisfied.  Thus Corollary~\rf{C293} implies subgame perfection (there seems no easier way to reach this conclusion).  In contrast, Figure~\rf{D518} depicts a strategy/value-function pair for Figure~\rf{D515}(iii)'s ``Bob'' game.  There lower-convergence fails; and admissibility, persistence, and piecewise-Nashness do not imply subgame perfection.  This is consistent with Corollary~\rf{C293}.

\nssec{One-piece unimprovability}{C493}

\newcommand{\notequantone}{\footnote{Footnote~\rf{D268} applies here as well.  In this fashion, definition (\rf{D210}) can be equivalently quantified by $(∀t∈T,i∈I^t,\psi∈S^t)$, or by $(∀t∈T,i∈I^t,δ∈S^t_i)$, at the expense of more notation.}}

A strategy $s⋅∈⋅S$ is said to be {\em one-piece unimprovable} iff{\notequantone} \begin{gather}
\zz
(∀t∈T,i∈I,σ∈S)⋅u_i(R(\tOO(˙\tss˙)))⋅≥⋅u_i(R(\tOO(˙σ^t_i,˙s|_{\tJ⧷J^t_i}˙))), \label{D210}
\zz
\end{gather} Thus a strategy is one-piece unimprovable iff no player has a one-piece deviation $σ^t_i$ which can improve their utility in the event that the piece is reached.

Because of definition (\rf{D189}) for $⁅\tu⁆_{t∈T}$, we have that definition (\rf{D234}) for subgame perfection is equivalent to \begin{gather}
\zz
(∀t∈T,i∈I,σ∈S)⋅u_i(R(\tOO(˙\tss˙)))⋅≥⋅u_i(R(\tOO(˙\tsig_i,˙\tss_{-i}˙))). \label{D235}
\zz
\end{gather} By comparing (\rf{D210}) with (\rf{D235}), it is apparent that one-piece unimprovability differs from subgame perfection to the extent that one-piece unimprovability considers only one-piece deviations $σ^t_i$ while subgame perfection considers subgame deviations $\tsig_i$.  This proves the reverse direction of the following corollary.  Conversely, the corollary's proof shows the forward direction by appealing to the forward direction of Theorem~\rf{C292}.

\begin{ncrly}\label{D211} Suppose $(Q,u)$ is a game and $u$ is lower-convergent.  Consider a strategy $s$.  Then $s$ is one-piece unimprovable if and only if it is a subgame-perfect equilibrium. (Proof~\rf{D211p}.) \end{ncrly}

This result is related to Kaminski 2019, which studies backward induction.  Very roughly, its Theorem~2(b$⇒$a) shows that subgame perfection perpetuates backward, via one-piece unimprovability, toward the root of the tree.  As explained on Kaminski 2019, page 12, this result has limited use in infinite-horizon games because there, subgame perfection at later subroots is typically no easier than subgame perfection at earlier subroots.

There are also two related papers concerning one-{\em shot} unimprovability, which is the absence of a utility-increasing deviation at any one information set (this is implied by one-piece unimprovability because every information set is in some piece).  First, Hendon, Jacobsen, and Sloth 1996 goes well beyond Corollary~\rf{D211} by deriving the sequential rationality of a mixed strategy from one-shot unimprovability.  However, their brief discussion of infinite-horizon games relies upon Fudenberg and Tirole 1991's (page~110) concept of ``continuity at infinity'', which is stronger than lower-convergence since it [a] implies upper-convergence and [b] is a special kind of uniform continuity.  Second, Al\'os-Ferrer and Ritzberger 2017 derive subgame perfection from one-shot unimprovability under an assumption very similar to lower-convergence.  However, it assumes perfect information, which is not assumed here (as discussed in Section~\rf{C270}).  Thus Corollary~\rf{D211} provides a new result about unimprovability.\nocite{FudenT91book}

\appendix \sloppy

\section{Preliminaries}\label{C378}\showit
\markb{\sc Appendix \rf{C378}. Preliminaries}

\begin{lemma}\label{C966} 
Suppose $Q$ is a form.  Then the following hold.  \begin{tlist}
\yl{C988} $\ZZ$ is nonempty.
\yl{C989} $(∀Z∈\ZZ)$ $|Z|⋅≥⋅2$.
\end{tlist} \end{lemma}

\begin{pf} {\em (\rf{C988})}.  [Step~0] Note $r$'s definition (\rf{D186}) implies  $r⋅∈⋅W$.  Thus there is $x_1$ such that $⁅r˛x_1⁆⋅∈⋅\pj{WY}(Q)$.  [Step~1] If $x_1⋅∉⋅W$, then $x_1⋅∈⋅Y⧷W$, so $⎨r,x_1⎬⋅∈⋅\ZZf$ and the argument is complete.  Otherwise $x_1⋅∈⋅W$, so there is $x_2$ such that $⁅x_1˛x_2⁆⋅∈⋅\pj{WY}(Q)$.  [Step~2] If $x_2⋅∉⋅W$, then $x_2⋅∈⋅Y⧷W$, so $⎨r,x_1,x_2⎬⋅∈⋅\ZZf$ and the argument is complete.  Otherwise $x_2⋅∈⋅W$, so there is $x_3$ such that $⁅x_2˛x_3⁆⋅∈⋅\pj{WY}(Q)$.  By inspection, similar steps either terminate at some $⎨r,x_1,x_2,...x_ℓ⎬⋅∈⋅\ZZf$ or continue indefinitely.  If they continue indefinitely, $⎨r,x_1,x_2,...⎬⋅∈⋅\ZZi$. 

{\em (\rf{C989})}.  Take a run $Z⋅∈⋅\ZZ$.  Since $\ZZ$'s definition (\rf{D281}) implies $r⋅∈⋅Z$, it suffices to show that $Z⋅≠⋅⎨r⎬$.  Toward that end, suppose $Z = ⎨r⎬$.  Then $\ZZf$'s definition (\rf{D279}) implies that $Z$ is a path from $r$ to itself and that $r⋅∈⋅Y⧷W$.  The latter (doubly) contradicts $r⋅∈⋅W⧷Y$, which follows from $r$'s definition (\rf{D186}). \end{pf}

\begin{lemma}\label{D486} Suppose $Q$ is a form and $x˙∈˙X$.  Then the following hold. \begin{tlist}
\yl{D495} $R(x)$ is equal to the path in $(X,\pj{WY}(Q))$ from $r$ to $x$.
\yl{D496} $R(x)$ is finite and linearly ordered by $≼$.
\end{tlist} \end{lemma} 

\begin{pf} {\em (\rf{D495})}. Take a node $x⋅∈⋅X$.  Streufert 2023p Lemma~\rf{D146} implies the existence and uniqueness of the path in the out-tree $(X,\pj{WY}(Q))$ from $r$ to $x$.  Call this path $P$.  Note that the path definition (\rf{D493}) implies $P$ is a set.  Thus by $R$'s definition (\rf{D206}), it suffices to show that $P = ⎨x_*|x_*≼x⎬$.

For the forward direction, consider some $x_*⋅∈⋅P$.  Then since $x_*$ is on the path from $r$ to $x$, there is a path from $x_*$ to $x$, which by $≼$'s definition (\rf{D359}) implies $x_*⋅≼⋅x$.  For the reverse direction, consider some $x_*$ such that $x_*⋅≼⋅x$.  Then $≼$'s definition (\rf{D359}) implies there is a path from $x_*$ to $x$.  Meanwhile a second application of Streufert 2023p Lemma~\rf{D146} implies there is a path from $r$ to $x_*$, which by concatenation implies there is a path from $r$ to $x$ which contains $x_*$.  This implies $x_*⋅∈⋅P$ because $P$ is the unique path from $r$ to $x$.  

{\em (\rf{D496})}.  This follows from part (\rf{D495}), path definition (\rf{D493}), and $≼$'s definition (\rf{D359}). \end{pf}

\begin{lemma}\label{C907} \mbox{Suppose $Q$ is a form and $N˙⊆˙X$.  Then the following hold.} \begin{tlist}
\yl{C909} $R(N)⋅∈⋅\ZZf$ iff $\max˙N$ exists and is in $Y⧷W$.
\yl{C910} $R(N)⋅∈⋅\ZZi$ iff $N$ is an infinite subset of a path in $(X,\pj{WY}(Q))$.
\end{tlist}\end{lemma}

\begin{pf} {\em (\rf{C909})'s forward direction}.  Suppose $R(N)⋅∈⋅\ZZf$.  Then $\ZZf$'s definition (\rf{D279}) implies there is \lic{D490} $y⋅∈⋅Y⧷W$ such that \il{D489} $R(N) = R(y)$.  In steps, \rf{D489} implies $R(N)⋅⊆⋅R(y)$, which by $R(N)$'s definition implies $⨆⎨R(x)|x∈N⎬⋅⊆⋅R(y)$, which by inspection implies $(∀x∈N)$ $R(x)⋅⊆⋅R(y)$, which by $R(x)$'s definition (\rf{D206}) implies $(∀x∈N)$ $x⋅∈⋅R(y)$, which by the $R(y)$'s definition (\rf{D206}) implies \il{D497} $(∀x∈N)$ $x⋅≼⋅y$.

Meanwhile, \rf{D489} also implies $R(N)⋅⊇⋅R(y)$, which by $R(N)$'s definition implies $⨆⎨R(x)|x∈N⎬⋅⊇⋅R(y)$, which by $R(y)$'s definition (\rf{D206}) implies $⨆⎨R(x)|x∈N⎬⋅∋⋅y$, which implies there is $x_*⋅∈⋅N$ such that $R(x_*)⋅∋⋅y$, which by $R(x)$'s definition (\rf{D206}) implies $x_*⋅≽⋅y$, which by \rf{D497} implies $x_* = y$, which by $x_*⋅∈⋅N$ implies $y⋅∈⋅N$, which by \rf{D497} again implies $\max˙N = y$, which by \rf{D490} suffices.  

{\em (\rf{C909})'s reverse direction}.  Suppose $\max˙N$ exists and is in $Y⧷W$.  Then $\ZZf$'s definition (\rf{D279}) implies $R(\max˙N)⋅∈⋅\ZZf$.  Thus it suffices to show $R(\max˙N) = R(N)$.  For the forward inclusion, $\max˙N⋅∈⋅N$ implies $R(\max˙N)⋅⊆⋅⨆⎨R(x)|x∈N⎬$, which by $R(N)$'s definition implies $R(\max˙N)⋅⊆⋅R(N)$.  For the reverse inclusion, $R$'s definition (\rf{D206}) implies that $(∀x∈N)$ $R(x)⋅⊆⋅R(\max˙N)$, which by inspection implies $⨆⎨R(x)|x∈N⎬⋅⊆⋅R(\max˙N)$, which by $R(N)$'s definition implies $R(N)⋅⊆⋅R(\max˙N)$. 

{\em (\rf{C910})'s forward direction}.  Suppose $R(N)⋅∈⋅\ZZi$.  Then $\ZZi$'s definition (\rf{D280}) implies \il{D491} $R(N)$ is an infinite path.  This and $N⋅⊆⋅R(N)$ imply $N$ is a subset of a path.  Thus it remains to show that $N$ is infinite.  Toward that end, suppose $N$ were finite.  In steps, \rf{D491} and the definition of $≼$ (\rf{D359}) imply that $R(N)$ is linearly ordered, which by $N⋅⊆⋅R(N)$ implies $N$ is linearly ordered, which by the assumed finiteness of $N$ implies $\max˙N$ exists, which by $R$'s definition (\rf{D206}) implies $(∀x∈N)$ $R(x)⋅⊆⋅R(\max˙N)$, which by inspection implies $⨆⎨R(x)|x∈N⎬⋅⊆⋅R(\max˙N)$, which by $R(N)$'s definition implies \il{D498} $R(N)⋅⊆⋅R(\max˙N)$.  But, $R(\max˙N)$ is finite by Lemma~\rf{D486}(\rf{D496}), which by \rf{D498} implies $R(N)$ is finite, which contradicts \rf{D491}.

{\em (\rf{C910})'s reverse direction}.  Suppose $N$ is an infinite subset of a path.  Then $≼$'s definition (\rf{D359}) implies that \li{D499} $N$ is linearly ordered.  Further, Lemma~\rf{D486}(\rf{D496}) implies that every node has finitely many predecessors, and thus \rf{D499} implies that $\min˙N$ exists.  Therefore, \rf{D499} implies there is a bijection $⎨0˛1˛...⎬⋅∋⋅m \mapsto x_m⋅∈⋅N$ such that $x_0 = \min˙N$ and $(∀m≥1)$ $x_{m-1}⋅≺⋅x_m$.  Thus $R$'s definition (\rf{D206}) implies that \il{D501} $(∀m≥1)$ $R(x_{m-1})⋅⊊⋅R(x_m)$, and that \begin{align}
\zz
R(N) =&⋅R(x_0)⋅⋃⋅⨆_{m≥1}⋅(R(x_m)⧷R(x_{m-1})) \nt
=&⋅R(x_0)⋅⋃⋅⨆_{m≥1}[(R(x_m)⧷R(x_{m-1}))⋃⎨x_{m-1}⎬].\notag
\zz
\end{align} Consider the sets on the right-hand side.  Lemma~\rf{D486}(\rf{D495}) implies that $R(x_0)$ is the path from $r$ to $x_0$.  Further, Lemma~\rf{D486}(\rf{D495}) and \rf{D501} imply that $(∀m≥1)$ $(R(x_m)⧷R(_{m-1}))⋃⎨x_{m-1}⎬$ is the nontrivial path from $x_{m-1}$ to $x_m$.  Thus the equality implies that $R(N)$ is the concatenation of an infinite collection of nontrivial paths beginning from $r$.  Hence $R(N)$ is an infinite path from $r$, which by $\ZZi$'s definition (\rf{D280}) implies $R(N)⋅∈⋅\ZZi$. \end{pf}

\begin{lemma}\label{D200} Suppose $Q$ is a form and $t⋅∈⋅T$.  Then the following hold. \begin{tlist}
\yl{D201} $\tY⧷\tW⋅⊆⋅Y⧷W$.
\yl{D202} $(∀N∈\tZZ)$ $R(N)⋅∈⋅\ZZ$.
\end{tlist}\end{lemma}

\begin{pf} {\em (\rf{D201})}.  Take a subform endnode $y⋅∈⋅\tY⧷\tW$.  Then Streufert 2023p, Lemma~\rf{C366}(\rf{C457},\rf{C458}), implies \ilc{D203} $y⋅∈⋅Y$, \il{D204} $t⋅≺⋅y$, and \il{D205} not ($y⋅∈⋅W$ and $t⋅≼⋅y$).  By logical manipulation, \rf{D205} implies $y \not\in W$ or $t \not\preccurlyeq y$, which by \rf{D204} implies $y \not\in W$.  Thus \rf{D203} implies $y⋅∈⋅Y⧷W$. 

{\em (\rf{D202})}.  Take a subform run $N⋅∈⋅\tZZ$.  On the one hand, suppose $N⋅∈⋅\tZZ_{\f{ft}}$.  Then the forward direction of Lemma~\rf{C907}(\rf{C909}), applied to the subform $\tQ$, implies that $\max˙N$ exists and is in $\tY⧷\tW$.  Thus part~(\rf{D201}) implies that $\max˙N$ exists and is in $Y⧷W$.  Hence the reverse direction of Lemma~\rf{C907}(\rf{C909}) implies $R(N)⋅∈⋅\ZZf$.  

On the other hand, suppose $N⋅∈⋅\tZZ_{\f{inft}}$.  Then definition (\rf{D280}), applied to the subform $\tQ$, implies that $N$ is an infinite path in the subform out-tree $(\tX,\pj{WY}(\tQ))$.  Thus $\tQ⋅⊆⋅Q$ implies that $N$ is an infinite subset of a path in the out-tree $(X,\pj{WY}(Q))$.  Hence the reverse direction of Lemma~\rf{C907}(\rf{C910}) implies $R(N)⋅∈⋅\ZZi$. \end{pf}

\newcommand{\notehighlight}{\footnote{\label{D215}It may be helpful to highlight a critical chain of reasoning.  The definition (\rf{D188}) of a subroot $t$ eventually leads to Lemmas~\rf{C459}(\rf{C460}) and \rf{C469}(\rf{C483}).  The first extends Streufert 2023p Lemma~\rf{C366}(\rf{C987})'s result that each $\tQ = ⨆_{j∈\tJ}Q_j$.  The second uses the first to show that each $Q^t = ⨆_{j∈J^t}Q_j$.  These parallel equations show that each $\tQ$ and each $Q^t$ is the union of a subcollection of the slice partition of the pentaform~$Q$.  Streufert 2023p Corollary~\rf{C579} makes it easy to show that such unions are themselves pentaforms.  Thereby each $\tQ$ and each $Q^t$ is shown to be a pentaform, and these are the subforms $\tQ$ and the piece forms~$Q^t$ on which this paper is built.  Details are in Streufert 2023p Proposition~\rf{C576} for subforms (repeated as Proposition~\rf{C892} here), and in Proposition~\rf{C282} here for piece forms.}} 

\begin{lemma}\label{C459} Suppose $Q$ is a form.  Then the following hold. \begin{tlist}
\yl{C460} $(∀t∈T)$ $⁅Q_j⁆_{j∈\tJ}$ is an injectively indexed partition of $\tQ$.{\notehighlight} 
\yl{C461} $(∀t∈T)$ $⁅W_j⁆_{j∈\tJ}$ is an injectively indexed partition of $\tW$.
\yl{C481} $(∀t∈T)$ $⁅Y_j⁆_{j∈\tJ}$ is an injectively indexed partition of $\tY$.
\end{tlist} \end{lemma}

\begin{pf} {\em (\rf{C460})}.  Take $t⋅∈⋅T$.  In the whole form $Q$, the definition (\rf{D212}) of $⁅Q_j⁆_{j∈J}$ implies that distinct situations $j_1$ and $j_2$ in $J$ have disjoint nonempty slices $Q_{j_1}$ and $Q_{j_2}$.  Thus $\tJ⋅⊆⋅J$ implies that distinct situations $j_1$ and $j_2$ in $\tJ$ have disjoint nonempty slices $Q_{j_1}$ and $Q_{j_2}$, and that the indexing function $\tJ⋅∋⋅j \mapsto Q_j$ is injective.  Thus it remains to show that $⨆_{j∈\tJ}Q_j = \tQ$.  This holds by Streufert 2023p Lemma~\rf{C366}(\rf{C987}).

{\em (\rf{C461})}.  Take $t⋅∈⋅T$.  In the whole form $Q$, Streufert 2023p Proposition~\rf{D328} implies that $⁅W_j⁆_{j∈J}$ is an injectively indexed partition of $W$.  Thus since $\tJ⋅⊆⋅J$, it remains to show that $⨆_{j∈\tJ}W_j = \tW$.  This follows by projection from $⨆_{j∈\tJ}Q_j = \tQ$, which holds by part (\rf{C460}).

{\em (\rf{C481})}.  Take $t⋅∈⋅T$.  In the whole form $Q$, Streufert 2023p Lemma~\rf{D368} implies that $⁅Y_j⁆_{j∈J}$ is an injectively indexed partition of $Y$.  Thus since $\tY⋅⊆⋅Y$, it remains to show that $⨆_{j∈\tJ}Y_j = \tY$.   This follows by projection from $⨆_{j∈\tJ}Q_j = \tQ$, which holds by part (\rf{C460}). \end{pf}

\begin{lemma}\label{D297}  Let $\rQ$ be the cry-wolf example defined in (\rf{D270})--(\rf{C855}).  Then $\rQ$ is a (penta)form with root $⎨⎬$.  \end{lemma}

\begin{pf}  The lemma follows from Claim~\rf{D452}.\footnote{This proof builds the pentaform $\rQ$ via the ``layering'' technique of Streufert 2023p Section~\rf{C564}.  More specifically, $⨆_{t∈⎨\f6˛\f7˛\f8⎬^ℓ}\rQ^t$ is the union of the ``layer'' with index $ℓ⋅≥⋅0$.} \begin{cllist}

\yl{D450} {\em $(∀t∈\rT)⋅\rQ^t$ is a block (Streufert 2023p (\rf{D464})) whose start- and end-node sets are $⎨t⎬$ and $⎨˙t⊕b˙|˙b∈⎨\f4˛\f5˛\f6˛\f7˛\f8⎬˙⎬$.}  Take $t⋅∈⋅T$.  $\rQ^t$ is a block because its definition (\rf{D271}) is like the definition of the example pentaform $\RQ$ in the table of Streufert 2023p Figure~\rf{B225}, and because every pentaform is a block by the block definition (Streufert 2023p (\rf{D464})).  Finally, the same definition (\rf{D271}) implies $\rW^t⧷\rY^t = ⎨t⎬$ and $\rY^t⧷\rW^t = ⎨˙t⊕b˙|˙b∈⎨\f4˛\f5˛\f6˛\f7˛\f8⎬˙⎬$ .

\yl{D461} {\em $⎨\rQ^t|t∈\rT⎬$ is weakly separated (Streufert 2023p (\rf{D418})).}  Assume $t_1⋅∈⋅\rT$ and $t_2⋅∈⋅\rT$ are such that $t_1⋅≠⋅t_2$.  It suffices to show [a] $\rY^{t_1}⋂\rY^{t_2}⋅≠⋅∅$, [b]~$\rW^{t_1}⋂\rW^{t_2}⋅≠⋅∅$, and [c] $\rJ^{t_1}⋂\rJ^{t_2}⋅≠⋅∅$.
To see [a], suppose $y⋅∈⋅\rY^{t_1}⋂\rY^{t_2}$.  Then definition (\rf{D271}) implies there is  $b⋅∈⋅⎨\f1˛\f2˛\f3˛\f4˛\f5˛\f6˛\f7˛\f8⎬$ such that $y$ equals both $t_1⊕b$ and $t_2⊕b$.  Hence $t_1 = t_2$, which contradicts the assumption $t_1⋅≠⋅t_2$.  
For [b], suppose $w⋅∈⋅\rW^{t_1}⋂\rW^{t_2}$.  If $w$ ends in $\f1$, $\f2$, or $\f3$, then definition (\rf{D271}) implies there is $b⋅∈⋅⎨\f1,\f2,\f3⎬$ such that $w$ equals both $t_1⊕b$ and $t_2⊕b$, which implies $t_1 = t_2$, which contradicts the assumption $t_1⋅≠⋅t_2$.  Otherwise, definition (\rf{D271}) implies that $w$ equals both $t_1$ and $t_2$, which implies $t_1 = t_2$, which again contradicts the assumption $t_1⋅≠⋅t_2$.  
For [c], note definition (\rf{D271}) implies $\rJ^{t_1}⋅⊆⋅\PP(\rW^{t_1})$ and $\rJ^{t_2} ⋅⊆⋅\PP(\rW^{t_2})$.  Thus [b] implies [c].  

\yl{D463} {\em $(∀ℓ≥0)⋅⎨\rQ^t|t∈⎨\f6˛\f7˛\f8⎬^ℓ⎬$ is strongly separated (Streufert 2023p (\rf{D419})).}  Take $ℓ⋅≥⋅0$, and assume $t_1⋅∈⋅⎨\f6˛\f7˛\f8⎬^ℓ$ and $t_2⋅∈⋅⎨\f6˛\f7˛\f8⎬^ℓ$ are such that $t_1˙≠⋅t_2$.  Since $\rJ^{t_1}⋂\rJ^{t_2} = ∅$ by Claim~\rf{D461}, it suffices to show that $\rX^{t_1}⋂\rX^{t_2} = ∅$.  Toward that end, suppose $x⋅∈⋅\rX^{t_1}⋂\rX^{t_2}$.  By definition (\rf{D271}), $x⋅∈⋅\rX^{t_1}$ implies \begin{gather}
\zz
\text{[1a]}⋅x = t_1⋅⋅⋅\text{or}⋅⋅⋅\text{[1b]}⋅x⋅∈⋅⎨˙t_1⊕b˙|˙b∈⎨\f1˛\f2˛\f3˛\f4˛\f5˛\f6˛\f7˛\f8⎬˙⎬. \notag
\zz
\end{gather}  Similarly, $x⋅∈⋅\rX^{t_2}$ implies \begin{gather}
\zz
\text{[2a]}⋅x = t_2⋅⋅⋅\text{or}⋅⋅⋅\text{[2b]}⋅x⋅∈⋅⎨˙t_2⊕b˙|˙b∈⎨\f1˛\f2˛\f3˛\f4˛\f5˛\f6˛\f7˛\f8⎬˙⎬. \notag
\zz
\end{gather}  The case [1a]-[2a] implies $x$ equals both $t_1$ and $t_2$, which implies $t_1 = t_2$, which contradicts the assumption $t_1⋅≠⋅t_2$.  The case [1b]-[2b] implies there is $b⋅∈⋅⎨\f1˛\f2˛\f3˛\f4˛\f5˛\f6˛\f7˛\f8⎬$ such that $x$ equals both $t_1⊕b$ and $t_2⊕b$, which implies $t_1 = t_2$, which contradicts the assumption $t_1⋅≠⋅t_2$.  The case [1a]-[2b] implies there is $b⋅∈⋅⎨\f1˛\f2˛\f3˛\f4˛\f5˛\f6˛\f7˛\f8⎬$ such that $x$ is equal to both $t_1$ and $t_2⊕b$, which implies that the length of $t_1$ is one more than the length of $t_2$, which contradicts the assumption that both belong to $⎨\f6˛\f7˛\f8⎬^ℓ$.  Finally, the case [1b]-[2a] is similar.

\yl{D449} {\em $(∀ℓ≥0)⋅⨆_{t∈⎨\f6˛\f7˛\f8⎬^ℓ}˙\rQ^t$ is a block whose start- and end-node sets are} \begin{gather}
\zz
⎨\f6˛\f7˛\f8⎬^ℓ⋅⋅\text{and}⋅⋅⎨˙t⊕b˙|˙t∈⎨\f6˛\f7˛\f8⎬^ℓ,˙b∈⎨\f4˛\f5˛\f6˛\f7˛\f8⎬˙⎬.\notag
\zz
\end{gather} Take $ℓ⋅≥⋅0$.  Claims~\rf{D450} and \rf{D463} imply $⎨\rQ^t|t∈⎨\f6˛\f7˛\f8⎬^ℓ⎬$ is a strongly separated collection of blocks.  Hence Streufert 2023p Proposition~\rf{D415}(\rf{D417}) implies that its union is a block with start-node set $⨆_{t∈⎨\f6˛\f7˛\f8⎬^ℓ}˙(\rW^t⧷\rY^t)$ and end-node set $⨆_{t∈⎨\f6˛\f7˛\f8⎬}˙(\rY^t⧷\rW^t)$.  By Claim~\rf{D450}, the former is $⎨\f6˛\f7˛\f8⎬^ℓ$ and the latter is $⎨˙t⊕b˙|˙t∈⎨\f6˛\f7˛\f8⎬^ℓ,˙⎨\f4˛\f5˛\f6˛\f7˛\f8⎬˙⎬$

\yl{D451} {\em $(∀m≥0)⋅⨆_{0≤ℓ≤m}˙⨆_{t∈⎨\f6˛\f7˛\f8⎬^ℓ}˙\rQ^t$ is a block whose start-node set is $⎨⎨⎬⎬$ and whose end-node set includes $⎨\f6˛\f7˛\f8⎬^{m+1}$.}  This will be proved by induction on $m$.  At the initial step ($m{=}0$), the double union reduces to $\rQ^{⎨⎬}$, which by Claim~\rf{D450} at $t=⎨⎬$ is a block whose start-node set is $⎨⎨⎬⎬$, and whose end-node set is $⎨˙⎨⎬⊕b˙|˙b∈⎨\f4˛\f5˛\f6˛\f7˛\f8⎬˙⎬$ which by inspection includes $⎨\f6˛\f7˛\f8⎬$.  The inductive step (for $m≥1$) will be proved by applying Streufert 2023p Proposition~\rf{D415}(\rf{D416}) at\begin{gather}
\zz
Q^1 = ⨆_{0≤ℓ≤m-1}˙⨆_{t∈⎨\f6˛\f7˛\f8⎬^ℓ}˙\rQ^t⋅⋅\text{and}⋅⋅Q^2 = ⨆_{t∈⎨\f6˛\f7˛\f8⎬^m}˙\rQ^t. \notag
\zz
\end{gather} Note that the indices in these definitions are distinct in the sense that\begin{gather}
\zz
⨆_{0≤ℓ≤m{-}1}⎨\f6˛\f7˛\f8⎬^ℓ⋅\text{and}⋅⎨\f6˛\f7˛\f8⎬^m⋅\text{are disjoint}. \label{D462}
\zz
\end{gather} Further, the inductive hypothesis (that is, the claim statement with $m{-}1$ replacing $m$) implies \subi \begin{gather}
\zz
Q^1⋅\text{is a block}, \label{D455} \\
W^1⧷Y^1 = ⎨⎨⎬⎬,⋅\text{and} \label{D456} \\
Y^1⧷W^1⋅⊇⋅⎨\f6˛\f7˛\f8⎬^m. \label{D457}
\zz
\end{gather} \subo Meanwhile, Claim~\rf{D449} at $ℓ{=}m$ implies \subi \begin{gather}
\zz
Q^2⋅\text{is a block}, \label{D458} \\
W^2⧷Y^2 = ⎨\f6˛\f7˛\f8⎬^m,⋅\text{and} \label{D459} \\
Y^2⧷W^2 = ⎨˙t⊕b˙|˙t∈⎨\f6˛\f7˛\f8⎬^m,˙b∈⎨\f4˛\f5˛\f6˛\f7˛\f8⎬˙⎬. \label{D460}
\zz
\end{gather} \subo Both $Q^1$ and $Q^2$ are blocks by (\rf{D455}) and (\rf{D458}).  Also, $⎨Q^1,Q^2⎬$ is a weakly separated by (\rf{D462}) and Claim~\rf{D461}.  Also, $Q^1$'s start nodes are distinct from $Q^2$'s end nodes by (\rf{D456}) and (\rf{D460}).  Thus Streufert 2023p Proposition~\rf{D415}(\rf{D416}) implies $Q^1⋃Q^2$ is a block whose start-node set is the union of\begin{gather}
\zz
W^1⧷Y^1⋅\text{and}⋅(W^2⧷Y^2)⧷(Y^1⧷W^1), \notag
\zz
\end{gather} and whose end-node set is the union of \begin{gather}
\zz
(Y^1⧷W^1)⧷(W^2⧷Y^2)⋅\text{and}⋅Y^2⧷W^2. \notag
\zz
\end{gather} The former is $⎨⎨⎬⎬$ by (\rf{D456}), (\rf{D459}), and (\rf{D457}).  The latter includes $⎨\f6˛\f7˛\f8⎬^{m+1}$ by (\rf{D460}).

\yl{D452} {\em $\rQ$ is a pentaform with root $⎨⎬$.}  The block definition (Streufert 2023p (\rf{D464})) implies that a pentaform is equivalent to a block with exactly one start node.  So Claim~\rf{D451} implies that \begin{gather}
\zz
⁅˙⨆_{0≤ℓ≤m}˙⨆_{t∈⎨\f6˛\f7˛\f8⎬^ℓ}˙\rQ^t⋅⁆_{m≥0}\notag
\zz
\end{gather} is an expanding sequence of pentaforms which share the root $⎨⎬$.  Thus Streufert 2023p Proposition~\rf{D454} implies that the union of these pentaforms is a pentaform with root $⎨⎬$.  By inspection, this union is equal to $⨆_{ℓ≥0}˙⨆_{t∈⎨\f6˛\f7˛\f8⎬^ℓ}˙\rQ^t$, which by $\rT$'s definition (\rf{D270}) is equal to $⨆_{t∈\rT}˙\rQ^t$, which by $\rQ$'s definition (\rf{C855}) is equal to $\rQ$. \end{cllist} \unskipcl \end{pf}

\section{For Piece Forms}\label{C377}\showit
\markb{\sc Appendix \rf{C377}. For Piece Forms}

\begin{lemma}
\label{C463} Suppose $Q$ is a form and $t⋅∈⋅T$.  Then the following hold. \begin{tlist}
\yl{C464} $Q^t = ⎨˙⁅i˛j˛w˛a˛y⁆∈Q˙|˙t≼w,˙(∄t\hi∈T)t≺t\hi≼w˙⎬$.
\yl{C465} $\pj{WY}(Q^t) = ⎨˙⁅w{,}y⁆∈\pj{WY}(Q)˙|˙t≼w,˙(∄t\hi∈T)t≺t\hi≼w˙⎬$.
\yl{C466} $W^t = ⎨˙w∈W˙|˙t≼w,˙(∄t\hi∈T)t≺t\hi≼w˙⎬$. 
\yl{C467} $W^t⋂T = ⎨t⎬$. 
\yl{C468} $Y^t = ⎨˙y∈Y˙|˙p(y)∈W^t˙⎬$.
\yl{D221} $t⋅∉⋅Y^t$.
\end{tlist} \end{lemma}

\begin{pf} {\em (\rf{C464})}. Note \begin{align}
\zz
Q^t 
&= \tQ⋅⧷⋅⋃_{t≺t\hi∈T}˙^{t\hi}Q \nt
&= ⎨˙⁅i˛j˛w˛a˛y⁆∈Q˙|˙t≼w˙⎬⋅⧷⋅⨆_{t≺t\hi∈T}˙⎨˙⁅i˛j˛w˛a˛y⁆∈Q˙|˙t\hi≼w˙⎬ \nt
&= ⎨˙⁅i˛j˛w˛a˛y⁆∈Q˙|t≼w˙⎬⋅⧷⋅⎨˙⁅i˛j˛w˛a˛y⁆∈Q˙|˙(∃t\hi∈T)t≺t\hi≼w˙⎬ \nt 
&= ⎨˙⁅i˛j˛w˛a˛y⁆∈Q˙|˙t≼w,˙(∄t\hi∈T)t≺t\hi≼w˙⎬, \notag
\zz
\end{align} where the first equality holds by the piece-form definition (\rf{D192}), the second equality holds by several applications of the subform definition (\rf{D190}), and the third and fourth equalities hold by rearrangement.

{\em (\rf{C465},\rf{C466})}.  These follow from part (\rf{C464}) by projection.

{\em (\rf{C467})}.  For the reverse direction, note that $T$'s definition (\rf{D188}) implies $t⋅∈⋅W$, and that inspection implies $t⋅≼⋅t$ and $(∄t\hi∈T)$ $t≺t\hi≼t$.  Thus part (\rf{C466})'s characterization of $W^t$ implies $t⋅∈⋅W^t$, which by the assumption $t⋅∈⋅T$ implies $t⋅∈⋅W^t⋂T$.  For the forward direction, consider an arbitrary $t_*⋅∈⋅W^t⋂T$.  Then the assumption $t_*⋅∈⋅W^t$ and part (\rf{C466})'s characterization of $W^t$ imply \lic{B824} $t⋅≼⋅t_*$ and \il{B825} $(∄t\hi∈T)$ $t⋅≺⋅t\hi⋅≼⋅t_*$.  Further, the assumption $t_*⋅∈⋅T$ and \rf{B825} imply that $t⋅≺⋅t_*⋅≼⋅t_*$ is false, which implies that $t⋅≺⋅t_*$ is false, which by \rf{B824} implies that $t = t_*$. 

{\em (\rf{C468})}.  It suffices to justify the three equalities in \begin{align}
\zz
Y^t
&= ⎨˙y∈Y˙|˙(∃w∈W)˙⁅w{,}y⁆∈\pj{WY}(Q),˙t≼w,˙(∄t\hi∈T)t≺t\hi≼w˙⎬ \nt
&= ⎨˙y∈Y˙|˙p(y)∈W,˙t≼p(y),˙(∄t\hi∈T)t≺t\hi≼p(y)˙⎬ \nt
&= ⎨˙y∈Y˙|˙p(y)∈W^t˙⎬. \notag
\zz
\end{align} The first equality holds by part (\rf{C464}) and projection.  The second equality holds because, for any $y⋅∈⋅Y$, axiom \rf{pwy} (Definition~\rf{D375}) and $p$'s definition (\rf{D374}) imply that $p(y)$ is the only element of $W$ to satisfy $⁅w˛y⁆⋅∈⋅\pj{WY}(Q)$.  The third equality holds because part (\rf{C466})'s characterization of $W^t$ implies that, for any $y⋅∈⋅Y$, that $p(y)⋅∈⋅W^t$ iff it satisfies $p(y)⋅∈⋅W$, $t⋅≼⋅p(y)$, and $(∄t\hi∈T)$ $t⋅≺⋅t\hi⋅≼⋅p(y)$.  

{\em (\rf{D221})}. Suppose $t⋅∈⋅Y^t$.  Then part~(\rf{C468})'s characterization of $Y^t$ implies $p(t)⋅∈⋅W^t$, which by part~(\rf{C466})'s characterization of $W^t$ implies $t⋅≼⋅p(t)$.  This contradicts the general fact that $(∀y∈Y)$ $p(y)⋅≺⋅y$.  \end{pf}

\begin{lemma}\label{C469} Suppose $Q$ is a form.  Then the following hold. \begin{tlist}
\yl{C483} $(∀t∈T)⋅Q^t = ⨆_{j∈J^t}Q_j$.  (Footnote~\rf{D215} on page \pageref{D215} provides context.)
\yl{C484} $(∀t∈T)⋅W^t = ⨆_{j∈J^t}W_j$.
\yl{C485} $(∀t∈T)⋅Y^t = ⨆_{j∈J^t}Y_j$.
\yl{C486} $⁅J^t⁆_{t∈T}$ is an injectively indexed partition of $J$. 
\end{tlist}\end{lemma}

\begin{pf} The lemma holds by Claims \rf{C482}, \rf{D503}, \rf{D509}, and \rf{C480}. \begin{cllist}

\yl{B906} {\em $(∀t∈T)$ $Q^t = ⋃⎨˙Q_j˙|˙j∈\tJ⧷⋃_{t≺t\hi∈T}˙\bJ{t\hi}˙⎬$.}  To show this, consider a subroot $t⋅∈⋅T$.  Then \begin{align}
\zz
Q^t 
&=⋅\tQ⋅⧷⋅⋃_{t≺t\hi∈T}˙\bQ{t\hi} \nt
&=⋅⋃_{j∈\tJ}˙Q_j⋅⧷⋅⋃_{t≺t\hi∈T}˙⋃_{j∈\bJ{t\hi}}˙Q_j \nt
&=⋅⋃⎨˙Q_j˙|˙j∈\tJ˙⎬⋅⧷⋅⋃⎨˙Q_j˙|˙j∈⋃_{t≺t\hi∈T}˙\bJ{t\hi}˙⎬ \nt
&=⋅⋃⎨˙Q_j˙|˙j∈\tJ⧷⋃_{t≺t\hi∈T}˙\bJ{t\hi}˙⎬, \notag
\zz
\end{align} where the first equality is $Q^t$'s definition (\rf{D192}), the second holds by several applications of Lemma~\rf{C459}(\rf{C460}), the third holds by rearrangement, and the fourth holds because $⁅Q_j⁆_{j∈J}$ is an injectively indexed partition by Lemma~\rf{C459}(\rf{C460}) at its $t$ equal to $r$.  

\yl{C470} $(∀t∈T)$ $J^t = \tJ⧷⋃_{t≺t\hi∈T}˙\bJ{t\hi}$.  To show this, take a subroot $t⋅∈⋅T$.  Then Claim~\rf{B906} implies $Q^t = ⋃⎨˙Q_j˙|˙j∈\tJ⧷⋃_{t≺t\hi∈T}˙\bJ{t\hi}˙⎬$, which by projection implies $J^t = ⨆⎨˙⎨j⎬˙|˙j∈\tJ⧷⋃_{t≺t\hi∈T}˙\bJ{t\hi}˙⎬$, which by simplification implies $J^t = \tJ⧷⋃_{t≺t\hi∈T}˙\bJ{t\hi}$. 

\yl{C482} {\em $(∀t∈T)⋅Q^t = ⨆_{j∈J^t}Q_j$.}  To see this, take a subroot $t⋅∈⋅T$.  Then $Q^t$ by Claim~\rf{B906} equals $⨆⎨˙Q_j˙|˙j∈\tJ⧷⋃_{t≺t\hi∈T}˙\bJ{t\hi}˙⎬$, which by Claim~\rf{C470} equals $⨆⎨˙Q_j˙|˙j∈J^t˙⎬$.   

\yl{D503} {\em $(∀t∈T)⋅W^t = ⨆_{j∈J^t}W_j$.}  This follows from Claim~\rf{C482} by projection.

\yl{D509} {\em $(∀t∈T)⋅Y^t = ⨆_{j∈J^t}Y_j$.}  This follows from Claim~\rf{C482} by projection.

\yl{C474} {\em $(∀t∈T)$ $J^t$ is nonempty.}  Take a subroot $t⋅∈⋅T$.  Lemma~\rf{C463}(\rf{C467}) implies $W^t$ is nonempty, which by abbreviation (\rf{C858}) implies $π_W(Q^t)$ is nonempty, which implies $Q^t$ is nonempty, which implies $π_J(Q^t)$ is nonempty, which by abbreviation (\rf{C858}) implies $J^t$ is nonempty.

\yl{D504} {\em Suppose $t_1∈T$ and $t_2∈T$ satisfy $t_1 \not\preccurlyeq t_2$ and $t_2 \not\preccurlyeq t_1$.  Then $\bJ{t_1}⋂\bJ{t_2} = ∅$.}  Because $t_1$ and $t_2$ are nodes in the out-tree $(X,\pj{WY}(Q))$, the assumptions $t_1 \not\preccurlyeq t_2$ and $t_2 \not\preccurlyeq t_1$ imply there is no decision node $w⋅∈⋅W$ such that $t_1⋅≼⋅w$ and $t_2⋅≼⋅w$.  Thus Streufert 2023p Lemma~\rf{C366}(\rf{C457}) implies $\bW{t_1}⋂\bW{t_2} = ∅$, which implies there is no $j⋅∈⋅J$ whose information set $W_j$ satisfies $W_j⋅⊆⋅\bW{t_1}⋂\bW{t_2}$, which by the indexed partition of Lemma~\rf{C459}(\rf{C461}) implies $\bJ{t_1}˙⋂˙\bJ{t_2} = ∅$. 

\yl{C471} {\em $(∀t_1∈T,t_2∈T)$ $t_1⋅≠⋅t_2$ implies $J^{t_1}⨅J^{t_2} = ∅$.} To show this, suppose $t_1$ and $t_2$ are distinct subroots.  Mechanically, \lic{B805} $t_1⋅≼⋅t_2$, \li{B806} $t_2⋅≼⋅t_1$, or \li{B807} $t_1 \not\preccurlyeq t_2$ and $t_2 \not\preccurlyeq t_1$  (it is irrelevant whether the cases are mutually exclusive).  First suppose \rf{B807}.  Then Claim~\rf{D504} implies $\bJ{t_1}⋂\bJ{t_2} = ∅$.  Meanwhile Claim~\rf{C470} implies $J^{t_1}⋅⊆⋅\bJ{t_1}$ and $J^{t_2}⋅⊆⋅\bJ{t_2}$.  Thus $J^{t_1}⋂J^{t_2} = ∅$.

Second suppose \rf{B805} or \rf{B806}.  Without loss of generality, assume \rf{B805}.  Then the assumed distinctness of $t_1$ and $t_2$ implies $t_1⋅≺⋅t_2$, which mechanically implies \tts{$*$}{D216} $\bJ{t_2}⋅⊆⋅⋃_{t_1≺t\hi∈T}˙\bJ{t\hi}$.  Meanwhile, two applications of Claim~\rf{C470} imply\begin{gather}
\zz
J^{t_1} = \bJ{t_1}⧷⋃_{t_1≺t\hi∈T}˙\bJ{t\hi}⋅\text{and}⋅
J^{t_2} = \bJ{t_2}⧷⋃_{t_2≺t\hi∈T}˙\bJ{t\hi}\notag
\zz
\end{gather} The second implies $J^{t_2}⋅⊆⋅\bJ{t_2}$, which by \rf{D216} implies $J^{t_2}⋅⊆⋅⋃_{t_1≺t\hi∈T}˙\bJ{t\hi}$, which by the first implies $J^{t_1}$ and $J^{t_2}$ are disjoint.

\yl{C480} {\em $⁅J^t⁆_{t∈T}$ is an injectively indexed partition of $J$.} Claim~\rf{C474} showed that each $J^t$ is nonempty.  Thus Claim \rf{C471} implies that distinct $t_1$ and $t_2$ have disjoint nonempty $J^{t_1}$ and $J^{t_2}$, which in turn implies that the indexing function $T⋅∋⋅t \mapsto J^{t}$ is injective.  Thus it remains to show that $⨆_{t∈T}J^t = J$.

For the forward inclusion, take a subroot $t⋅∈⋅T$.  Then $Q^t⋅⊆⋅Q$ implies $J^t⋅⊆⋅J$.  For the reverse inclusion, take any situation $j⋅∈⋅J$.  It suffices to show there is a subroot $t_*⋅∈⋅T$ such that $j⋅∈⋅J^{t_*}$.  To begin, note $j⋅∈⋅J$ implies there is $⁅i˛w˛a˛y⁆$ such that $⁅i˛j˛w˛a˛y⁆⋅∈⋅Q$.  Lemma~\rf{D486}(\rf{D496}) implies that $R(w) = ⎨x|x≼w⎬$ is finite and linearly ordered.  Thus $R(w)⋂T = ⎨x∈T|x≼w⎬$ is finite and linearly ordered.  Further, $R(w)⋂T$ is nonempty because it contains $r$.  Thus we may let $t_*$ be its maximum.  Then $t_*⋅∈⋅T$, $t_*⋅≼⋅w$, and $(∄t\hi∈T)$ $t_*⋅≺⋅t\hi⋅≼⋅w$.  Thus Lemma~\rf{C463}(\rf{C464}) implies that $⁅i˛j˛w˛a˛y⁆⋅∈⋅Q^{t_*}$.  Hence $j⋅∈⋅J^{t_*}$. \end{cllist} \unskipcl \end{pf}

\begin{npf}[{{\bf for Proposition~\rf{C282}}}]\label{C282p}  Take a subroot $t⋅∈⋅T$.  The proposition follows from Claims \rf{D218} and \rf{D220}. \begin{cllist}

\yl{D217} $W^t⧷⎨t⎬⋅⊆⋅Y^t$.  To see this, take $w⋅∈⋅W^t⧷⎨t⎬$.  Since $w⋅∈⋅W^t$, Lemma~\rf{C463}(\rf{C466})'s characterization of $W^t$ implies \ilc{C382} $t⋅≼⋅w$ and \il{D506} $(∄t\hi∈T)$ $t⋅≺⋅t\hi⋅≼⋅w$.  Since $w⋅≠⋅t$,  \rf{C382} can be strengthened to \il{D507} $t⋅≺⋅w$, and thus \il{C476} $w⋅∈⋅Y$.  

Since the domain and range of $p$ are $Y$ and $W$, \rf{C476} implies that $p(w)$ exists and satisfies $p(w)⋅∈⋅W$.  Further, \rf{D507} implies $t⋅≼⋅p(w)$, and \rf{D506} implies $(∄t\hi∈T)$ $t⋅≺⋅t\hi⋅≼⋅p(w)$.  Thus Lemma~\rf{C463}(\rf{C466})'s characterization of $W^t$ implies $p(w)⋅∈⋅W^t$.  Thus \rf{C476} and Lemma~\rf{C463}(\rf{C468})'s characterization of $Y^t$ imply $w⋅∈⋅Y^t$. 

\yl{D219} $W^t⧷Y^t = ⎨t⎬$.  The forward inclusion follows from Claim~\rf{D217}.  For the reverse inclusion, Lemma~\rf{C463}(\rf{C467}) implies $t⋅∈⋅W^t$ and Lemma~\rf{C463}(\rf{D221}) implies $t⋅∉⋅Y^t$. 

\yl{D218} {\em $Q^t$ is a pentaform.}  Lemma~\rf{C469}(\rf{C483}) shows $Q^t = ⨆⎨Q_j|j∈J^t⎬$, which by $J^t⋅⊆⋅J$ is the union of a subcollection of $Q$'s slice partition $⎨Q_j|j∈J⎬$.  Also, Claim~\rf{D219} implies axiom \rf{pr}. Thus Streufert 2023p Corollary~\rf{C579}(\rf{D414}) implies that $Q^t$ is a (penta)form.  

\yl{D220} {\em The root of $Q^t$ is $t$.}  This follows from Claim~\rf{D219} and the root definition (\rf{D186}). \end{cllist}\unskipcl\end{npf}

\begin{npf}[{{\bf for Proposition~\rf{C277}}}]\label{C277p} Part~(\rf{C279}) repeats Lemma~\rf{C469}(\rf{C486}).  Thus parts (\rf{C278}), (\rf{C280}), and (\rf{C281}) remain.

For part (\rf{C278}), note that $⁅Q_j⁆_{j∈J}$ is an injectively indexed partition of $Q$ by Lemma~\rf{C459}(\rf{C460}) at its $t$ equal to $r$, and that $⁅J^t⁆_{t∈T}$ is an injectively indexed partition of $J$ by part (\rf{C279}).  Thus $⁅⨆_{j∈J^t}Q_j⁆_{t∈T}$ is an injectively indexed partition of $Q$.  This suffices since $(∀t∈T)$ $⨆_{j∈J^t}Q_j = Q^t$ by Lemma~\rf{C469}(\rf{C483}).

Part (\rf{C280}) for $W$ follows similarly from Lemma~\rf{C459}(\rf{C461}) and Lemma~\rf{C469}(\rf{C484}).

Part (\rf{C281}) for $Y$ follows similarly from Lemma~\rf{C459}(\rf{C481}) and Lemma~\rf{C469}(\rf{C485}). \end{npf}

\begin{lemma}\label{C407} Suppose $Q$ is a form and $t⋅∈⋅T⧷⎨r⎬$.  Then there is $t\lo⋅∈⋅T$ such that $t⋅∈⋅Y^{t\lo}⧷W^{t\lo}$. \end{lemma}

\begin{pf} In steps, the assumption $t⋅∈⋅T⧷⎨r⎬$, by $T$'s definition (\rf{D188}), implies $t⋅∈⋅W⧷⎨r⎬$, which by $X$'s definition (\rf{D182}) implies $t⋅∈⋅X⧷⎨r⎬$, which by the identity $X⧷⎨r⎬ = Y$ (Streufert 2023p Lemma~\rf{C637}(\rf{C636})) implies $t⋅∈⋅Y$, which by Proposition~\rf{C277}(\rf{C281})'s partition implies there is $t\lo⋅∈⋅T$ such that \tts{$*$}{D224} $t⋅∈⋅Y^{t\lo}$.  

Thus it suffices to show that $t⋅∉⋅W^{t\lo}$.  Toward that end, suppose $t⋅∈⋅W^{t\lo}$.  Then the assumption $t⋅∈⋅T$ and the identity $W^{t\lo}⋂T = ⎨t\lo⎬$ (Lemma~\rf{C463}(\rf{C467})) imply $t = t\lo$.  Thus \rf{D224} implies $t⋅∈⋅Y^t$, which contradicts the general fact that $t⋅∉⋅Y^t$ (Lemma~\rf{C463}(\rf{D221})). \end{pf}

\begin{npf}[{{\bf for Proposition~\rf{C390}}}]\label{C390p} By inspection, each set in the collection \linebreak $⎨⎨r⎬⎬⋃⎨Y^t⧷W^t≠∅|t∈T⎬$ is nonempty.  Also, since $r⋅∉⋅Y$ by $r$'s definition (\rf{D186}), and since $⁅Y^t⁆_{t∈T}$ is an injectively indexed partition of $Y$ by Proposition~\rf{C277}(\rf{C281}), the collection $⎨⎨r⎬⎬⋃⎨Y^t⧷W^t≠∅|t∈T⎬$ is pairwise disjoint.  Thus it remains to show that the union of the collection is equal to $T⋃(Y⧷W)$.

For the reverse inclusion, take $x⋅∈⋅T⋃(Y⧷W)$, so that $x$ is either a subroot or a whole-form endnode.  First suppose $x⋅∈⋅Y⧷W$.  Then $x⋅∈⋅Y$, so Proposition~\rf{C277}(\rf{C281}) implies there is $t⋅∈⋅T$ such that $x⋅∈⋅Y^t$.  Also, $x⋅∉⋅W$, so Proposition~\rf{C277}(\rf{C280}) implies $x⋅∉⋅W^t$.  Hence $x⋅∈⋅Y^t⧷W^t$.  Second suppose $x⋅∈⋅T$.  If $x = r$, we have $x⋅∈⋅⎨r⎬$ and the argument is complete.  Otherwise $x⋅≠⋅r$, so Lemma~\rf{C407} implies there is $t\lo⋅∈⋅T$ such that $t⋅∈⋅Y^{t\lo}⧷W^{t\lo}$.

For the forward inclusion, take a node $x$ in the union of the collection.  Equivalently, suppose $x⋅∈⋅⎨r⎬˙⋃˙⨆⎨Y^t⧷W^t|t∈T⎬$.  It must be shown that $x⋅∈⋅T⋃(Y⧷W)$.  If $x = r$, the general fact (\rf{D266}) that $r⋅∈⋅T$ implies $x⋅∈⋅T$, which completes the argument.  So assume there is $t⋅∈⋅T$ such that \ilc{C411} $x⋅∈⋅Y^t⧷W^t$.  If $x⋅∉⋅W$, then $x⋅∈⋅Y^t$ implies $x⋅∈⋅Y^t⧷W$, which by $Y^t⋅⊆⋅Y$ implies $x⋅∈⋅Y⧷W$, which completes the argument.  So assume \il{C529} $x⋅∈⋅W$, that is, that $x$ is a decision node.  The sequel will use \rf{C411} and \rf{C529} to show that $x$ is a subroot, that is, an element of $T$.\footnote{Intuitively, \rf{C411} says that $x$ is an endnode of the piece $Q^t$ and \rf{C529} implies that $x$ is a decision node of another piece $Q^{t\hi}$ (it could be shown that $t⋅≺⋅t\hi$).  It will be shown that $x$ is the root of $Q^{t\hi}$, which implies $x = t\hi$, which serves to prove $x⋅∈⋅T$ because $t\hi⋅∈⋅T$.}

Proposition~\rf{C277}(\rf{C280}) shows $⁅W^{t_*}⁆_{t_*∈T}$ is an injectively indexed partition of $W$.  This partition is used in each of the following three sentences.  First, \rf{C529} and the partition imply there is $t\hi⋅∈⋅T$ such that $x⋅∈⋅W^{t\hi}$.  Second, \rf{C411} implies $x⋅∉⋅W^t$, which by $x⋅∈⋅W^{t\hi}$ and the partition's injective index implies $t\hi⋅≠⋅t$.  Third, \rf{C411} implies $x⋅∈⋅Y^t$, which by $p^t{:}Y^t→W^t$ implies $p(x)⋅∈⋅W^t$, which by $t\hi⋅≠⋅t$ and the injectively indexed partition implies $p(x)⋅∉⋅W^{t\hi}$.  For the sequel, it suffices to remember from the first step that \il{C526} $t\hi⋅∈⋅T$ and \il{C412} $x⋅∈⋅W^{t\hi}$, and from the third step that \il{C415} $p(x)⋅∉⋅W^{t\hi}$. 

This paragraph will show that $x⋅∉⋅Y^{t\hi}$.  Because $t\hi⋅∈⋅T$ by \rf{C526}, we may consider the piece form $Q^{t\hi}$.  General definition (\rf{D374}) implies $Q^{t\hi}$'s immediate-predecessor function is $p^{t\hi} = \pj{YW}(Q^{t\hi})$, which by Lemma~\rf{C463}(\rf{C465}) is a restriction of the whole form's immediate-predecessor function $p = \pj{YW}(Q)$.  Now suppose $x⋅∈⋅Y^{t\hi}$ held.  Then $p^{t\hi}{:}Y^{t\hi}→W^{t\hi}$ implies $p^{t\hi}(x)⋅∈⋅W^{t\hi}$, which by the previous sentence implies $p(x)⋅∈⋅W^{t\hi}$, which contradicts \rf{C415}.

The previous paragraph and \rf{C412} imply $x⋅∈⋅W^{t\hi}⧷Y^{t\hi}$.  Meanwhile, Proposition~\rf{C282} implies that the root of $Q^{t\hi}$ is $t\hi$, which by the root definition (\rf{D186}) implies $W^{t\hi}⧷Y^{t\hi} = ⎨t\hi⎬$, which by the previous sentence implies $x = t\hi$, which by \rf{C526} implies $x⋅∈⋅T$. \end{npf}

\begin{npf}[{{\bf for Proposition \rf{C391}}}]\label{C391p}  The definitions (\rf{D282}) of $\ZZf$, $\ZZi$, and $\ZZ$ imply $\ZZ = \ZZf⋃\ZZi$ and $\ZZf⋂\ZZi = ∅$.  Thus the cases (a) $R(N)⋅∉⋅\ZZ$, (b) $R(N)⋅∈⋅\ZZf$, and (c) $R(N)⋅∈⋅\ZZi$ are exhaustive and mutually exclusive.  Hence Claims \rf{C911}--\rf{C913} suffice. \begin{cllist}

\yl{D229} {\em $N$ is finite iff $\max˙N$ exists.} This holds since $N$ is a path by assumption, and since every node has a finite number of predecessors by Lemma~\rf{D486}(\rf{D496}).

\yl{C911} {\em Suppose $R(N)⋅∉⋅\ZZ$.  Then $N$ is finite, and $\max˙N$ exists and is in $T$.} The assumption $R(N)⋅∉⋅\ZZ$ implies $R(N)⋅∉⋅\ZZf$ and $R(N)⋅∉⋅\ZZi$.  Since $R(N)⋅∉⋅\ZZi$, and since $N$ is a path by assumption, Lemma~\rf{C907}(\rf{C910}) implies that \ilc{D185} $N$ is finite.  Thus Claim~\rf{D229} implies that \il{C915} $\max˙N$ exists.  Thus, since $R(N)⋅∉⋅\ZZf$, Lemma~\rf{C907}(\rf{C909}) implies that \il{C914} $\max˙N⋅∉⋅Y⧷W$.  

Because of \rf{D185} and \rf{C915}, it remains to show that $\max˙N$ is in $T$.  In other words, it remains to show that $\max˙N$ is a subroot.  Since the proposition assumes $N⋅∈⋅\ZZ^t$, and since $N$ is finite by \rf{D185}, the general definition of a run (\rf{D282}) implies that $\max˙N⋅∈⋅Y^t⧷W^t$.  Thus Proposition~\rf{C390} implies that $\max˙N⋅∈⋅T⋃(Y⧷W)$.  Thus \rf{C914} implies that $\max˙N⋅∈⋅T$. 

\yl{C912} {\em Suppose $R(N)⋅∈⋅\ZZf$.  Then $N$ is finite, and $\max˙N$ exists and is in $Y⧷W$.}  Lemma~\rf{C907}(\rf{C909}) implies $\max˙N$ exists and is in $Y⧷W$.  This suffices by Claim~\rf{D229}.

\yl{C913} {\em Suppose $R(N)⋅∈⋅\ZZi$.  Then $N$ is infinite, and $\max˙N$ does not exist.}  Lemma~\rf{C907}(\rf{C910}) implies $N$ is infinite.  This suffices by Claim~\rf{D229}. \end{cllist} \unskipcl \end{npf}

\begin{lemma}\label{C453} Suppose $Q$ is a form, $s⋅∈⋅S$, and $t⋅∈⋅T$.  Then \begin{gather}
\zz
R(\tOO(\tss)) = \mip
R(^{\max˙O^t(s^t)}O(\bss{\max˙O^t(s^t)}))⋅\text{if}⋅\max˙O^t(s^t)⋅\text{exists and is in $T$} \\[.6ex]
\quad R(O^t(s^t))⋅\text{otherwise}\rule{9.5ex}{0ex} \mop.
\notag
\zz
\end{gather} \end{lemma}

\begin{pf} First consider the case where $\max˙O^t(s^t)$ exists and is in $T$.  Then the piece run $O^t(s^t)$ is immediately succeeded by the subform run $\bO{\max˙O^t(s^t)}(\bss{\max˙O^t(s^t)})$.  Thus, because both $s^t$ and $\bss{\max˙O^t(s^t)}$ are restrictions of $\tss$,
\begin{gather}
\zz
\tOO(\tss) = O^t(s^t)⋅⋃⋅^{\max˙O^t(s^t)}O(\bss{\max˙O^t(s^t)}). \notag
\zz
\end{gather} This implies\begin{gather}
\zz
R(⋅\tOO(\tss)⋅) = R(⋅O^t(s^t)⋅⋃⋅^{\max˙O^t(s^t)}O(\bss{\max˙O^t(s^t)})⋅), \notag
\notag
\zz
\end{gather} which implies\begin{gather}
\zz
R(⋅\tOO(\tss)⋅) = R(⋅^{\max˙O^t(s^t)}O(\bss{\max˙O^t(s^t)})⋅) \notag
\notag
\zz
\end{gather} by $R$'s definition (\rf{D206}) and the fact that each node in the piece run $O^t(s^t)$ weakly precedes the nodes in the subsequent subform run $\bO{\max˙O^t(s^t)}(\bss{\max˙O^t(s^t)})$. 

Second consider the ``otherwise'' case.  Then Proposition~\rf{C391} at the piece run $N = O^t(s^t)$ implies that $R(O^t(s^t))⋅∈⋅\ZZf⋃\ZZi = \ZZ$.  Meanwhile, Lemma~\rf{D200}(\rf{D202}) at the subform run $N = \tOO(\tss)$ implies that $R(\tOO(\tss))⋅∈⋅\ZZ$.  Thus both $R(O^t(s^t))$ and $R(\tOO(\tss))$ are whole-form runs that go through $t$.  Thus since $s^t$ is a restriction of $\tss$, $R(O^t(s^t)) = R(\tOO(\tss))$.\footnote{Although it can be shown that $O^t(s^t) = \tOO(\tss)$, it cannot be shown that $s^t = \tss$.  In particular, it is possible that the domain of the piece strategy $s^t$ is a proper subset of the domain of the subform strategy $\tss$ because the piece form $Q^t$ is a proper subset of the subform $\tQ$.  This happens whenever $Q^t$ has an endnode which is a subroot.  This is consistent with $\max˙O^t(s^t)$ being a whole-form endnode and also with $O^t(s^t)$ being infinite (bear in mind that $O^t(s^t)$ is just one of $Q^t$'s piece runs).} \end{pf}

\section{For Games}\label{C406}\showit
\markb{\sc Appendix \rf{C406}. For Games}

\begin{lemma}\label{C431} Suppose $(Q,u)$ is a game and $(s,v)$ is authentic.  Then, $v$ is admissible, and $(s,v)$ is persistent. \end{lemma}

\begin{pf} {\em Admissibility}. By definition (\rf{C921}), it suffices to show \begin{gather}
\zz
(∀t∈T,k∈K)⋅\inf⎨u_k(Z\lo)|t∈Z\lo∈\ZZ⎬⋅≤⋅v_k(t)⋅≤⋅\sup⎨u_k(Z\hi)|t∈Z\hi∈\ZZ⎬.\notag
\zz
\end{gather} Take a subroot $t⋅∈⋅T$ and a stakeholder $k⋅∈⋅K$.  Authenticity's definition (\rf{C923}) implies that the value $v_k(t)$ equals $u_k(R(^tO(^ts)))$.  Thus, since $R(\tOO(\tss))$ is a run in $\ZZ$ which contains $t$, the value $v_k(t)$ belongs to $⎨u_k(Z_*)|t∈Z_*∈\ZZ⎬$.  The inequalities follow.

{\em Persistence}. By definition (\rf{C922}), it suffices to show \begin{gather}
\zz
(∀t∈T)⋅v(t) = \mip
v(\max˙O^t(s^t))⋅\text{if}⋅\max˙O^t(s^t)⋅\text{exists and is in $T$} \\[.6ex]
\rule{1.1ex}{0ex}u(R(O^t(s^t)))⋅\text{otherwise}\rule{21.5ex}{0ex} \mop.
\notag
\zz
\end{gather} Take a subroot $t⋅∈⋅T$.  In the first case, the value profile $v(t)$ by authenticity (\rf{C923}) is equal to $u(R(^tO(^ts)))$, which by Lemma~\rf{C453} is equal to $u(R(^{\max˙O^t(s^t)}O(^{\max˙O^t(s^t)}s)))$, which by authenticity is equal to $v(\max˙O^t(s^t))$.  In the second case, $v(t)$ by authenticity is equal to $u(R(^tO(^ts)))$ which by Lemma~\rf{C453} is equal to $u(R(O^t(s^t)))$.  \end{pf}

\begin{lemma}\label{C405} Suppose $(Q,u)$ is a game and $Z⋅∈⋅\ZZf$.  Then $u$ is both upper- and lower-convergent at $Z$. \end{lemma}

\begin{pf} The assumption $Z⋅∈⋅\ZZf$ implies $\max˙Z$ exists.  To show that $u$ is lower-convergent at $Z$, fix a stakeholder $k$.  It will be argued that \begin{gather}
\zz
\lim_{x∈Z}⋅\inf⎨u_k(Z\lo)|x∈Z\lo∈\ZZ⎬
= \inf⎨u_k(Z\lo)|\max˙Z∈Z\lo∈\ZZ⎬ \nt
= \inf⎨u_k(Z)⎬
= u_k(Z).  \notag
\zz
\end{gather} The first equality holds by the existence of $\max˙Z$.  The second holds since $Z$ itself is the only run $Z\lo⋅∈⋅\ZZ$ which contains $\max˙Z$.  The third holds because the set is a singleton.  A similar argument implies upper-convergence (replace ``lower'' with ``upper'', replace ``inf'' with ``sup'', and let ``max'' remain).  \end{pf}

\begin{lemma}\label{D284} Consider a game $(Q,u)$.  Then \ttr{a}{D288} $(∀Z∈\ZZ,k∈K)$ the limit in (\rf{D287}) exists.  Further, \ttr{b}{D289} upper-convergence fails iff \begin{gather}
\zz
(∃Z∈\ZZ,k∈K)⋅\text{\normalfont lim}_{x∈Z}⋅\text{\normalfont sup}⎨u_k(Z\hi)|x∈Z\hi∈\ZZ⎬ > u_k(Z). \label{D287}
\zz
\end{gather} \end{lemma}

\begin{pf} (\rf{D288}) follows from Claim~\rf{D290}.  (\rf{D289}) follows from the two claims and upper-convergence's definition (\rf{C924}).  \begin{cllist}

\yl{D290} {\em $(∀Z∈\ZZ,k∈K)⋅\text{\normalfont lim}_{x∈Z}⋅\text{\normalfont sup}⎨u_k(Z\hi)|x∈Z\hi∈\ZZ⎬$ exists.}  To see this, take a run $Z⋅∈⋅\ZZ$ and a stakeholder $k⋅∈⋅K$.  Next consider two nodes $x$ and $x_+$ in $Z$ such that $x⋅≺⋅x_+$.  Then any run $Z\hi⋅∈⋅\ZZ$ through $x_+$ also goes through $x$.  In other words,  $⎨Z\hi|x∈Z\hi∈\ZZ⎬$ $⊇$ $⎨Z\hi|x_+∈Z\hi∈\ZZ⎬$.  This implies $\sup⎨u_k(Z\hi)|x∈Z\hi∈\ZZ⎬$ $≥$ $\sup⎨u_k(Z\hi)|x_+∈Z\hi∈\ZZ⎬$.  Therefore $\text{sup}⎨u_k(Z\hi)|x∈Z\hi∈\ZZ⎬$ is weakly decreasing in $x$, which implies that the limit in (\rf{D287}) exists.  

\yl{D291} {\em $(∀Z∈\ZZ,k∈K)⋅\text{\normalfont lim}_{x∈Z}⋅\text{\normalfont sup}⎨u_k(Z\hi)|x∈Z\hi∈\ZZ⎬⋅≥⋅u_k(Z)$.}  To see this, take a run $Z⋅∈⋅\ZZ$ and a stakeholder $k⋅∈⋅K$.  By Claim~\rf{D290}, the limit exists.  Further, for any node $x⋅∈⋅Z$, the set $⎨Z\hi|x∈Z\hi∈\ZZ⎬$ contains $Z$, which by inspection implies $\sup⎨u_k(Z\hi)|x∈Z\hi∈\ZZ⎬$ is weakly above $u_k(Z)$.  Therefore the limit is weakly above $u_k(Z)$. \end{cllist}\unskipcl\end{pf}

\begin{lemma}\label{D292} Consider a game $(Q,u)$.  Then \ttr{a}{D294} $(∀Z∈\ZZ,k∈K)$ the limit in (\rf{D293}) exists.  Further, \ttr{b}{D295} lower-convergence fails iff \begin{gather}
\zz
(∃Z∈\ZZ,k∈K)⋅\text{\normalfont lim}_{x∈Z}⋅\text{\normalfont inf}⎨u_k(Z\lo)|x∈Z\lo∈\ZZ⎬ < u_k(Z). \label{D293}
\zz
\end{gather} \end{lemma}

\begin{pf} This is proved as Lemma~\rf{D284} was proved.  Replace upper-convergence (\rf{C924}) with lower-convergence (\rf{C925}), sup with inf, $Z\hi$ with $Z\lo$, $>$ and $≥$ with $<$ and $≤$, ``decreasing'' with ``increasing'', and ``above'' with ``below''. \end{pf}

\begin{ndef}[{{\bf Subroot sequence}}]\label{C929} Suppose $Q$ is a form, $s⋅∈⋅S$, and $t_0⋅∈⋅T$.  Then the {\em subroot sequence from $t_0$ via $s$} is the sequence $⁅t_m⁆_{m∈M}$ defined recursively by the given $t_0$, by $t_1 = \max˙O^{t_0}(s^{t_0})$, by $t_2 = \max˙O^{t_1}(s^{t_1})$, and so on, either [a] indefinitely or [b] until an $ℓ⋅≥⋅0$ for which it is not the case that $\max˙O^{t_ℓ}(s^{t_ℓ})$ exists and is in $T$.  (To be clear, $M = ⎨0˛1˛...⎬$ in case~[a], and $M = ⎨0˛1˛...ℓ⎬$ in case~[b].) \end{ndef}

\begin{lemma}\label{C441} Suppose $(Q,u)$ is a game, $s⋅∈⋅S$, and $t_0⋅∈⋅T$.  Let $⁅t_m⁆_{m∈M}$ be the subroot sequence from $t_0$ via $s$ (Definition \rf{C929}).  Then the following hold. \begin{tlist}
\yl{D370} $(∀m∈M)$ $O^{t_m}(s^{t_m})⋅∈⋅\ZZ^{t_m}$.
\yl{C931} $(∀m∈M⧷⎨0⎬)$ $t_{m-1}⋅≺⋅t_m$.
\yl{C451} $(∀m∈M⧷⎨0⎬)$ $R(^{t_{m-1}}O(^{t_{m-1}}s)) = R(^{t_m}O(^{t_m}s))$.
\yl{C448} $(∀m∈M)$ $R(^{t_0}O(^{t_0}s)) = R(^{t_m}O(^{t_m}s))$.
\yl{C429} $R(^{t_0}O(^{t_0}s))⋅⊇⋅⎨t_m|m∈M⎬$.
\yl{C445} Suppose $M$ is infinite and $u$ is upper-convergent.  Then \begin{gather} 
\zz
(∀k∈K)⋅\lim_{m→∞}˙\sup⎨u_k(Z\hi)|t_m∈Z\hi∈\ZZ⎬ = u_k(R(^{t_0}O(\bss{t_0}))), \notag
\zz
\end{gather} 
\yl{C446} Suppose $M$ is infinite and $u$ is lower-convergent.  Then \begin{gather}
\zz
(∀k∈K)⋅\lim_{m→∞}˙\inf⎨u_k(Z\lo)|t_m∈Z\lo∈\ZZ⎬ = u_k(R(^{t_0}O(\bss{t_0}))). \notag
\zz
\end{gather} \end{tlist} \end{lemma}

\begin{pf} {\em (\rf{D370})}.  Take an index $m⋅∈⋅M$.  The subroot sequence's Definition~\rf{C929} implies $t_m⋅∈⋅T$.  Thus the restriction $s^{t_m}$ is in $S^{t_m}$ by footnote \rf{D510} (Section~\rf{C896}).  This suffices because $O^{t_m}{:}S^{t_m}→\ZZ^{t_m}$ by the definition of $⁅O^t⁆_{t∈T}$ (Section~\rf{C896}).

{\em (\rf{C931})}.  Take an index $m⋅∈⋅M⧷⎨0⎬$.  The subroot sequence's Definition~\rf{C929} implies \lic{C967} $\max˙O^{t_{m-1}}(s^{t_{m-1}}) = t_m$.  Part (\rf{D370}) implies $O^{t_{m-1}}(s^{t_{m-1}})$ is a run in $\ZZ^{t_{m-1}}$, which by the general nontriviality of runs (Lemma~\rf{C966}(\rf{C989})) implies $t_{m-1}⋅≺⋅\max˙O^{t_{m-1}}(s^{t_{m-1}})$, which by \rf{C967} implies $t_{m-1}⋅≺⋅t_m$.

{\em (\rf{C451})}. Take an index $m⋅∈⋅M⧷⎨0⎬$.  The subroot sequence's Definition~\rf{C929} implies \lic{D511} $\max˙O^{t_{m-1}}(s^{t_{m-1}})$ exists and is in $T$, and \li{C968} $\max˙O^{t_{m-1}}(s^{t_{m-1}}) = t_m$.  Now consider this first case of Lemma~\rf{C453}, with $t$ there equal to $t_{m-1}$ here.  This case is relevant because of \rf{D511}, and it implies \begin{gather}
\zz
R(^{t_{m-1}}O(^{t_{m-1}}s)) = R(^{\max˙O^{t_{m-1}}(s^{t_{m-1}})}O(^{\max˙O^{t_{m-1}}(s^{t_{m-1}})}s)). \notag
\zz
\end{gather} This suffices because of \rf{C968}.

{\em (\rf{C448})}. For $m = 0$, the result holds vacuously.  For $m⋅∈⋅M⧷⎨0⎬$, apply part (\rf{C451}) $m$ times.  

{\em (\rf{C429})}. Take an index $m⋅∈⋅M$.  Then $t_m$ by inspection belongs to $^{t_m}O(^{t_m}s)$, which by inspection is included in $R(^{t_m}O(^{t_m}s))$, which by part~(\rf{C448}) equals $R(^{t_0}O(^{t_0}s))$.

{\em (\rf{C445})}. Since $M$ is infinite by this part's assumption, Definition~\rf{C929} implies $M = ⎨0˛1˛...⎬$.  Thus parts (\rf{C931}) and (\rf{C429}) imply that \tts{$*$}{D231} $⁅t_m⁆^∞_{m∈0}$ is an infinite and strictly monotonic sequence of nodes in the run $R(^{t_0}O(^{t_0}s))$.  Now take a stakeholder $k⋅∈⋅K$.  Upper-convergence (\rf{C924}) at the run $Z = R(^{t_0}O(^{t_0}s))$ implies\begin{gather}
\zz
\lim_{x∈R(^{t_0}O(^{t_0}s))}⋅\sup⎨u_k(Z\hi)|x∈Z\hi∈\ZZ⎬ = u_k(R(^{t_0}O(\bss{t_0}))), \notag
\zz
\end{gather} which by \rf{D231} implies this part's conclusion (footnote~\rf{D232} on page \pageref{D232} can be helpful if convergence over directed sets is unfamiliar). 

{\em (\rf{C446})}. This is proved as part~(\rf{C445}).  More specifically, replace ``upper'' with ``lower'', and ``sup'' with ``inf''.  \end{pf}

\begin{lemma}\label{C930} Suppose $(Q,u)$ is a game, $s⋅∈⋅S$, and $v$ is a value function which is persistent for $s$.  Further, take $t_0⋅∈⋅T$ and let $⁅t_m⁆_{m∈M}$ be the subroot sequence from $t_0$ via $s$ (Definition~\rf{C929}).  Then if $M$ is finite,\footnote{It is possible that the subroot sequence $⁅t_m⁆_{m∈M}$ is finite even though the path $^{t_0}O(\bss{t_0})$ is infinite.  Relatedly, Lemma~\rf{C930} is applied to finite paths in the proof of Lemma~\rf{C926}, and then to arbitrary paths in Proof~\rf{C290p}, Claim~\rf{C421}. } $v(t_0) = u(R(^{t_0}O(\bss{t_0})))$. \end{lemma}

\begin{pf} Since $M$ is finite, there is $ℓ⋅≥⋅0$ such that $M = ⎨0˛1˛...ℓ⎬$.  It suffices to prove by induction that $(∀m∈M)$ $v(t_m) = u(R(^{t_m}O(^{t_m}s)))$.  For the initial step ($m{=}ℓ$), note that the subroot sequence's Definition~\rf{C929} and the above definition of $ℓ$ together imply that it is not the case that $\max˙O^{t_ℓ}(s^{t_ℓ})$ exists and is in $T$.  Thus $v(t_ℓ)$ by the second case of persistence (\rf{C922}) equals $u(R(O^{t_ℓ}(s^{t_ℓ})))$, which by the second case of Lemma~\rf{C453} equals $u(R(^{t_ℓ}O(^{t_ℓ}s)))$.  

For the inductive step ($m{<}ℓ$), note that the definition of $t_m$ (Definition~\rf{C929}) implies that $\max˙O^{t_m}(s^{t_m})$ exists and is in $T$.  Thus $v(t_m)$ by the first case of persistence (\rf{C922}) is equal to $v(\max˙O^{t_m}(s^{t_m}))$, which by the definition of $t_{m+1}$ (Definition~\rf{C929}) is equal to $v(t_{m+1})$, which by the inductive hypothesis is equal to $u(R(^{t_{m+1}}O(^{t_{m+1}}s)))$, which by Lemma~\rf{C441}(\rf{C451}) [with $m$ there equal to $m{+}1$ here] is equal to $u(R(^{t_m}O(^{t_m}s)))$. \end{pf}

\begin{lemma}\label{C926} Suppose $(Q,u)$ is a game and $\ZZ = \ZZf$.  Then persistence and authenticity are equivalent, and either implies admissibility. \end{lemma}

\begin{pf} Lemma~\rf{C431} implies that authenticity implies admissibility and persistence (the assumption $\ZZ = \ZZf$ is not used).  Thus it suffices to show that persistence implies authenticity.  Toward that end, assume that $(s,v)$ is persistent.  By authenticity's definition (\rf{C923}), it suffices to show that $(∀t_0∈T)$ $v(t_0) = u(R(^{t_0}O(\bss{t_0})))$.  

Take $t_0⋅∈⋅T$ and let $⁅t_m⁆_{m∈M}$ be the subroot sequence from $t_0$ via $s$ (Definition~\rf{C929}).  The assumption $\ZZ = \ZZf$ implies that the path $^{t_0}O(\bss{t_0})$ is finite, which by Lemma~\rf{C441}(\rf{C429}) implies that the subroot sequence $⁅t_m⁆_{m∈M}$ is finite, which by Lemma~\rf{C441}(\rf{C931}) implies that $M$ is finite, which by Lemma~\rf{C930} implies $v(t_0) = u(R(^{t_0}O(\bss{t_0})))$. \end{pf}

\begin{npf}[{{\bf for Theorem~\rf{C290}}}]\label{C290p} Lemma~\rf{C431} shows authenticity implies admissibility and persistence.  To show the converse, suppose $v$ is admissible and $(s,v)$ is persistent.  By authenticity's definition (\rf{C923}), it suffices to show that $(∀t_0∈T)$ $v(t_0) = u(R(^{t_0}O(\bss{t_0})))$.  Toward that end, take $t_0⋅∈⋅T$ and let $⁅t_m⁆_{m∈M}$ be the subroot sequence from $t_0$ via $s$ (Definition~\rf{C929}).  Then the result follows from Claims~\rf{C421} and \rf{C426}. \begin{cllist} 

\yl{C421} {\em If $M$ is finite, then $v(t_0) = u(R(^{t_0}O(\bss{t_0})))$}.  This follows from Lemma~\rf{C930} (admissibility and upper- and lower-convergence play no role here).

\yl{C971} {\em $(∀m∈M)$ $v(t_0) = v(t_m)$}.  To see this, take an index $m⋅∈⋅M$.  It suffices to show by induction that $(∀n∈⎨0˛1˛...m⎬)$ $v(t_n) = v(t_m)$.  The initial step ($n{=}m$) holds vacuously.  For the inductive step ($n{<}m$), note that the definition of $⁅t_m⁆_{m∈M}$ (Definition~\rf{C929}) implies that $\max˙O^{t_n}(s^{t_n})$ exists and is in $T$.  Thus $v(t_n)$ by the first case of persistence (\rf{C922}) is equal to $v(\max˙O^{t_n}(s^{t_n}))$, which by the definition of $t_{n+1}$ (Definition~\rf{C929}) is equal to $v_k(t_{n+1})$, which by the inductive hypothesis is $v_k(t_m)$.

\yl{C425} ⋅\begin{gather}
\zz
(∀k∈K,m∈M)⋅\inf⎨u_k(Z\lo)|t_m∈Z\lo∈\ZZ⎬⋅≤⋅v_k(t_0)⋅≤⋅\sup⎨u_k(Z\hi)|t_m∈Z\hi∈\ZZ⎬. \notag
\zz
\end{gather} For this, take a stakeholder $k⋅∈⋅K$ and an index $m⋅∈⋅M$.  Admissibility (\rf{C921}) implies\begin{gather}
\zz
\inf⎨u_k(Z\lo)|t_m∈Z\lo∈\ZZ⎬⋅≤⋅v_k(t_m)⋅≤⋅\sup⎨u_k(Z\hi|t_m∈Z\hi∈\ZZ⎬.\notag
\zz
\end{gather} Claim~\rf{C971} implies $v_k(t_m) = v_k(t_0)$.

\yl{C426} {\em If $M$ is infinite, $v(t_0) = u(R(^{t_0}O(\bss{t_0})))$}.  Suppose $M$ is infinite.  It suffices to show that $(∀k∈K)$ $v_k(t_0) = u_k(R(^{t_0}O(\bss{t_0})))$.  Toward that end, take a stakeholder $k⋅∈⋅K$.  Claim~\rf{C425} implies that the value $v_k(t_0)$ is between Claim~\rf{C425}'s lower and upper bounds.  By Lemma~\rf{C441}(\rf{C445},\rf{C446}), these bounds both converge to the utility $u_k(R(^{t_0}O(\bss{t_0})))$.  Hence $v_k(t_0) = u_k(R(^{t_0}O(\bss{t_0})))$. \end{cllist} \unskipcl \end{npf}

\begin{lemma}\label{C435} Suppose $(Q,u)$ is a game, $(s,v)$ is authentic, and $t⋅∈⋅T$.  Then $(∀σ∈S)$ $u(R(\tOO(σ^t,s|_{\tJ⧷J^t}))) = u^t_v(O^t(σ^t))$.\footnote{[a] At the expensive of more notation, this can be quantified by $(∀\psi∈S^t)$ with $\psi$ replacing $σ^t$ in the equality (a similar alternative is discussed in footnote~\rf{D268} on page \pageref{D268}). [b] Intuitively, Lemma~\rf{C435} shows that the utility from obeying strategy $s$ after piece $t$ can be found by inserting the value function $v$ after piece $t$.} \end{lemma}

\begin{pf} Lemma~\rf{C453}, with its $s$ being $(σ^t,s|_{J⧷J^t})$, implies\begin{gather}
\zz
R(\tOO(˙\bbS{t}{(σ^t,s|_{J⧷J^t})})) = \label{D512} \\
\mip
R(\bO{\max˙O^t(σ^t)}(\bbS{\max˙O^t(σ^t)}{(σ^t,s|_{J⧷J^t})}))⋅\text{if}⋅\max˙O^t(σ^t)⋅\text{exists and is in $T$} \\
\rule{2.2ex}{0ex} R(O^t(σ^t))⋅\text{otherwise} \mop. \notag
\zz
\end{gather} Two simplifications can be made.  On the left-hand side, $\bbS{t}{(σ^t,s|_{J⧷J^t})}$ by restriction definition (\rf{D191}) is $(σ^t,s|_{J⧷J^t})|_{\tJ}$˙, which reduces to $(σ^t,s|_{\tJ⧷J^t})$.  On the right-hand side, $\bbS{\max˙O^t(σ^t)}{(σ^t,s|_{J⧷J^t})}$ by restriction definition (\rf{D191}) is $(σ^t,s|_{J⧷J^t})|_{\bJ{\max˙O^t(σ^t)}}$˙, which by restriction definition (\rf{D193}) is $(σ|_{J^t},s|_{J⧷J^t})|_{\bJ{\max˙O^t(σ^t)}}$˙, which by $J^t˙⋂˙\bJ{\max˙O^t(σ^t)} = ∅$ reduces to $s|_{\bJ{\max˙O^t(σ^t)}}$˙, which by restriction definition (\rf{D191}) is $\bss{\max˙O^t(σ^t)}$.  

By the previous two sentences, and by applying $u$ to both sides of (\rf{D512}), we find that $u(R(\tOO(σ^t,s|_{\tJ⧷J^t})))$ is equal to \begin{gather}
\zz
\mip u(R(\bO{\max˙O^t(σ^t)}(\bss{\max˙O^t(σ^t)})))⋅\text{if}⋅\max˙O^t(σ^t)⋅\text{exists and is in $T$} \\
u(R(O^t(σ^t)))⋅\text{otherwise}\rule{7.1ex}{0ex} \mop, \notag 
\zz
\end{gather} which, by authenticity's definition (\rf{C923}) with its $t$ being $\max˙O^t(σ^t)$, is equal to\begin{gather}
\zz
\mip v(\max˙O^t(σ^t))⋅\text{if}⋅\max˙O^t(σ^t)⋅\text{exists and is in $T$} \\
u(R(O^t(σ^t)))⋅\text{otherwise}\rule{20.5ex}{0ex} \mop, \notag 
\zz
\end{gather} which, by $u^t_v$'s definition (\rf{C972}) with its $N$ being $O^t(σ^t)$, is equal to $u^t_v(O^t(σ^t))$. \end{pf}

\pagebreak

\begin{lemma}\label{D242} Suppose $(Q,u)$ is a game and $(s,v)$ is authentic.  Then the following hold. \begin{tlist}
\yl{D243} $(∀t∈T)$ $u(R(\tOO(\tss))) = u^t_v(O^t(s^t))$.  
\yl{D244} $(∀t∈T,i∈I,σ∈S)$ $u(R(\tOO(σ^t_i,\tss|_{\tJ⧷J^t_i})))$ $=$ $u^t_v(O^t(σ^t_i,s^t_{-i}))$. \end{tlist}\end{lemma}

\begin{pf} For (\rf{D243}), apply Lemma~\rf{C435}, with its $σ$ equal to $s$.  For (\rf{D244}), apply Lemma~\rf{C435} again, this time with its $σ$ equal to $(σ^t_i,s|_{J⧷J^t_i})$. \end{pf}

\begin{lemma}\label{C434} Suppose $(Q,u)$ is a game and $s$ is a subgame-perfect equilibrium.  Define $v$ by $(∀t∈T)$ $v(t) = u(R(^tO(\tss)))$.  Then $(s,v)$ is authentic and piecewise-Nash. \end{lemma}

\begin{pf} Authenticity (\rf{C923}) follows immediately from the lemma's definition of $v$.  To show piecewise-Nashness, suppose $(s,v)$ is not piecewise-Nash (\rf{D223}).  Then there are $t⋅∈⋅T$, $i⋅∈⋅I$, and $σ⋅∈⋅S$ such that \begin{align}
\zz
u^t_{v,i}(O^t(s^t)) < u^t_{v,i}(O^t(σ^t_i,s^t_{-i})). \label{D245}
\zz
\end{align} Since authenticity has already been shown, the assumptions of Lemma~\rf{D242} are met.  Thus (\rf{D245}) and Lemma~\rf{D242}'s two conclusions imply \begin{gather}
\zz
u_i(R(\tOO(\tss))) < u_i(R(\tOO(σ^t_i,\tss|_{\tJ⧷J^t_i}))). \notag
\notag
\zz
\end{gather} Hence the definition (\rf{D189}) of the subgame utility function $\tu$ implies \begin{gather}
\zz
\tu_i(\tOO(\tss)) < \tu_i(\tOO(σ^t_i,\tss|_{\tJ⧷J^t_i})). \notag
\zz
\end{gather} This violates the definition (\rf{D234}) for subgame perfection, in contradiction to the lemma's assumptions. \end{pf}

\begin{npf}[{{\bf for Theorem~\rf{C292}}}]\label{C292p} Lemma~\rf{C434} shows that if $s$ is a subgame-perfect equilibrium, then there is a value function $v$ such that $(s,v)$ is authentic and piecewise-Nash.  To show the converse, suppose that $(s,v)$ is authentic and piecewise-Nash.  By definition (\rf{D234}) for subgame perfection, it suffices to show \begin{gather}
\zz
(∀τ_0∈T,i∈I,σ∈S)⋅\bu{τ_0}_i(\bO{τ_0}(⋅\bss{τ_0}⋅))⋅≥⋅ \bu{τ_0}_i(\bO{τ_0}(˙\bsig{τ_0}_i,{\bss{τ_0}_{-i}}˙)). \notag
\zz
\end{gather} By definition (\rf{D189}) for $\bu{τ_0}$, this is equivalent to \begin{gather}
\zz
(∀τ_0∈T,i∈I,σ∈S)⋅u_i(R(\bO{τ_0}(⋅\bss{τ_0}⋅)))⋅≥⋅
u_i(R(\bO{τ_0}(⋅\bsig{τ_0}_i,\bss{τ_0}_{-i}⋅))). \notag 
\zz
\end{gather} To prove this, take a subroot $τ_0⋅∈⋅T$, a player $i⋅∈⋅I$, and an alternative $σ⋅∈⋅S$.  The theorem then follows from Claims \rf{C440} and \rf{C439} below.

Before entertaining the claims, use general Definition~\rf{C929} to let \begin{gather}
\zz
⁅τ_m⁆_{m∈M}⋅\text{be the subroot sequence from $τ_0$ via $(˙σ_i,s_{-i}˙)$}.  \label{D249}
\zz
\end{gather} Note that the subroot sequence $⁅τ_m⁆_{m∈M}$ is derived via the alternative $(σ_i,s_{-i})$ rather than via the original $s$ (the Greek-ness of the notation $⁅τ_m⁆_{m∈M}$ is meant to emphasize this).  Also note that the notation accommodates multiple cases.  First, the finiteness of the original path $\bO{τ_0}(\bss{τ_0})$ is unrelated to the finiteness of the alternative path $\bO{τ_0}(\bsig{τ_0}_i,\bss{τ_0}_{-i})$.  Second, the finiteness of the alternative path $\bO{τ_0}(\bsig{τ_0}_i,\bss{τ_0}_{-i})$ implies the finiteness of the subroot sequence $⁅τ_m⁆_{m∈M}$, but not conversely.  (The implication holds because [a] the cardinality of the subroot sequence equals the cardinality of its image $⎨τ_m|m∈M⎬$ [by Lemma~\rf{C441}(\rf{C931})] and [b] this image is a subset of the alternative path $\bO{τ_0}(\bsig{τ_0}_i,\bss{τ_0}_{-i})$ [by Lemma~\rf{C441}(\rf{C429})].) \begin{cllist}

\yl{C974} {\em $(∀m∈M⧷⎨0⎬)$ $u_i(R(\bO{τ_{m-1}}(\bss{τ_{m-1}})))$ $≥$ $u_i(R(\bO{τ_m}(\bss{τ_m})))$.  (For intuition, note that $R(^{τ_m}O(\bss{τ_m}))$ is the full run that goes from $r$ to $τ_0$, then from $τ_0$ to $τ_m$ via the alternative $(σ_i,s_{-i})$, and then obeys the original $s$ thereafter.  So roughly, this claim states that it is weakly better to start obeying $s$ after $τ_{m-1}⋅≺⋅τ_m$ rather than to wait and start obeying $s$ after $τ_m$.)}  

To show this, take a subroot index $m⋅∈⋅M⧷⎨0⎬$.  Then definition (\rf{D249}) for the alternative subroot sequence implies \begin{gather}
\zz
\max˙O^{τ_{m-1}}(˙σ^{τ_{m-1}}_i,˙s^{τ_{m-1}}_{-i})⋅
\text{exists and is in $T$, and} \label{D246} \\
τ_m = \max˙O^{τ_{m-1}}(˙σ^{τ_{m-1}}_i,˙s^{τ_{m-1}}_{-i}). \label{D247}
\zz
\end{gather} It suffices to show that \begin{align}
\zz
u_i(R(\bO{τ_{m-1}}(\bss{τ_{m-1}}))) 
=&⋅u^{τ_{m-1}}_{v,i}(O^{τ_{m-1}}(s^{τ_{m-1}})) \nt 
≥&⋅u^{τ_{m-1}}_{v,i}(O^{τ_{m-1}}(⋅σ^{τ_{m-1}}_i,˙s^{τ_{m-1}}_{-i}⋅)) \nt
=&⋅v_i(\max˙O^{τ_{m-1}}(⋅σ^{τ_{m-1}}_i,˙s^{τ_{m-1}}_{-i}⋅)) \nt
=&⋅v_i(τ_m) \nt
=&⋅u_i(R(^{τ_m}O(\bss{τ_m}))). \notag
\zz
\end{align} The first equality holds by authenticity and Lemma~\rf{D242}(\rf{D243}), with $t$ there equal to $τ_{m-1}$ here.  The inequality holds by definition (\rf{D223}) for piecewise-Nashness, with $t$ there equal to $τ_{m-1}$ here.  The second equality holds by (\rf{D246}) and the first case of definition (\rf{C972}) for $⁅u^t_v⁆_{t∈T}$, with $t$ and $N$ there equal to $τ_{m-1}$ and $O^{t_{m-1}}(σ^{t_{m-1}}_i,s^{t_{m-1}}_{-i})$ here.  The third equality holds by (\rf{D247}), and finally, the fourth equality holds by definition (\rf{C923}) for authenticity.

\yl{D250} {\em If $ℓ = \max˙M$, $u_i(R(\bO{τ_ℓ}(\bss{τ_ℓ})))⋅≥⋅ u_i(R(^{τ_ℓ}O(\bsig{τ_ℓ}_i,\bss{τ_ℓ}_{-i})))$. (Roughly, if the alternative subroot sequence terminates at $τ_ℓ$, then it is weakly better to obey $s$ after $τ_ℓ$ than to not obey $s$ after $τ_ℓ$.)} 

Suppose $ℓ = \max˙M$.  Then definition (\rf{D249}) for the alternative subroot sequence implies\begin{gather}
\zz
\text{not}\big( \max˙O^{τ_ℓ}(σ^{τ_ℓ}_i,s^{τ_ℓ}_{-i})⋅\text{exists and is in}⋅T\big). \label{D248} 
\zz
\end{gather} This will be used to argue that\begin{align}
\zz
u_i(R(\bO{τ_ℓ}(\bss{τ_ℓ}))) 
=&⋅u^{τ_ℓ}_{v,i}(O^{τ_ℓ}(s^{τ_ℓ})) \nt
≥&⋅u^{τ_ℓ}_{v,i}(O^{τ_ℓ}(σ^{τ_ℓ}_i,s^{τ_ℓ}_{-i})) \nt
=&⋅u_i(R(O^{τ_ℓ}(σ^{τ_ℓ}_i,s^{τ_ℓ}_{-i})) \nt
=&⋅u_i(R(˙^{τ_ℓ}O(\bsig{τ_ℓ}_i,\bss{τ_ℓ}_{-i}))). \notag
\zz
\end{align} The first equality holds by authenticity and Lemma~\rf{D242}(\rf{D243}), with $t$ there equal to $τ_ℓ$ here.  
The inequality holds by definition (\rf{D223}) for piecewise-Nashness, with $t$ there equal to $τ_ℓ$ here.  
The second equality holds by (\rf{D248}) and the second case of definition (\rf{C972}) for $⁅u^t_m⁆_{t∈T}$, with $t$ and $N$ there equal to $τ_ℓ$ and $O^{τ_ℓ}(σ^{τ_ℓ}_i, s^{τ_ℓ}_{-i})$.  
The third equality holds by (\rf{D248}) and the second case of Lemma~\rf{C453}, with $t$ and $s$ there equal to $τ_ℓ$ and $(σ_i,s_{-i})$ here.

\yl{C440} {\em If $M$ is finite, $u_i(R(\bO{τ_0}(\bss{τ_0})))$ $≥$ $u_i(R(\bO{τ_0}(\bsig{τ_0}_i,{\bss{τ_0}_{-i}})))$.  (This completes the proof if $M$ is finite.)}  

Assume $M$ is finite, and let $ℓ = \max˙M$.  Then \begin{align}
\zz
u_i(R(\bO{τ_0}(\bss{τ_0})))⋅ 
≥&⋅u_i(R(\bO{τ_ℓ}(\bss{τ_ℓ}))) \nt
≥&⋅u_i(R(^{τ_ℓ}O(\bsig{τ_ℓ}_i,\bss{τ_ℓ}_{-i}))) \nt
=&⋅u_i(R(\bO{τ_0}(\bsig{τ_0}_i,\bss{τ_0}_{-i}))). \notag
\zz
\end{align} The first inequality holds by $ℓ$ applications of Claim~\rf{C974}.  The second inequality is Claim~\rf{D250}.  Finally, the equality holds by applying Lemma~\rf{C441}(\rf{C448}) with $s$, $t_0$, and $m$ there equal to $(σ_i,s_{-i})$, $τ_0$, and $ℓ$ here (this is a relatively straightforward property of the alternative subroot sequence).

\yl{D252} {\em Suppose $M$ is infinite and $u_i(R(\bO{τ_0}(\bss{τ_0}))) <
u_i(R(\bO{τ_0}(˙\bsig{τ_0}_i,{\bss{τ_0}_{-i}}˙)))$.  Then there is an $m⋅∈⋅M$ such that $u_i(R(\bO{τ_0}(\bss{τ_0}))) < u_i(R(^{τ_m}O(\bss{τ_m})))$.  (Intuitively, if the alternative is better, there comes a subroot $τ_m$ in the alternative subroot sequence after which reverting to the original does not wreck the alternative.)}˙\footnote{\label{D519}Claim~\rf{D252} relies upon the theorem's assumption of lower-convergence.  For example, consider the one-player example of Figure~\rf{D518}, in which the player Bob's utility function is not lower-convergent.  The remainder of this note will show how Claim~\rf{D252} fails in this example.  \par
Let the original $s = s_{\f{Bob}}$ be the strategy of Bob always choosing $\f{out}$, let the alternative $σ = σ_{\f{Bob}}$ be the strategy of Bob always choosing $\f{in}$, and consider the subroot $τ_0 = r = \f5$.  Then the alternative subroot sequence is $τ_0 = \f5$, $τ_1 = \f6$, $τ_2 = \f7$, and so on.  As assumed by Claim~\rf{D252}, $M$ is infinite, and the original utility\begin{gather}
\zz
u_{\f{Bob}}(R(\bO{τ_0}(\bss{τ_0}))) = u_{\f{Bob}}(⎨\f5,\f{5^*}⎬) = -1 \notag
\zz
\end{gather} is less than the alternative utility\begin{gather}
\zz
u_{\f{Bob}}(R(\bO{τ_0}(\bsig{τ_0}))) = u_{\f{Bob}}(⎨\f5,\f6,\f7,...⎬) = 0.\notag
\zz
\end{gather} Yet, it is always possible to wreck the alternative by reverting to the original. Informally, the original chooses $\f{out}$ and yields utility $-1$ regardless of how long the original is delayed.  Formally, for any $m⋅∈⋅M = ⎨0˛1˛2˛...⎬$, we have $u_{\f{Bob}}(R(\bO{τ_m}(\bss{τ_m}))) = u_{\f{Bob}}(⎨\f5,\f6,\f7,...˙(\f5{+}{m}),(\f5{+}m)^{\f{*}}⎬) = -1$, which does not exceed the original utility $u_{\f{Bob}}(R(\bO{τ_0}(\bss{τ_0}))) = -1$.  Thus Claim~\rf{D252}'s conclusion fails.}

Consider Lemma~\rf{C441}, with $s$ and $⁅t_m⁆_{m∈M}$ there equal to $(σ_i,s_{-i})$ and $⁅τ_m⁆_{m∈M}$ here.  As assumed by the lemma, $⁅τ_m⁆_{m∈M}$ is the subroot sequence from $τ_0$ via $(σ_i,s_i)$.  Then, because of lower-convergence, the lemma's part (\rf{C446}) at $k = i$ implies \begin{gather}
\zz
\lim_{m_*→∞}˙\inf⎨u_i(Z\lo)|τ_{m_*}∈Z\lo∈\ZZ⎬ = u_i(R(\bO{τ_0}(\bsig{τ_0}_i,\bss{τ_0}_{-i}))). \label{D269}
\zz
\end{gather} By the claim's assumption, the original utility $u_i(R(\bO{τ_0}(\bss{τ_0})))$ is less than the alternative utility on (\rf{D269})'s right-hand side.  Hence (\rf{D269}) implies there is $m⋅∈⋅M$ such that\begin{gather}
\zz
\inf⎨u_i(Z\lo)|τ_m∈Z\lo∈\ZZ⎬ > u_i(R(\bO{τ_0}(\bss{τ_0}))). \notag
\zz
\end{gather} By inspection $τ_m⋅∈⋅{^{τ_m}O}(\bss{τ_m})$, 
which by $R$'s definition implies $τ_m⋅∈⋅R({^{τ_m}O}(\bss{τ_m}))$, 
which by Lemma~\rf{D200}(\rf{D202}) implies $R({^{τ_m}O}(\bss{τ_m}))⋅∈⋅⎨Z\lo|τ_m∈Z\lo∈\ZZ⎬$, 
which by applying $u_i$ implies $u_i(R({^{τ_m}O}(\bss{τ_m})))⋅∈⋅⎨u_i(Z\lo)|τ_m∈Z\lo∈\ZZ⎬$, 
which by inspection implies $u_i(R({^{τ_m}O}(\bss{τ_m})))$ $≥$ $\inf˙⎨u_i(Z\lo)|τ_m∈Z\lo∈\ZZ⎬$.  
Thus the previous sentence implies the claim's conclusion: \begin{gather}
\zz
u_i(R({^{τ_m}O}(\bss{τ_m})) > u_i(R(\bO{τ_0}(\bss{τ_0}))). \notag
\zz
\end{gather} 

\yl{C439} {\em If $M$ is infinite, $u_i(R(\bO{τ_0}(\bss{τ_0})))$ $≥$ $u_i(R(\bO{τ_0}(\bsig{τ_0}_i,{\bss{τ_0}_{-i}})))$. (This completes the proof if $M$ is infinite.)} 

Suppose $M$ is infinite and $u_i(R(\bO{τ_0}(\bss{τ_0}))) <
u_i(R(\bO{τ_0}(˙\bsig{τ_0}_i,{\bss{τ_0}_{-i}}˙)))$.  Then Claim~\rf{D252} implies there is $m⋅∈⋅M$ such that $u_i(R(\bO{τ_0}(\bss{τ_0}))) < u_i(R(^{τ_m}O(\bss{τ_m})))$.  But $m$ applications of Claim~\rf{C974} imply that $u_i(R(\bO{τ_0}(\bss{τ_0})))$ $≥$ $ u_i(R(^{τ_m}O(\bss{τ_m})))$.  These two inequalities contradict.

\end{cllist} \unskipcl \end{npf}

\begin{npf}[{{\bf for Corollary~\rf{D211}}}]\label{D211p} The reverse direction is shown in the paragraph before the corollary statement.  For the forward direction, suppose $s$ is one-piece unimprovable.  Then definition (\rf{D210}) implies \begin{gather}
\zz
(∀t∈T,i∈I,σ∈S)⋅⋅u_i(R(\tOO(⋅\tss⋅)))⋅≥⋅u_i(R(\tOO(⋅σ^t_i,˙s|_{\tJ⧷J^t_i}⋅))). \label{D237}
\zz
\end{gather} Define the value function $v{:}T→\smash{Ṛ^K}$ by $v(t) = u(R(\tOO(\tss)))$.   Then $(s,v)$ is authentic (\rf{C923}).  Consequently, Lemma~\rf{D242}(\rf{D243}) implies \begin{gather}
\zz
(∀t∈T,i∈I)⋅⋅u_i(R(\tOO(⋅\tss⋅))) = u^t_{v,i}(O^t(⋅s^t⋅)), \notag 
\zz
\end{gather} and Lemma~\rf{D242}(\rf{D244}) implies \begin{gather}
\zz
(∀t∈T,i∈I,σ∈S)⋅⋅u_i(R(\tOO(⋅σ^t_i,⋅s|_{\tJ⧷J^t_i}⋅))) = u^t_{v,i}(O^t(⋅σ^t_i,⋅s^t_{-i}⋅)). \notag 
\zz
\end{gather} These two equalities can be used to replace, respectively, the left- and right-hand sides of (\rf{D237}).  The result is \begin{gather}
\zz
(∀t∈T,i∈I,σ∈S)⋅⋅u^t_{v,i}(O^t(⋅s^t⋅))⋅≥⋅u^t_{v,i}(O^t(⋅σ^t_i,⋅s^t_{-i}⋅)). \notag
\zz
\end{gather} Thus $(s,v)$ is piecewise-Nash (\rf{D223}).  Therefore, since $(s,v)$ has been shown to be authentic, the forward direction of Theorem~\rf{C292} implies that $s$ is subgame-perfect. \end{npf}

\eput

\begin{picture}(0,0)
  \put(1.5,1.4){\color{white} \rule{35ex}{2.5ex}}   
  \put(2.4,1.6){\SMALL\sc Appendix A. Preliminaries} \end{picture}

%% file: common-dp.tex

\listfiles 
\usepackage{
  amssymb, 
  amsmath, 
  latexsym, 
  graphpap, 
  graphics, 
  verbatim, 
  color, 
  natbib,
  har2nat, 
  layout, 
  amsbsy, 
  calc, 
  pstricks, 
  pst-plot, 
  pst-pdf, 
  pstricks-add, 
  xr, 
  mathdots, 
  enotez, 
  lipsum,
  stmaryrd 
 }

\usepackage[bottom]{footmisc} 

\usepackage[inline]{enumitem} 




\newcommand{\ifdraft}[1]{} 
\newcommand{\ifdrafttext}[1]{} 
\newcommand{\iffinal}[1]{#1} 
\newcommand{\ifallkeys}[1]{} 
\newcommand{\ifbeamer}[1]{} 



  \providecommand{\ifmain}[1]{#1} 
  \newcommand{\ifsuba}[1]{}
  \newcommand{\ifsubb}[1]{}
  \newcommand{\ifsubc}[1]{}
  \newcommand{\ifsubd}[1]{}
  \newcommand{\subadir}[1]{}
  \newcommand{\subbdir}[1]{}
  \newcommand{\subcdir}[1]{}
  \newcommand{\subddir}[1]{}
  
  
  \newcommand{\eput}{ \ifmain{\showbib \end{document}}  }






\newcommand{\capp}{}
\newcommand{\sella}[1]{}
\newcommand{\sellb}[1]{}
\newcommand{\sellc}[1]{}
\newcommand{\selld}[1]{}
\newcommand{\selle}[1]{}
\newcommand{\sellf}[1]{}
\newcommand{\sellg}[1]{}
\newcommand{\sellh}[1]{}
\newcommand{\selli}[1]{}
\newcommand{\sellj}[1]{}





\newtheorem{lemma}{\indent Lemma}
\newtheorem{nthm}[lemma]{\indent Theorem}
\newtheorem{ndef}[lemma]{\indent Definition}
\newtheorem{ncrly}[lemma]{\indent Corollary}
\newtheorem{prop}[lemma]{\indent Proposition}


\theoremstyle{definition} 
\newtheorem{pfa}[lemma]{\indent Proof}
\newenvironment{npf} 
  { \begin{pfa} }
  { \hfill $\Box$ \end{pfa} }


\theoremstyle{remark} 

\newtheorem{absatz}[lemma]{\hspace{-1ex}\tt} 
\newcommand{\alabel}{}
\newcommand{\amark}{}
\newcommand{\ai}[2][]{\setcounter{equation}{\thelemma}
  \begin{absatz}\label{#2}\amark \renewcommand{\alabel}{#2}
  \begin{subequations} {\em#1}} 
\newcommand{\ao}{\end{subequations} \end{absatz}}

\newtheorem*{pfb}{\indent Proof}
\newenvironment{pf} 
  { \begin{pfb} }
  { \hfill $\Box$ \end{pfb}  }

\newtheorem{exa}[lemma]{\indent Question} 
\newtheorem{soln}[lemma]{\indent Answer to Question}
\newcommand{\exercomment}{} 

\newcommand{\ifbothexeransr}[1]{}
  { \begin{exa} \label{#1} \exercomment} 
  { \end{exa} }
  { \end{soln} } 



\newcounter{blistno} %
  { \begin{list} %
      { (\arabic{blistno}) } 
      { \setlength{\leftmargin}{0mm} 
        \setlength{\rightmargin}{0mm} %
        \setlength{\itemindent}{-1mm} 
        \setlength{\labelwidth}{0mm} %
        \setlength{\labelsep}{0mm} %
        \setlength{\itemsep}{1mm} 
        \setlength{\topsep}{1mm} 
        \usecounter{blistno}
        \nopagebreak } } %
  { \end{list} }

\newcounter{clistno} %
  { \begin{list} %
      {  \arabic{clistno}. } 
      {  \setlength{\leftmargin}{0mm} 
        \setlength{\rightmargin}{0mm} %
        \setlength{\itemindent}{3mm} 
        \setlength{\labelwidth}{0mm} %
        \setlength{\labelsep}{0mm} %
        \setlength{\itemsep}{2mm} 
        \setlength{\topsep}{2mm} 
        \usecounter{clistno}
        \nopagebreak } }%
  { \end{list} }    

\newcommand{\tlistindent}{0mm}
\newcounter{tlistno} %
\newenvironment{tlist} %
  { \begin{list} %
      {\indent(\alph{tlistno}) } %
      { \setlength{\leftmargin}{0mm} %
        \setlength{\rightmargin}{0mm} %
        \setlength{\itemindent}{\tlistindent} 
        \setlength{\labelwidth}{0mm} %
        \setlength{\labelsep}{0mm} %
        \setlength{\itemsep}{0mm} %
        \setlength{\topsep}{0mm} %
        \usecounter{tlistno} } } %
  { \end{list} } 

\newcounter{cllistno} %
\newenvironment{cllist} %
  { \begin{list} %
      {\emph{Claim \arabic{cllistno}:} } 
      { \setlength{\leftmargin}{0mm} %
        \setlength{\rightmargin}{0mm} %
        \setlength{\itemindent}{\parindent} %
        \setlength{\labelwidth}{0mm} %
        \setlength{\labelsep}{0mm} %
        \setlength{\itemsep}{1.2mm}
        \setlength{\topsep}{1.2mm}%
        \usecounter{cllistno} } } %
  { \end{list} }

\newcommand{\unskipcl}{\vspace{-5.2mm}} 

\newcommand{\itemi}{ \begin{itemize} }
\newcommand{\itemo}{ \end{itemize} }
\newcommand{\yl}[1]{\item\label{#1}}



\newcommand{\ili}{\begin{enumerate*}[label=$\lbrack$\alph*$\rbrack$]}
\newcommand{\ilr}{\begin{enumerate*}[resume*]}
\newcommand{\ilo}{\end{enumerate*}}
\newcommand{\ilitem}[1]{\item\label{#1}\ilo\hspace{-1mm}\ifallkeys{\hspace{10mm}}}
\newcommand{\ilc}[1]{\ili\ilitem{#1}}
\newcommand{\il}[1]{\ilr\ilitem{#1}}

\newcommand{\lii}{\begin{enumerate*}[label=$\lbrack$\arabic*$\rbrack$]}
\newcommand{\lir}{\begin{enumerate*}[resume*]}
\newcommand{\lio}{\end{enumerate*}}
\newcommand{\liitem}[1]{\item\label{#1}\lio\hspace{-1mm}\ifallkeys{\hspace{10mm}}}
\newcommand{\lic}[1]{\lii\liitem{#1}}
\newcommand{\li}[1]{\lir\liitem{#1}}


\newcommand{\ttc}[2]{\begin{enumerate*}[label={#1}]\item\label{#2}
  \end{enumerate*}\hspace{-0.8ex}\ifallkeys{\hspace{10mm}}}


\newcommand{\ttr}[2]{{(\ttc{#1}{#2})}}
\newcommand{\tts}[2]{{\ttc{[#1]}{#2}}}
\newcommand{\ttz}[2]{{\ttc{#1}{#2}}}



\newcommand{\zz}{}
\newcommand{\subi}{\begin{subequations}}
\newcommand{\subo}{\end{subequations}}
\newcommand{\nt}{\notag \\ }

\newenvironment{maneq}{%
  \setlength{\arraycolsep}{.1ex}
  
  \begin{array}{lc} \rule{5ex}{0ex} & \rule{79ex}{0ex} \\[-3ex] }{%
  \end{array}}
\newcommand{\mqi}{\begin{maneq}}
\newcommand{\mqo}{\end{maneq}}

\newcommand{\ttt}[2]{\tag*{#1}\label{#2}} 

\newcommand{\ci}{\par\begin{center}}
\newcommand{\co}{\end{center}\par\noindent}


\newcommand{\mi}{ \left[ \begin{matrix} }
\newcommand{\mo}{ \end{matrix} \right] }
\newcommand{\miz}{ \left. \begin{matrix} }
\newcommand{\moz}{ \end{matrix} \right. }
\newcommand{\mip}{ \left( { \ } \begin{matrix} }
\newcommand{\mop}{ \end{matrix} { \ } \right) }
\newcommand{\mib}{ \left[ { \ } \begin{matrix} }
\newcommand{\mob}{ \end{matrix} { \ } \right] }
\newcommand{\smi}{ \left[ \begin{smallmatrix} } 
\newcommand{\smo}{ \end{smallmatrix} \right] }
\newcommand{\casei}{ \left\{ \begin{matrix} } 
\newcommand{\caseo}{ \end{matrix} \right. }

\newcommand{\mpi}{ \begin{minipage} }
\newcommand{\mpo}{ \end{minipage} }
\newcommand{\mpic}{\begin{minipage}{4cm}\begin{center}}
\newcommand{\mpoc}{\end{center}\end{minipage}}



\newcommand{\datestmp}{%
  $ \the\year \backslash \the\month \backslash \the\day \backslash
    \the\time $ }

\newcommand{\filestamp}[1]{} 


\newcommand{\markb}[1]{\markboth{#1}{#1}}

\newcommand{\nssec}[2]{%
  \vspace{.4mm} 
  \subsection{#1}\label{#2}\hspace{0mm}%
  \par\vspace{1.6mm}
  \begin{picture}(0,0)
  \put(-.50,.33){\color{white} \rule{85ex}{4ex}}
  \put(-.4,.53){\color{black} \sc \rf{#2}. #1}
  \end{picture}
  \par\vspace{-4.5mm}}

 \newcommand{\ifssec}[1]{#1} 
 \providecommand{\ifssec}[1]{}
\ifssec{

  }

\allowdisplaybreaks  

\newcommand{\toenote}[1]{}


\newcommand{\nichts}[1]{} 
\newcommand{\showbib}{%
  \markb{\sc References}
  \bibliographystyle{\tpath/ecta/econometrica}
  \bibliography{\tpath/bibtex/streuf,\tpath/bibtex/others,References}}  


\definecolor{darkred}{rgb}{.7,0,0}
  
\definecolor{darkgreen}{rgb}{0,.7,0}
  
\definecolor{darkblue}{rgb}{0,0,.97}

\definecolor{igray}{gray}{.87}


\newcommand{\rf}[1]{\ref{#1}} 



 
\newcommand{\ct}[1]{\ensuremath{\mathbf{#1}}}

 







  \renewcommand{\inf}{\text{\normalfont{inf}}}
  \renewcommand{\sup}{\text{\normalfont{sup}}}
  \renewcommand{\lim}{\text{\normalfont{lim}}}

  \renewcommand{\max}{\text{\normalfont max}} 
  \renewcommand{\min}{\text{\normalfont min}}


  \newcommand{\nnexists}{\raisebox{0.15ex}{/}\hspace{-1.15ex}\exists}


  
  \newcommand{\dra}{\rotatebox[origin=c]{180}{$\Lsh$}} 

  \newcommand{\lo}{_{\vartriangle}}
  \newcommand{\hi}{_{\triangledown}}




 













  \newcommand{\rA}{\dot{A}}

  \newcommand{\rI}{\dot{I}} 
  \newcommand{\rJ}{\dot{J}}
  \newcommand{\rK}{\dot{K}}

  \newcommand{\rQ}{\smash{\dot{Q}}}
  \newcommand{\rR}{\dot{R}} 
   
  \newcommand{\rT}{\dot{T}} 
   \newcommand{\ru}{\dot{u}}
   
  \newcommand{\rW}{\dot{W}}
  \newcommand{\rX}{\dot{X}} 
  \newcommand{\rY}{\dot{Y}} 
  
  \newcommand{\rZZ}{\dot{\ZZ}}



  \newcommand{\RQ}{\smash{\ddot{Q}}}







   \newcommand{\ga}{{\tilde{a}}}
   \newcommand{\gc}{{\tilde{c}}}

  \newcommand{\ggr}{\tilde{r}}




  \newcommand{\HH}{{ \mathcal H}}
  
  \newcommand{\PP}{{ \mathcal P}}

  
  \newcommand{\ZZ}{{ \mathcal Z}}
  \newcommand{\ZZf}{ \mathcal Z_{\mathsf{ft}}}
  \newcommand{\ZZi}{\mathcal Z_{\mathsf{inft}}}


  \newcommand{\eR}{\bar{\mathbb R}} 




  \newcommand{\fa}{{\mathsf{a}}}
  
  \newcommand{\fc}{{\mathsf{c}}}
  
  \newcommand{\fe}{{\mathsf{e}}}
  \newcommand{\ff}{{\mathsf{f}}}

  \newcommand{\f}[1]{\mathsf{#1}} 

  \newcommand{\PB}{\mathsf{P}}

  \newcommand{\SB}{\mathsf{S}}




  \newcommand{\bbA}[2]{{^{#1}\mspace{-4mu}{#2}}}
  \newcommand{\bbI}[2]{{^{#1}\mspace{-2.5mu}{#2}}}
  \newcommand{\bbJ}[2]{{^{#1}\mspace{-4mu}{#2}}}
  
  \newcommand{\bbQ}[2]{{^{#1}\mspace{-1.5mu}{#2}}}
  \newcommand{\bbO}[2]{{^{#1}\mspace{-1.5mu}{#2}}}
  \newcommand{\bbS}[2]{{^{#1}\mspace{-2mu}{#2}}}
  \newcommand{\bbss}[2]{{^{#1}\mspace{-2mu}{#2}}}
  \newcommand{\bbW}[2]{{^{#1}\mspace{-1mu}{#2}}}
  \newcommand{\bbY}[2]{{^{#1}\mspace{-1mu}{#2}}}
  \newcommand{\bbu}[2]{{^{#1}\mspace{-2mu}{#2}}}
  \newcommand{\bbZZ}[2]{{^{#1}\mspace{-3mu}{#2}}}
  \newcommand{\bA}[1]{\bbA{#1}{A}}
  \newcommand{\bI}[1]{\bbI{#1}{I}}
  \newcommand{\bJ}[1]{\bbJ{#1}{J}}
  
  \newcommand{\bO}[1]{\bbO{#1}{O}}
  \newcommand{\bQ}[1]{\bbQ{#1}{Q}}
  \newcommand{\bS}[1]{\bbS{#1}{S}}
  \newcommand{\bss}[1]{\bbss{#1}{s}}
  \newcommand{\bsig}[1]{\bbss{#1}{σ}}
  \newcommand{\bW}[1]{\bbW{#1}{W}}
  \newcommand{\bY}[1]{\bbY{#1}{Y}}
  \newcommand{\bu}[1]{\bbu{#1}{u}}
  \newcommand{\bZZ}[1]{\bbZZ{#1}{\ZZ}}
  \newcommand{\tA}{{\bA{t}}}
  \newcommand{\tI}{{\bI{t}}}
  \newcommand{\tJ}{{\bJ{t}}}
  \newcommand{\tQ}{{\bQ{t}}}
  \newcommand{\tOO}{{\bO{t}}}
  \newcommand{\tSS}{{\bS{t}}}
  \newcommand{\tss}{{\bss{t}}}
  \newcommand{\tsig}{{\bsig{t}}}
  \newcommand{\tu}{{\bu{t}}}
  \newcommand{\tW}{{\bW{t}}}
  \newcommand{\tX}{{^t\mspace{-3mu}X}} 
  \newcommand{\tY}{{\bY{t}}} 
  \newcommand{\tZZ}{{\bZZ{t}}}



\usepackage[utf8]{inputenc}
\usepackage{amssymb}
\usepackage{newunicodechar}
\newcommand{\nuc}[2]{\newunicodechar{#1}{#2}} 



  \nuc{α}{{ \alpha}} 
  \nuc{β}{{ \beta}}
  \nuc{γ}{{ \gamma}} \nuc{Γ}{{ \varGamma}}
  \nuc{δ}{{ \delta}} \nuc{Δ}{{ \varDelta}} 
  \nuc{ε}{{ \varepsilon}}
  \nuc{ζ}{{ \zeta}}
  \nuc{η}{{ \eta}}
  \nuc{θ}{{ \theta}} \nuc{Θ}{{ \varTheta}}
  \nuc{ι}{{ \iota}}
  \nuc{κ}{{ \kappa}}
  \nuc{λ}{{ \lambda}} \nuc{Λ}{{ \varLambda}}
  \nuc{μ}{{ \mu}}
  \nuc{ν}{{ \nu}}
  \nuc{ξ}{{ \xi}} \nuc{Ξ}{{ \varXi}}
  \nuc{π}{{ \pi}} \nuc{Π}{{ \varPi}}
  \nuc{ρ}{{ \rho}} 
  \nuc{σ}{{ \sigma}} \nuc{Σ}{{ \varSigma}}
  \nuc{τ}{{ \tau}}
  \nuc{υ}{{ \upsilon}} \nuc{Υ}{{ \varUpsilon}}
  \nuc{φ}{{ \phi}} \nuc{Φ}{{ \varPhi}}
  \nuc{χ}{{ \chi}} 
  \nuc{ψ}{{ \psi}} \nuc{Ψ}{{ \varPsi}}
  \nuc{ω}{{ \omega}} \nuc{Ω}{{ \varOmega}}

  \nuc{≠}{{ \ne}}
  \nuc{∾}{{ \sim}} 
  \nuc{≈}{{ \approx}}
  \nuc{≡}{{ \equiv}} 
  \nuc{∈}{{ \in}} \nuc{∉}{{ \notin}}
  \nuc{∋}{{ \ni}} \nuc{∌}{{ \not\ni}}   
  \nuc{⊊}{{ \subset}} \nuc{⊆}{{ \subseteq}} 
  \nuc{⊋}{{ \supset}} \nuc{⊇}{{ \supseteq}} 
  \nuc{≤}{{ \leq}}
  \nuc{≥}{{ \geq}}
  \nuc{≪}{{ \ll}}
  \nuc{≫}{{ \gg}}
  \nuc{≺}{{ \prec}} \nuc{≼}{{ \preccurlyeq}}
  \nuc{≻}{{ \succ}} \nuc{≽}{{ \succcurlyeq}}
  \nuc{‖}{{ \parallel}}  

  \nuc{∑}{{ \Sigma}} 
  \nuc{∫}{{ \textstyle \int\nolimits}}
  \nuc{∏}{{ \Pi}} 
  \nuc{×}{{ \times}} 
  \nuc{÷}{{ \div}} 
  \nuc{○}{{ \circ}}
  \nuc{⨅}{{\scalebox{1}[1.2]{\ensuremath{\cap}}}} 
  \nuc{⋂}{{\cap}}                                   
  \nuc{⨆}{{\scalebox{1}[1.2]{\ensuremath{\cup}}}} 
  \nuc{⋃}{{\cup}}                                   
  \nuc{⋱}{{ \setminus}} 
  \nuc{⧷}{{ \smallsetminus}} 
  \nuc{∧}{{ \wedge}} \nuc{⋀}{{ \textstyle \bigwedge\nolimits}}
  \nuc{∨}{{ \vee}} \nuc{⋁}{{ \textstyle \bigvee\nolimits}}
  \nuc{⊗}{{ \otimes}}
  \nuc{⊙}{{ \odot}} \nuc{⊛}{{ \textstyle \bigodot\nolimits}}
  \nuc{⊕}{{ \oplus}}
  \nuc{∗}{{ \star}} 
  \nuc{⁕}{{ \ast}} 

  \nuc{→}{{ \rightarrow}}  
  \nuc{⇉}{{ \rightrightarrows}} 
  \nuc{←}{{ \leftarrow}}
  \nuc{⟷}{{ \leftrightarrow}}
  \nuc{⇒}{{ \Rightarrow}}  
  \nuc{⇐}{{ \Leftarrow}} 
  \nuc{⇔}{{ \Leftrightarrow}} 
  \nuc{⟺}{{ \Leftrightarrow}} 
  \nuc{⇡}{{ \uparrow}}
  \nuc{⇣}{{ \downarrow}}

  \nuc{ℓ}{{ \ell}}
  \nuc{∂}{{ \partial}}
  \nuc{∞}{{ \infty}}
  \nuc{‴}{^{ \prime\prime\prime}} 
  \nuc{″}{^{ \prime\prime}} 
  \nuc{′}{^{ \prime}}
  \nuc{¬}{{ \neg}}
  \nuc{∀}{{ \forall}}
  \nuc{∃}{{ \exists}}
  \nuc{∆}{{ \triangle}} 
  \nuc{∄}{{ \nnexists}} 
  \nuc{∅}{{ \varnothing}}
  \nuc{□}{{ \square}}

  \nuc{⎨}{ \lbrace }
  \nuc{⎬}{ \rbrace }
  \nuc{⁅}{ \langle }
  \nuc{⁆}{ \rangle }

  \nuc{Ṛ}{{ \mathbb R}}
  \nuc{Ẓ}{{ \mathbb Z}}
  \nuc{⋅}{{ \ }} 
  \nuc{˙}{{\,}} 
  \nuc{˛}{{,}\mspace{.5mu}} 



\setlength{\unitlength}{1cm}

\newrgbcolor{maroon}{.5 0 0} 
\psset{gridcolor=green} 

\newcommand{\myoutergrid}[1]{\ifdraft{
  \put(-7.5,0){\color{brown} \framebox(7.5,#1){}}
  \put(-3.75,0){\color{brown} \framebox(7.5,#1){}}
  \put(0,0){\color{brown} \framebox(7.5,#1){}}} }
\psset{unit=4.3ex} 






